\documentstyle[seceq,supplement,epsf,wrapft,preprint]{ptptex}

\preprintnumber[3.5cm]{
KUCP-0104\\
KUNS 1435\\
HE(TH)97/02\\
OHU-PH 9702\\
YITP-97-12\\
hep-th/9704025}
\title{
Recent Developments of the Theory of Tunneling
}

\author{
Hideaki {\sc Aoyama},\footnote{The email addresses of the
authors are; {\sl aoyama@phys.h.kyoto-u.ac.jp,
harano@gauge.scphys.kyoto-u.ac.jp, 
kikuchi@yukawa.kyoto-u.ac.jp,
dai@phys.h.kyoto-u.ac.jp, 
msato@yukawa.kyoto-u.ac.jp,} and  
{\sl shinya} {\sl @phys.h.kyoto-u.ac.jp,} respectively.}  
Toshiyuki {\sc Harano}$^{\dagger}$,
Hisashi {\sc Kikuchi}$^{\dagger\dagger}$,\\
Ikuo {\sc Okouchi}$^{\ddagger}$,
Masatoshi {\sc Sato}$^{\ddagger\ddagger}$\footnote{Address after April 1, 1997:
{\it Physics Department, University of Tokyo, Tokyo 113, Japan}}
and 
Shinya {\sc Wada}$^{\ddagger}$}

\inst{
Faculty of Integrated Human Studies, Kyoto University, Kyoto 606-01, Japan
\\ \vspace{1mm}
$^\dagger$Department of Physics, Kyoto University, Kyoto 606-01, Japan
\\ \vspace{1mm}
$^{\dagger\dagger}$Ohu University, Koriyama 963, Japan 
\\ \vspace{1mm}
$^{\ddagger}$Graduate School of Human and Environmental Studies,\\
Kyoto University, Kyoto 606-01, Japan
\\ \vspace{1mm}
$^{\ddagger\ddagger}$Yukawa Institute for Theoretical Physics,
Kyoto University, Kyoto 606-01, Japan
}

\recdate{
\today
}

\abst{%
Path-integral approach in imaginary and complex time 
has been proven successful in treating the tunneling phenomena
in quantum mechanics and quantum field theories.
Latest developments in this field, the proper valley method
in imaginary time, its application to various quantum systems,
complex time formalism, asympton theory for the large order
analysis of the perturbation theory, are reviewed in a
self-contained manner.}

\begin{document}

\maketitle

\makeatletter
\def\@pnumwidth{1.55em}
\def\@tocrmarg {2.55em}
\def\@dotsep{4.5}
\setcounter{tocdepth}{3}

\def\tableofcontents{\section*{Contents\@mkboth{CONTENTS}{CONTENTS}}
 \@starttoc{toc}}
\def\l@part#1#2{\addpenalty{\@secpenalty}
 \addvspace{2.25em plus 1pt} \begingroup
 \@tempdima 3em \parindent \z@ \rightskip \@pnumwidth \parfillskip
-\@pnumwidth 
 {\large \bf \leavevmode #1\hfil \hbox to\@pnumwidth{\hss #2}}\par
 \nobreak \endgroup}
\def\l@section#1#2{\addpenalty{\@secpenalty} \addvspace{1.0em plus 1pt}
\@tempdima 1.5em \begingroup
 \parindent \z@ \rightskip \@pnumwidth 
 \parfillskip -\@pnumwidth 
 \bf \leavevmode \advance\leftskip\@tempdima \hskip -\leftskip #1\nobreak\hfil
\nobreak\hbox to\@pnumwidth{\hss #2}\par
 \endgroup}
\def\l@subsection{\@dottedtocline{2}{1.5em}{2.3em}}
\def\l@subsubsection{\@dottedtocline{3}{3.8em}{3.2em}}
\def\l@paragraph{\@dottedtocline{4}{7.0em}{4.1em}}
\def\l@subparagraph{\@dottedtocline{5}{10em}{5em}}
\makeatother
\tableofcontents
\newpage

\section{Introduction} 

Tunneling phenomena are unique to quantum systems
and play important roles in various fields.
To name a few of them, 
B+L violation of the standard electroweak model,
inflation in the early universe,
nuclear fission and fusion reactions, 
spontaneous nucleation in He$^3$/He$^4$ systems,
and chemical reactions are the area of immediate influence.
Although it can be readily understood in Schr\"odinger 
formalism in simple quantum mechanical systems, once
the degree of freedom increases the calculation becomes a challenge.
This situation is quite serious in field theories, which have
infinite degree of freedom.
Since the tunneling phenomena is essentially non-perturbative 
one, extracting relevant degree of freedom and doing reliable
calculation requires new formalism.

The imaginary time path-integral method has
been known to be effective in dealing with such a situation
for some time.
This is mainly because of the existence of the  classical configurations, 
instantons and bounces, around which the path-integral may be 
evaluated.\footnote{In some literature, the word ``instanton'' is
used for any the imaginary time classical configurations.
This, however, is avoided in this paper for the sake of clarity;
``instanton" and ``bounce" are distinguished as will be
seen in the rest of this paper.}
In theories with degenerate classical vacua,
one often has instanton solutions, which connects
the different (perturbative) vacua.
When the coupling is small enough, the dilute-gas approximation
for multi-instanton configurations
yields the energy splitting of the ground states by the tunneling.
Bounces are configurations that starts from the ground state
and ``bounces back" at the escape point and then comes back
to the original ground state.  A relevant analytical continuation
of the Gaussian integral around bounce solution yields 
the decay rate of the unstable states.
The power of the instanton method and the bounce method 
described so far are actually quite limited.
Here we will point out three of the problems encountered
by these methods.

The most obvious one is the fact that 
those calculations are limited only to the ground state.
In quantum mechanical context, 
if one wishes to obtain the energy splitting of the excited states in
the symmetric double-well, or if one wants the 
decay rate from higher excited states, one realizes the
current methods are not enough.
In field theoretical context, 
this difficulty became most apparent and serious
in the late 80's, when the baryon number and lepton
number violation processes in the electroweak standard 
model caught attention of some of the particle
physicists, in relation of the anticipated PP collider
in TeV region at that time.
The tunneling process induced by the colliding quarks 
results in the B+L violation through
the anomaly of the model.  Since the potential height 
is of order $M_{\rm W}/\alpha \sim $ few TeV, this process
is highly suppressed in the low energy region.
However when the initial energy of order TeV is available,
this process may occur at observable rate.
However, it was soon noticed that the usual instanton
calculation is not only invalid, but yields inconsistent
result at higher energies, violating unitarity.
Thus the necessity of the further development of the
formalism became apparent.

Another difficulty is a subtle one.
There are theories in which tunneling is physically possible,
but no classical solution exists.  
This happens for a specific potential in the scaler field theories
and in the gauge-Higgs system.
An elementary scaling argument shows that the action of any finite action
configuration can be lowered by shrinking it.
In other words, any configuration that is the 
stationary point of the action functional is point-like, which
denies any hope of naive application of the instanton formalism.

The third problem is actually a paradox.
In the quantum mechanics with asymmetric double-well potential,
the ordinary bounce method yields a decay rate.
This is due to the fact that the Gaussian integration of
the fluctuations is concerned only with the neighborhood
of the bounce solution.
On the other hand, there is no decay, since Hamiltonian
is hermitian; the energy spectrum is certainly affected
by the tunneling, but does remain real and discrete. 
Thus this instability is a ``fake''.

Motivated by these problems at various stages, 
we have developed several formalisms in the recent years.
The aim of the paper then is to present
these developments in a self-contained manner, together 
with several new results. 
We believe that the power of these formalisms
is not limited to the original contexts and thus hope to make these 
widely available to researchers in other fields as well.

The organization of this paper is as follows.
In section 2, we briefly review the B+L violation problem
in the standard model, which was the starting point of these
works.  The use of the dilute instanton-gas approximation,
its infamous ``unitarity problem", its solution by the
``interacting instanton method", the need for further
developments are explained in a concise manner.
The ultimate method in this line is ``the proper valley
method", described in section 3.  The presentation
here is in a general model, not limited to any particular
quantum mechanical or quantum field theoretical model.
Its advantage over the pre-existing methods, the streamline
method and the constrained instanton method are explained 
in detail. A numerical analysis that demonstrates the power
of the proper valley method is also presented.

The application of the proper valley method to various models
are presented in section 4.
The asymmetric double-well potential, the ``Fake instability"
problem is treated and solved in the proper valley method
in subsection 4.1.
False vacuum problem in the scalar field theory is analyzed
in subsection 4.2.
Unstable scalar field theory that does not possess any finite-size
instanton is studied in subsection 4.3, where we also compare our
results with the one from the constrained instanton method.
Similar analysis is done for the gauge-Higgs system in subsection 4.4.
Subsection 4.5 gives the application of the valley instanton
configuration obtained in subsection 4.4 to the actual
calculation of the cross section in the standard model.

In section 5 we describe a completely different approach, the
complex time method.  
The big advantage of this formalisms over the imaginary
time formalism is that the configuration goes through not only 
the forbidden region (just like instanton/bounce), 
but also the allowed region.  This implies the solution of
the fake instability problem, and calculation of the excited
states, among others.
However, the difficulty lies in the fact that the exact
formulation is not clear at all.
It is easy to construct complex-time configuration
by patching up the instanton and the real-time solutions
and then use those for evaluation by hand-waving argument.
However, just how exactly these configuration should contribute
to the path-integral remains a mystery, even
in a simple one-dimensional quantum mechanics.
What we present in this section is 
an analysis of the WKB calculation to show
that there is at least one formalism that works in the
one-dimensional quantum mechanical system.

In section 6, we discuss the problems of the perturbation theories
that arise due to the existence of the tunneling phenomena.
In models whose the perturbative series is not a Borel-summable series,
we expect that it is due to the fact that physical observables
are well defined only after the non-perturbative terms are added.
This interplay between the perturbative effects at large orders
and the non-perturbative effects has been a subject of the 
investigation for some time.  In this section we give some
of the new insights obtained recently. 
In subsection 6.1 we give a Borel analysis 
of this phenomenon using the idea of valley.
And then in subsection 6.2 we present the theory of the asympton, 
which is the configuration that fictitiously dominates 
the perturbative functional at large orders.
This provides us with an insight as to the original of
the non Borel-summability and the possible direction for the
converging calculations.

\section{Interacting instantons in the anomalous 
B and L violation} 

\def\be{\begin{equation}}
\def\ee{\end{equation}}
\def\bea{\begin{eqnarray}}
\def\eea{\end{eqnarray}}

The anomalous baryon number (B)  and lepton number (L) violation is 
one of the most fascinating processes predicted
in the electroweak standard model.
The prediction is based on the chiral anomaly and topological 
transition.
In the electroweak model the apparent conservation of
baryon number current \(j_B^\mu\) and the lepton number current \(j_L^
\mu\)
is, in fact, no longer valid due to the anomaly.
Their divergence  is  deduced from
the representation of the quarks or leptons under SU\(_{\rm L}\)(2) 
and 
U\(_{\rm Y}\)(1) group.
Since we have, in each generation, 
three SU(2) doublets that has Y = 1/3 for 
the left-handed quarks and six singlets with Y = 4/3  or $-$2/3
for the right-handed ones, the divergence of \(j_B^\mu\) is
\be
\partial_\mu j_B^\mu = {N_{\rm f}\over 32 \pi^2}
[g^2 W^a_{\mu\nu} \tilde W^{a\mu\nu} - g'^2 B_{\mu\nu} \tilde B^
{\mu\nu}], 
\label{eq:s2divjb}
\ee
where 
\bea
W^a_{\mu\nu} & = & \partial_\mu W^a_\nu - \partial_\nu W^a_\mu
- g \epsilon^{abc} W^b_\mu W^c_\nu,\\
B_{\mu\nu} & = & \partial_\mu B_\nu - \partial_\nu B_\mu ,
\eea
are the field strength of the SU(2)  \(W^a_\mu\) and
U(1)  \(B_\mu\) gauge fields, respectively;
\(g\) and \(g'\) are their coupling constants;
\(\tilde W^{a\mu\nu} \equiv (1/2)\epsilon^{\mu\nu\lambda\sigma} 
W^a_{\lambda\sigma}\),  and 
the \(N_{\rm f}\) is the number of the generation.
Interestingly the divergence of \(j_{\rm L}^\mu\)  is  exactly the 
same. Therefore the (B$-$L) current stays conserved.

The topological transition resides in the  SU(2) sector in the
model. Consider a trajectory 
in the finite energy configuration space
made of \(W^a_\mu(x)\) and the doublet Higgs \(\Phi(x)\) that 
starts from and ends up with the vacuum. 
We parametrize this as 
\[ W^a_\mu(0, {\bf x}) = W^a_\mu(T, {\bf x}) = 0,  \quad \Phi(0, {\bf 
x})
= \Phi(T, {\bf x}) = \left( \begin{array}{c} 0 \\ v\end{array}\right), 
\]
using a parameter \( t \in [ 0, T] \),
where \(v\) is the vacuum expectation value.
All the trajectories of this type are divided into topological classes 
that 
are labeled by  the winding number of the homotopy \(\Pi_3(S^3)\).
The integral along the trajectory,
\be 
N_{\rm w} \equiv  {g^2\over 32\pi^2} \int dt\, d^3x\, W^a_{\mu\nu} 
\tilde W^{a\mu\nu}, 
\ee
gives its winding number.
Combined with  Eq.~(\ref{eq:s2divjb}), this relation indicates that
the transition along the trajectory causes the 
B and L violation  by \(N_{\rm f} N_{\rm w} \) units.
 
The height of the potential barrier that the 
topological transition with \(N_{\rm w}= \pm 1\)
need to go over is of the order of \((4\pi/g^2) m_{\rm W}\).
This is the static energy \(E_{\rm sp}\) of the sphaleron,  
a static but unstable solution
of the equations of motion.\cite{rf:manton}
This large energy causes a strong suppression if the transition  takes
 place through 
quantum tunneling starting from the low-lying states near 
the vacuum.
Actually the WKB suppression factor in the amplitude 
is estimated\cite{rf:thooft} to be \(e^{-8\pi^2/g^2}\simeq e^{-180}\).

A naive expectation, however, leads us to the interesting possibility
that this strong suppression may be weakened if the energy of
the transition becomes comparable to \(E_{\rm sp}\).
Along this line, two issues have been discussed in literature;
the effect of the B and L violation on the baryogenesis of the
Universe\cite{rf:kuz} and the direct detection of the violation 
at a high energy collision.\cite{rf:ag}

Only  reliable estimates of the transition amplitudes in 
high energy processes can answer the question about these issues.
This is not, however, a simple task.
The difficulty lies in the fact that it is
non-perturbative: the  process
\be q + q \rightarrow 7 \bar q + 3 \bar l,  \ee
or, more generally, the one accompanied by the simultaneous
production of \(n_{\rm W}\) W bosons and \(n_{\rm H}\) Higgs bosons,
\be q + q \rightarrow 7 \bar q + 3\bar l + n_{\rm W} W + n_{\rm H}
\phi,  \label{eq:s2process}\ee
is the specific example of the transition,
but we cannot write down the relevant diagrams for this process
by using the Feynman rules in the electroweak standard model.

't Hooft is the first person to have estimated the 
amplitude.\cite{rf:thooft}
His tool was the imaginary time formalism of the path-integral; 
it can incorporate the topological transition  as the background
in the path-integral. 
Ringwald and Espinosa pushed the idea further;
they applied the path-integral to the evaluation of 
Green functions appropriate for
the scattering process (\ref{eq:s2process}).\cite{rf:ringwald,rf:espinosa}
Aoyama and Kikuchi improved their analysis by taking into account
multi-transitions;\cite{rf:ii}
this can resolve the unitarity problem pointed out  
in their single transition calculation.
The following of this section is a review of
this non-perturbative calculation according to the line mentioned 
above.
We are still at the middle of the way to the final answer and
a lot of related works have been published.
We are not going to look back all of these works here.
(For the other related works, see, for example, Ref.~\citen{rf:rubsha}.)

Let us start with Ringwald and Espinosa  calculation.
For simplicity, we take the SU(2)-Yang-Mills-Higgs-model with
a single generation of massless quarks and leptons.
All the essence of the calculation can be seen in this simple version.
We intend to evaluate the  B and L violating Green functions of 
the Heisenberg operator,
\bea
\lefteqn{ G(x_1,..., x_4; y_1,..., y_{n_W}; z_1,...,z_{n_H}) } 
\nonumber\\
&= &{ \langle {\rm vac}| q_1(x_1) q_2(x_2) q_3(x_3) l(x_4) 
W(y_1) ... W(y_{n_W}) \phi(z_1)...\phi(z_{n_H})
|{\rm vac} \rangle\over \langle {\rm vac}|{\rm vac} \rangle },
\label{eq:s2Gdef}
\eea
where \(q_1(x)\), \(q_2(x)\), and \(q_3(x)\) are SU(2) doublet quarks 
distinguished by their color,   \(l(x)\) the doublet lepton,
\(\phi(x)\) the physical Higgs boson.
We have  suppressed  the Lorentz and 
the SU(2) indices in Eq.~(\ref{eq:s2Gdef}).
We use the imaginary time formalism for the calculation, i.e.
the time evolution operator is  \(e^{-\tau H}\) (\(H\) is the 
Hamiltonian) 
instead of \(e^{-it H}\).
This Green function can be evaluated also in the path-integral
\bea 
\lefteqn{ G(x_1,..., x_4; y_1,..., y_{n_W}; z_1,..., z_{n_H}) } 
\nonumber\\
& = & {1\over Z} \int 
\prod_{f = q_1, q_2, q_3, l}[  df\,d\bar f ] [dW] [d\Phi] 
\exp\left\{ - S[ W, \Phi, q, l ]\right\}\nonumber\\
 && \times q_1(x_1) q_2(x_2) q_3(x_3) l(x_4) W(y_1) ...W(y_{n_W})
 \phi(z_1) ... \phi(z_{n_H}),
\label{eq:s2Gpath}\\
Z & = & \int 
\prod_{f = q_1, q_2, q_3, l}[  df\,d\bar f ] [dW] [d\Phi] 
\exp\left\{ - S[ W, \Phi, q, l ]\right\},
\eea
where \(S\) is the sum of the terms \(S_{\rm W}\),
\(S_\phi\), and \(S_{\rm f}\), evaluated respectively from the
Lagrangians,
\bea
{\cal L}_{\rm W} & = & {1\over 4} \left( W^a_{\mu\nu} \right)^2,\\
{\cal L}_{\phi} & = & \left| D_\mu \Phi \right |^2 + \lambda \left(
\left| \Phi \right|^2 - v^2 \right),\\
{\cal L}_{\rm f} & = & \sum_{\rm c}\bar q_{\rm c} D_\mu \bar 
\sigma_\mu 
q_{\rm c}
+ \bar l D_\mu \bar \sigma_\mu l.
\eea
In these equations, \(D_\mu\) is  the covariant derivative;
\be D_\mu = \partial_\mu + i g \left({\tau^a\over 2}\right) W^a_\mu, 
\ee
and \(\bar\sigma_\mu \equiv ( \vec\sigma, - i) \) is 
the two component Pauli matrices for the left-handed spinors%
\footnote{We use the  relation \(t = -i \tau\) of the imaginary time 
\(\tau\) and the real time \(t\) when we need to go back 
to Minkowski expression. Using the  other one \(t = i \tau\) as
in Refs.~\citen{rf:ii,rf:clec} is just the matter of convention.}.
Hereafter, the Greek indices are understood as being summed up in 
the Euclidean metric.

The usual perturbative calculation of the function (\ref{eq:s2Gpath}) 
around the vacuum is trivially zero.
We instead evaluate it in the background of the topological 
transition.
The anomaly equation (\ref{eq:s2divjb}) indicates it
should have \(|N_{\rm W}| = 1\).
In the standard model, we do not have a stationary configuration
of the action around which we construct an systematic expansion of
the Green function in powers of \(g\): 
any finite size configuration tends to shrink toward zero size.
To circumvent this,  we use the constrained
 instanton method.\cite{rf:Aff}
We first restrict the functional space by imposing the constraint
on the size \(\rho\) and evaluate
the path-integral over all fluctuations around the stationary
configuration in this restricted functional space.
At the end of the calculation, we integrate over  \(\rho\) to recover 
the 
whole functional space.
The constrained instanton is the name for the topological transition
stationary in the restricted functional space.
Hereafter in this section we simply call it instanton.
The bosonic component of the instanton in the
singular unitary gauge has the following form
\bea
{W_{\rm in}}^a_\mu & \simeq & {1\over g} \eta_{a\mu\nu}
\partial_\nu \ln \left( 1+ {\rho^2\over x^2} \right) \label
{eq:s2Win1},\\
\Phi_{\rm in} & \simeq & \left( {x^2\over x^2 + \rho^2}\right)^{1/2}
\left( \begin{array}{c} 0 \\ v\end{array}\right),
\eea
in the central region \( x \ll m_{\rm W}^{-1}, m_{\rm H}^{-1} \), and
\bea
{W_{\rm in}}^a_\mu & \simeq & {4\pi^2\rho^2 \over g} \eta_{a\mu\nu}
\partial_\nu G_{m_{\rm W}}(x) \label{eq:s2Win2},\\
\Phi_{\rm in} & \simeq & \left( 1 - 2\pi^2 \rho^2 G_{m_{\rm H}}(x) 
\right)
\left(\begin{array}{c} 0 \\ v\end{array}\right), \label{eq:s2Phiin2}
\eea
in the asymptotic region \( x \gg \rho \),
where we have used \(G_m\) for the four dimensional massive 
propagator,
\be (-\partial^2 + m^2) G_m(x) = \delta^4(x), \ee
and the 't Hooft symbol \(\eta_{a\mu\nu}\),
\be
\eta_{a\mu\nu} =  \cases{
\delta_{a\nu} & if $\mu = 4$;\cr
-\delta_{a\mu} & if $\nu = 4$; \cr
\epsilon_{a\mu\nu} & otherwise.}
\ee
Its action is approximately
\be S_{\rm in} \simeq {8\pi^2\over g^2} + \pi^2 \rho^2 v^2. \ee
In this background, the operator \(D\bar\sigma\) has a zero mode 
\(\psi^{(0)}\);
\be D_\mu\bar\sigma_\mu \psi^{(0)}(x) = 0. \label{eq:s2psi0}\ee
In the explicit notation with \(\alpha\) for the Lorentz index and
\(a\) for the SU(2) index,  it behaves as
\be \psi^{(0)}_{\alpha a}(x) \simeq 
{\rho\over \pi}{x_\mu (\sigma_\mu)_{\alpha \beta}
\epsilon_{a \beta} \over |x| (x^2+\rho^2)^{3/2} }
\label{eq:s2zeromode} \ee
at the central region and
\be \psi^{(0)}_{\alpha a}(x) 
\simeq (2\pi \rho)G_{{\rm F}\, \alpha \beta}(x)\epsilon_{\beta a} \ee
at the asymptotic region, where 
\(\sigma_\mu = ( \vec\sigma, i) \),
\be G_{{\rm F}\,\alpha\beta}(x) \equiv 
\partial_\mu (\sigma_\mu)_{\alpha\beta}
 G_{m=0}(x) \ee
is the fermionic  massless propagator, and 
\(\epsilon_{\beta a}\) is the two dimensional anti-symmetric matrix.
For the path-integral of the fermionic degrees of freedom,
we expand the fermion fields, collectively denoting them by \(f\), as
\bea f(x) & = & \sum_n \psi^{(n)}(x) a^f_n, \\
\bar f(x) & = & \sum_n \varphi^{(n)}{}^\dagger(x) \bar a^f_n,\eea
with Grassmann coefficients \(a_n\) and \(\bar a_n\) and 
eigenfunctions \(\psi^{(n)}\) and \(\varphi^{(n)}\); 
they obey eigenvalue equations
\bea
- D_\mu \sigma_\mu D_\nu\bar\sigma_\nu \psi^{(n)} & = & \lambda_n^2 
\psi^{(n)} \label{eq:s2eigenpsi},\\
- D_\mu \bar\sigma_\mu D_\nu\sigma_\nu \varphi^{(n)} & = & \lambda_n^2 
\varphi^{(n)}, \label{eq:s2eigenphi}
\eea
and are normalized as follows;
\be \int d^4x \psi^{(n)}{}^\dagger (x) \psi^{(m)}(x) = \delta_{m n},
\quad \int d^4x \varphi^{(n)}{}^\dagger (x) \varphi^{(m)}(x) = \delta_
{m n}.\ee
Note \(\psi^{(0)}\) is the one with \(\lambda = 0\).
The fermionic action \(S_{\rm f}\)  becomes diagonalized,
\be S_{\rm f} = \sum_{n\neq 0,\; f } \lambda_n
\bar a^f_n a^f_n, \label{eq:s2diaSf}\ee
because the nonzero modes of \(\varphi^{(n)}\) and \(\phi^{(n)}\) are
related  one-to-one,
\be \varphi^{(n)} = {1\over \lambda_n}(\bar\sigma_\mu D_\mu)
\psi^{(n)}. \ee
The zero mode \(\psi^{(0)}\) does not have the partner and
its coefficients \(a^f_0\) do not appear in Eq.~(\ref{eq:s2diaSf}).
The fields \(q_1\), ... \(l\) put in the integrand of Eq.~(\ref
{eq:s2Gpath})
provide a proper number of 
\(a^f_0\) for the non-vanishing Green function 
against the path-integral measure;
\be [df d\bar f] \propto da^{q_1}_ 0 da^{q_2}_ 0da^{q_3}_ 0da^l_0. \ee

We now have  all the necessary ingredients to evaluate 
the Green function (\ref{eq:s2Gpath}).
The fermion fields in the integrand  are now replaced with the 
zero mode.
For the bosonic fields, we can just put the background instanton,
Eqs.~(\ref{eq:s2Win1}--\ref{eq:s2Phiin2}),
to the leading order in \(g\):
the nonzero mode fluctuation around the background
will give smaller contribution 
as long as \( n_{\rm W}\) or \( n_{\rm H}\) is not so large%
\footnote{The nonzero mode fluctuations of W and \(\phi\) 
are of the order of \(g\) compared with the background. 
The next leading correction is obtained by replacing any two of 
the background with a propagator in the background;
there are about \( n_{\rm W}^2\) ways to pick up the two of W and 
\( n_{\rm H}^2\) ways for \(\phi\).
It is thus of the order of  \(n_{\rm W}^2 g^2\) or  \(n_{\rm H}^2 g^
2\) 
compared with the leading term.} 
and zero mode fluctuations, corresponding to the translation and
the SU(2) isospin rotation, can be taken care of  collective 
coordinate integral as usually done (see the Eq.~(\ref{eq:s2vertex}) 
below.)
The path-integral is carried out  by the Gaussian approximation
and results in the determinants from the bosonic and fermionic nonzero 
mode fluctuations.
We then perform the Fourier transformation on the resulting Green 
function \(G\)  to use the LSZ procedure to extract the proper vertex.
The external momenta have the magnitude of 
the particles masses and are smaller than the dominant instanton size
\( \rho \sim 1/v\).
Noticing that both the fermionic zero mode and the bosonic 
configuration 
of the instanton have the form of the propagator at the asymptotic 
region,
we obtain 
\be  \Gamma = \int {d\rho\over \rho^5} e^{-S_{\rm in}(\rho)}
\int dR \rho^2 D(\rho) \left[\prod_{f} \chi_{\alpha_f a_f}\right]
\left[\prod_{j=1}^{n_{\rm W}} (\beta^{a_j}_{\mu_j\nu}
ip_{j \nu})\right](- 2\pi^2\rho^2 v)^{n_H} \label{eq:s2vertex}\ee
for the vertex, where
\bea
\beta_{\mu\nu}^a &\equiv& {4\pi^2\rho^2\over g}R^{ab}\eta_{b\mu\nu},
\label{eq:s2beta}\\
\chi_{\alpha a} &\equiv& 2\pi\rho U_{ab}\epsilon_{\alpha b}, 
\label{eq:s2chi}
\eea
and \(p_j\) is the momentum of the $j$-th external W boson.
The integral measure \(dR\) represents  the functional integral
over the zero mode of the isospin rotation;
the corresponding variables \( R^{ab} \) and \(U_{ab}\) 
are related by \(U\tau^b U^\dagger = \tau^a R^{ab}\).
The integral over the translational zero mode has been done already
when we Fourier-transformed the Green function to arrive at
Eq.~(\ref{eq:s2vertex}), and  it provides the momentum conservation 
at the vertex.
The factor \(\rho^{2-5} D(\rho)\) is the product of the Jacobians for
the bosonic collective coordinates and the 
determinants of the bosonic and fermionic nonzero modes.

The main outcome of the above Ringwald and Espinosa analysis is 
that the so-obtained vertex (\ref{eq:s2vertex}) 
is point-like  independently of 
the momenta of external particles.
Thus the cross section  increases monotonically
as \(\sim E^{(2 + 2 n_{\rm W} + 2 n_{\rm H})}\)
as the energy \(E\) of the process increases.
It would eventually  violate the unitarity bound if one naively
extrapolate the result to high energy.
One may think that we could  predict nothing about the high energy
behavior of the amplitude from the calculation 
we have done above:
We have just used the asymptotic form of the instanton, while
it is the form at the  central region that determines
the  behavior at \(E > 1/\rho \sim v\).
This very plausible argument, however, is false.
As we have seen, the Green function to the leading order in \(g\) 
is basically the product of the instanton configurations.
The Lorentz invariance then forces its Fourier transform to
depend on the simple momentum squares \(p_i^2\) of the external 
particle, not on \(p_i\cdot p_j\) for example.
Then we can use the asymptotic forms since we will put the on-shell
condition \(p_i^2 = m^2\) after all when we apply the LSZ procedure.

The single instanton never disappears from
the functional space no matter how large we take the energy 
in the calculation.
Therefore, there must be other configurations that cancel it,
becoming also the same order of magnitude at the energy
of the unitarity violation.
Aoyama and Kikuchi have shown that the multi-instanton 
configurations take this task.\cite{rf:ii}
The interaction between the instantons plays the
important role in this cancelation.
With the interactions the instantons 
construct ``connected diagrams'' for  the
scattering  amplitude and in fact they have the property
that is required to recover the unitarity.
Let us now turn to this part.

We investigate the contribution of the 
\(n\)-instanton--\(\bar n\)-anti-instanton configuration,
\bea
W(x) & = & \sum_{i = 1}^n W_{\rm in}(x - w_i, R_i) + 
            \sum_{j=n+1}^{\bar n + n} \bar W_{\rm in}(x - w_j, R_j), 
\label{eq:s2multiconfW}\\
\phi(x) & = & \sum_{k = 1}^{\bar n + n} \phi_{\rm in}(x - w_k), 
\label{eq:s2multiconf}
\eea
to the Green function. 
In Eq.~(\ref{eq:s2multiconfW}), \(\bar W_{\rm in}\) represents 
the anti-instanton, whose form is
given by replacing \(\eta_{a\mu\nu}\) 
in Eqs.~(\ref{eq:s2Win1}) and (\ref{eq:s2Win2})
with  \(\bar \eta_{a\mu\nu}\) (\(\bar\eta_{a\mu\nu}\) is
minus of \(\eta_{a\mu\nu}\) for \(\mu = 4\) or \(\nu = 4\),
 otherwise the same.)
The parameters \(w_i\) and \(R_i\) are the location and isospin-
orientation 
of the i-th (anti-)instanton.
We intend to treat the parameters
\(w_i\), \(R_i\), and \(\rho_i\) as the collective coordinates
for the path-integral.
In the strict sense, we should extract out the canonical variables
for the fluctuations at each values of the collective coordinates
and carry out first the functional integral over the fluctuations, and
then integrate the resulting effective action over them.
This is, however,  impossible in practice.
We adopt two approximations.
First, we approximate the functional integral over the fluctuations
to the simple product of the determinants of each (anti-)instanton.
Although the configuration  is  not  stationary in any sense,
we do not expect this will cause a large error since the fluctuations
are suppressed by \(g\) (the coupling constant) compared with the 
collective coordinates.
Second, we approximate the action to the sum of the instanton
action \(S_{\rm in}\) of each (anti-)instanton and two-body
interaction of each pair evaluated at large separation.
These approximations are quite adequate if the relative 
distances \(|w_i-w_j|\) are larger than \(m_{\rm W}^{-1}\)
or \(m_{\rm H}^{-1}\),
but if some of them become smaller, we will miss some part of 
the exact contribution.
It will turn out that the contribution taken into account by 
the above procedure is sufficient to show the cancelation
of the single instanton contribution at the energy of unitarity 
violation.

The bosonic part of the interaction between 
the \(i\)-th instanton and \(j\)-th anti-instanton is defined by
\be S_{\rm b:int}(i,j) \equiv S[ W(i) + \bar W(j), \phi(i) + \phi(j)]
- 2 S_{\rm in} \ee
and readily evaluated by using the fact that the each 
background configuration satisfies  the equation of motion
in  the asymptotic region (see Ref.~\citen{rf:ii} for detail);
\bea S_{\rm b: int}(i,j) &=& - \beta^a_{\nu\lambda}(\rho_i, R_i)
{\partial \over \partial w_{i\lambda}}\bar\beta^a_{\nu\sigma}(\rho_j, 
R_j)
{\partial \over \partial w_{j\sigma}}G_{m_{\rm W}}(w_i - 
w_j)\nonumber\\
&&- (2\pi^2\rho_i^2 v)(2\pi^2\rho_j^2v) G_{m_{\rm H}}(w_i - w_j),
\label{eq:s2bint}
\eea
where \(\bar\beta\) is defined by (\ref{eq:s2beta}) with
\(\bar\eta\) substituting \(\eta\).
The interaction term for a pair of instantons or  anti-instantons is
written in a similar equation in which \(\beta\) or \(\bar\beta\)
are replaced correspondingly.
Regarding the fermionic part, the zero mode located at each instanton 
has now overlap with the other, which  also generates the
interaction. 
Starting from the expansion (we suppress the suffix 0 for the 
zero-modes
hereafter),
\bea 
f(x) & = & \sum_i a^{fi}\psi(x - w_i) + \mbox{(nonzero modes)}\\
\bar f (x) & = & \sum_j \bar a^{fj}\varphi^\dagger(x - w_j) 
+ \mbox{(nonzero modes)},
\eea
we obtain
\bea
S_{\rm f: int}(i,j) & \equiv & \bar a^{fj} a^{fi} 
\int dx\, \varphi^{\dagger}(j)
\bar\sigma_\mu \left(\partial_\mu + ig W_\mu(i) + ig \bar W_\mu(j)
\right)\psi(i) \nonumber\\
&= &  \bar a^{fj} a^{fi} 
\int dx\, \varphi^{\dagger}(j)(- \partial_\mu)\bar\sigma _\mu 
\psi(i) \nonumber\\
&= & - \bar a^{fj} a^{fi} \bar\chi_{a\alpha}G_{{\rm F}\alpha\beta}
(w_j - w_i)\chi_{\beta a}, \label{eq:s2fint}
\eea
where \(\bar\chi_{a \alpha} \equiv 2\pi\rho U^\dagger_{ba}
\epsilon^\dagger_{b\alpha}\) comes from the
fermionic zero mode \(\varphi\) at the anti-instanton
(its form is Eq.~(\ref{eq:s2zeromode}) with \(\bar\sigma\) for
\(\sigma\).)
We have also used the fact that the zero modes satisfy 
Eq.~(\ref{eq:s2psi0}) to go from the first to the second line.
Note that the interactions (\ref{eq:s2bint}) and (\ref{eq:s2fint})
have  the one-particle exchange type with the vertex exactly the 
same as the single instanton one for the relevant particle
production.  

\begin{wrapfigure}{c}{14cm}
\centerline{\epsfxsize=11cm \epsfbox{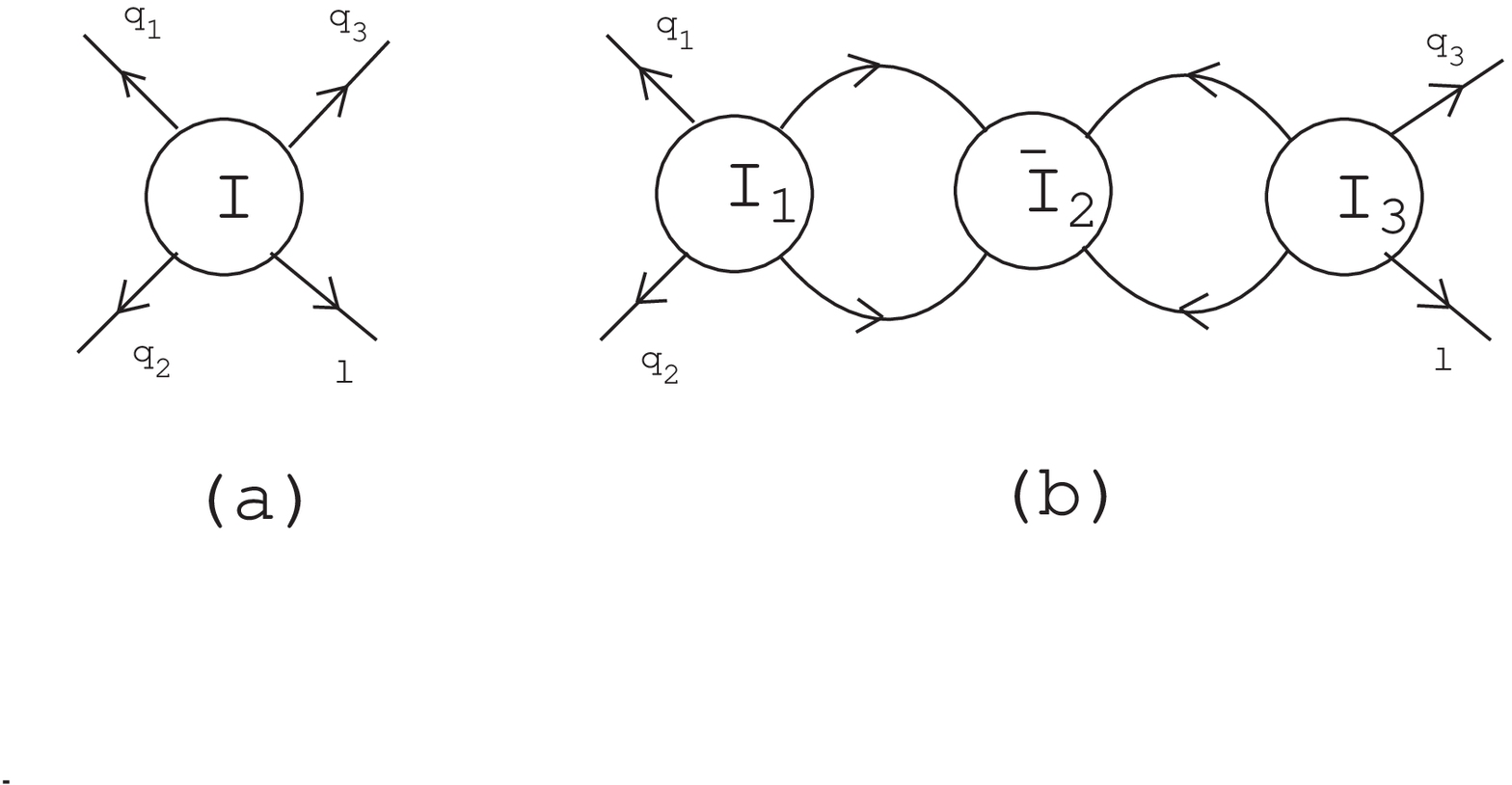}}
\vspace{-17mm}
\caption{The diagrams in the instanton expansion 
of the process \(\bar {q_1} \bar {q_2} \rightarrow q_3 l \).
}
\label{fg:s2fig}
\end{wrapfigure}

These interacting multi-instantons 
 contributes to the B and L violating Green function.
As an example let us see the contribution of two instantons,
labeled  1 and 3, and one anti-instanton, 2, to the process
\(\bar {q_1} \bar {q_2} \rightarrow q_3 l\) (See Fig.~\ref{fg:s2fig} 
(b).)

\vspace{5mm}

\vspace{-10mm}
The corresponding Green function is 
\bea \lefteqn{ G^{(3)}(x_1, ..., x_4) } \nonumber\\
& = & {1\over Z} \int [\prod_f dfd\bar f][dW][d\phi]
q_1(x_1)...l(x_4) \exp \left\{ - 3 S_{\rm in} 
- \sum S_{\rm b:int}(i,j)
- \sum S_{\rm f:int}(i,j)\right\}.\nonumber\\
\eea
To evaluate this,  we  expand the factor \(e^{ - S_{\rm int}}\) in 
powers
of \(S_{\rm int}\) and pick up nonzero terms.
The simplest one in \(G^{(3)}\) is
\bea G^{(3)} &\simeq& \left[ \prod_{i = 1}^3 \int {d\rho_i\, dR_i \, 
dw_i 
\over \rho_i^{5-2}} D(\rho_i)e^{ - S_{\rm in}(i)} \right]
\left[ \prod_{\rm f} \int da^{f 1}  d\bar a^{f 2} da^{f 3}\right]
\nonumber\\
&& \times a^{q_1 1} \psi(x_1 - w_1) a^{q_2 1} \psi(x_2 - w_1)
a^{q_3 3} \psi(x_3 - w_3 ) a^{l 3} \psi(x_4 - w_3) \nonumber\\ 
&& \times \bar a^{q_3 2} a^{q_3 1} \bar\chi(2) G_{\rm F}(w_2 - w_1) \chi(1)
\bar a^{l 2} a^{l 1} \bar\chi(2) G_{\rm F}(w_2 - w_1) \chi(1) 
\nonumber\\
&& \times \bar a^{q_1 2} a^{q_1 3} \bar\chi(2) G_{\rm F}(w_2 - w_3) \chi(3) 
\bar a^{q_2 2} a^{q_2 3} \bar\chi(2) S_{\rm F}(w_2 - w_3) \chi(3)
\nonumber\\
& \simeq & \left[ \prod_{i = 1}^3 \int {d\rho_i\, dR_i \, dw_i 
\over \rho_i^{5-2}} D(\rho_i)e^{ - S_{\rm in}(i)} \right] \nonumber\\
&&\times \left( G_{\rm F}(x_1 - w_1)\chi(1) \right)
\left( G_{\rm F}(x_2 - w_1)\chi(1) \right)
\left( G_{\rm F}(x_3 - w_3)\chi(3) \right)
\left( G_{\rm F}(x_4 - w_3)\chi(3) \right) \nonumber\\
&&\times \left( \bar\chi(2) G_{\rm F}(w_2 - w_1) \chi(1)\right)
\left( \bar\chi(2) G_{\rm F}(w_2 - w_1) \chi(1)\right)\nonumber\\
&& \times \left( \bar\chi(2) G_{\rm F}(w_2 - w_3) \chi(3)\right)
\left( \bar\chi(2) G_{\rm F}(w_2 - w_3) \chi(3)\right).
 \eea
This function has the same structure as the one obtained from 
the diagram (b) in Fig.~\ref{fg:s2fig} with the ``Feynman rule",
the vertices are understood as \(\Gamma\) in (\ref{eq:s2vertex}) or
\(\bar\Gamma\) defined for the anti-instanton with
the proper replacements of \(\beta\) and \(\chi\), and 
the lines are the free  propagators.
The similar analysis can be extended  for the other terms that contain
 any powers of \(S_{\rm int}\).
This way we will get the instanton expansion for the Green function.
The Green function of a given process gets the contribution from
all possible multi-instanton sectors in which \(n - \bar n\) is fixed
by the degree of B and L violation.
Each term in the given sector is obtained by expanding
\(e^{ - S_{\rm int}}\) in \( S_{\rm int}\).

We can show the terms generated by the instanton expansion 
is exactly the same as  those generated by the following
effective Lagrangian;\cite{rf:ii}
\bea L & = & \sum_{f} \bar f \partial_\mu \bar\sigma_\mu f + {1\over 
2}
[(\partial_\mu \phi)^2 + m_{\rm H}^2 \phi^2] \nonumber\\
&&+ {1\over 4} (\partial_\mu W^a_\nu - \partial_\nu W^a_\mu)^2 + 
{1\over 2} {m_{\rm W}}^2 W^a_\mu{}^2
+ V + \bar V, \eea
where
\bea V & = & \int {d\rho dR D(\rho)\over \rho^{5-2}} 
\left[\prod_f \bar f(x) \chi \right]
\exp[ - S_{\rm in} + ( - 2\pi^2\rho^2 v) \phi(x) +
\beta^a_{\mu\nu}\partial_\mu W^a_\nu ], \nonumber\\
\bar V & = & \int {d\rho dR D(\rho)\over \rho^{5-2}} 
\left[\prod_f \bar \chi f(x) \right]
\exp[ - S_{\rm in} + ( - 2\pi^2\rho^2 v) \phi(x) +
\bar\beta^a_{\mu\nu}\partial_\mu W^a_\nu ].
\eea
The single instanton amplitude now turns  out to be
the leading term of this expansion.
If one moves back to the Minkowski metric, this Lagrangian
becomes hermitian.
This indicates the non-leading terms  in this expansion give 
the contribution that cancel exactly the unitarity violating 
high energy behavior of the single instanton amplitude.
We can use for example the K-matrix method to show this 
cancelation.\cite{rf:ii} 
[The K-matrix method provides us with a systematic way to
pick up the diagrams in  different orders of the instanton expansion
so that their sum holds the unitarity.
In the leading order of  the K-matrix method, 
the internal momenta of loop diagrams are set on-shell
and we can see the  cancelation without suffering from
the ultra-violet divergence that the instanton expansion
has in the non-leading terms.]
In this respect, the unitarity problem that exists in the single 
instanton sector is  resolved.

When returning back to the initial problem, 
 the reliable estimate of the B and L violating
amplitude, we note that the instanton expansion
constructed so far is not sufficient.
Once one wants to evaluate the amplitude to the non-leading
orders, he encounters pathological ultra-violet divergences 
coming from the loop integrals of internal particles
that connect the instanton vertices.
The cure must come from the more rigid evaluation
of the action of the multi-instanton configuration.
Remember the approximation we adopt for the calculation of the 
interaction is  adequate only for well-separated instantons,
not for those with  small separations.
The interaction in the latter configuration 
must be important for the loop integrals of large momenta.
We believe the proper valley method that is introduced and
explained in the following sections is the most promising  to this 
end.
To choose the proper collective coordinates for  the functional
integral is indispensable in the multi-instanton calculation
of the path-integral.
The proper valley method has the potential to tell us the way 
even for those configurations in which a small-separated 
instanton--anti-instanton pair is merging into the vacuum.

\section{Proper valley method}\label{sec:NV}

\def\shiki#1#2{\begin{eqnarray} #1 \label{eq:#2} 
        \end{eqnarray}}
\def\eqa#1{(\ref{eq:#1})}

In order to describe the 
proper valley method in the most general context, we
will discretize all the bosonic degrees of freedom and 
denote then by $\phi_i$.  The subscript $i$ stands for
coordinate labels in quantum mechanics, particle species,
vector and tensor indices and any internal quantum numbers,
as well as the space-time coordinates $x_\mu$.
Translation to the continuum indices is trivial; sums over indices
should be understood as integrals with suitable measure, the 
derivatives as functional derivatives.\footnote{Inclusion 
of fermionic degree of freedom is done
in the background of the valley configurations, as will
be seen later in Section 4.}
These variables $\phi_i$ are the ``coordinates'' of the
functional space, which allows some intuitive understanding
of the equations.
The bosonic action is written as a function of these variables;
$S = S(\phi_i).$
In this notation, the equation of motion is written as
$\partial_i S =0$, where 
$\partial_j \equiv \partial / \partial \phi_j$.
We assume that the metric of the functional space is
trivial in these variables.  Otherwise, the metric should be
inserted in the following equations in a straightforward manner.

The proper valley equation\footnote{This method in the manner
described here was proposed by two of the current authors,
H.~Aoyama and K.~Kikuchi in Ref.~\citen{rf:pvalley}.
Later they found that the same equation had been written down
by P.~G.~Silvestrov in Ref.~\citen{rf:silv},
and well before that (in different contexts) in 
Ref.~\citen{rf:originals}. 
(Some very early literatures of general discussion of the
contour-lines in geological context 
are found in Ref.~\citen{rf:oldoriginals}.)
Furthermore, this method is apparently useful 
in the theoretical chemistry.\cite{rf:quapp}
The authors thank Dr.~Silvestrov, Dr.~Quapp 
and S.~Takagi at Tohoku University for letting us know
of these references.}
is the following;
\shiki{
D_{ij} \partial_j S = \lambda \partial_i S, 
}{nvmdef}
where summation over the repeated indices are assumed
and 
\shiki{
D_{ij} \equiv \partial_i\partial_j S.
}{ddef}
Since \eqa{nvmdef} has a parameter $\lambda$, it defines a 
one-dimensional trajectory in the $\phi$-space.
(The cases when we have a finite region with $\lambda=0$
can be treated without any problems. 
This will be explained later in this section.
For the moment, we assume $\lambda \ne 0$.)
The solution of the equation of motion apparently
satisfies the proper valley equation \eqa{nvmdef}.
In this sense the proper valley equation is an
extension of the (field) equation of motion.

According to the proper valley equation \eqa{nvmdef}, 
the parameter $\lambda$ in the right-hand side of
\eqa{nvmdef} is one of the eigenvalues of the matrix $D_{ij}$.
Therefore the proper valley equation requires that the 
``gradient vector'' $\partial_j S$ be parallel to the
eigenvector of $D_{ij}$ with the eigenvalue $\lambda$.
A question arises which eigenvalue of $D_{ij}$ we should 
choose for \eqa{nvmdef}.
As we will show later, the proper valley method 
converts the eigenvalue $\lambda$ to a collective coordinate,
by completely removing $\lambda$ from the Gaussian integration and 
introducing the valley trajectory parameter instead.
Thus the question is which eigenvalue ought to be 
converted to a collective coordinate for the given theory.
This really depends on the theory in question.
Therefore we will address this issue later as we investigate
various models.
Here we will simply mention that 
the general guideline is to choose $\lambda$ to be the
eigenvalue with the smallest absolute value, {\it i.e.,} the 
pseudo-zero mode, or the negative eigenvalue, 
for which the Gaussian integration converges badly or diverges.  
In some models, the eigenvalue changes from a negative value to
(still negative and) pseudo-zero value along the valley.
In such a case, proper valley equation will be shown to 
be able to deal with both types of eigenvalues.

The proper valley equation \eqa{nvmdef} can be interpreted 
within a framework
of the variational method: Let us rewrite it as the following;
\shiki{
\partial_i \left(
{1 \over 2}\left( \partial_j S \right)^2 - \lambda S 
\right) =0.
}{lagra}
This allows an interpretation 
that the norm of the gradient vector is extremized
under the constraint $S =$ const, where $\lambda$ plays the
role of the Lagrange multiplier.
Since such a point is found at each hypersurface of constant action,
the solutions of the proper valley equation form 
a line in the functional space.
This is an alternative explanation for the existence of the valley line. 
In addition, we require that the norm be {\sl minimized}.
Since the valley with maximum norm is disjoint from the one
with minimum norm, this can be easily done by choosing the
appropriate valley line.
We are therefore defining the valley to be the trajectory 
that is tangent to the most gentle direction.
This is a plausible definition, suitable for the word ``valley''.

\begin{wrapfigure}{c}{14cm}
\centerline{
\epsfxsize=10cm
\epsfbox{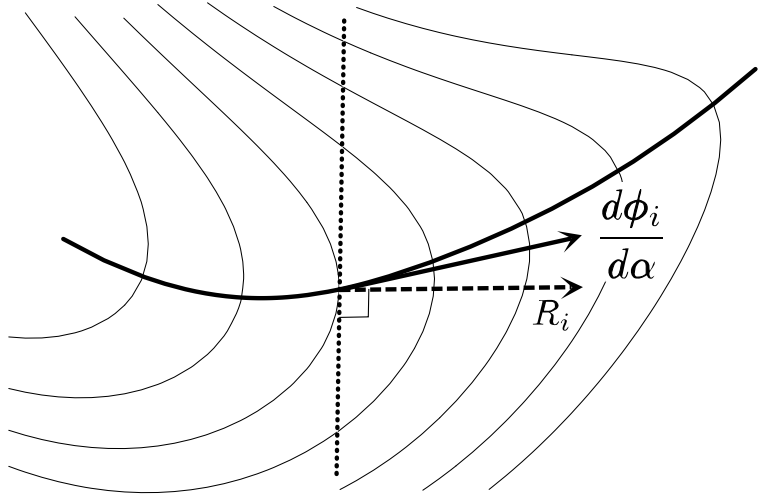}
}
\caption{A two-dimensional model of the functional space.
The thin solid lines denote the contours of the action $S$.
The thick solid line is the valley line. The
gradient vector $R_i$ and the tangent vector 
$d \phi_i / d\alpha$ are denoted by the broken and 
solid arrows, respectively.
The vertical dotted line is the subspace chosen for the 
one-loop integral.}
\label{fig:valillust}
\end{wrapfigure}

What is important, however, is not its intuitive
interpretation, but the evaluation of the relevant 
functional integral.
It is carried out along the valley line in the following manner:
Let us parametrize the valley line by a parameter $\alpha$
and denote the solutions of \eqa{nvmdef} as $\phi(\alpha)$.
The integration over $\alpha$ is to be carried out exactly,
while for other directions the one-loop (or higher order)
approximation is applied.
We are to change the integration variables
($\phi_i$) to $\alpha$ and a subspace of $(\phi_i)$,
which is determined uniquely by the following argument:
We first expand the action in terms of the fluctuation 
$\tilde\phi = \phi - \phi(\alpha)$,
\shiki{
S(\phi) = S(\phi(\alpha)) + \partial_i S (\phi(\alpha)) 
\tilde\phi_i
+ {1 \over 2}D_{ij} (\phi(\alpha)) 
\tilde\phi_i \tilde\phi_j
+ \cdots .}{sexpand}
The first order derivative term in the above is {\sl not} equal to zero,
for $\phi(\alpha)$ is {\it not} a solution of the equation of motion.
In such a case, the $\hbar$-expansion is no longer the loop
expansion, as the tree contribution floods the expansion.
In order to avoid this, 
we force this term to vanish by the choice of the subspace.
It is most conveniently done by the Faddeev-Popov technique.
We introduce the FP determinant $\Delta(\phi(\alpha))$ by 
the following identity;
\shiki{
\int d\alpha\ \delta(\tilde\phi_i R_i )
\Delta(\phi(\alpha))=1,
}{fpdef}
where $R_i (\phi(\alpha)) \equiv \partial_i S / \sqrt{ (\partial S)^2}$
is the normalized gradient vector.
From the above definitions, we find the FP determinant to be
the following;
\shiki{
\Delta(\phi(\alpha)) = \left| {d\phi_i(\alpha) \over d\alpha}
\{R_i - \partial_i R_j \tilde\phi_j \}
\right| .
}{dfres}
In the one-loop approximation, the second term in the curly bracket
can be neglected and the remaining term is simply
the cosine of the angle between
the gradient vector and the vector tangent to the trajectory.
This is simply a Jacobian factor, because
the trajectory is not necessarily orthogonal to the chosen subspace.
This situation is illustrated in Fig.~\ref{fig:valillust}.

Let us evaluate the the path-integral for the vacuum-to-vacuum
transition amplitude,
\shiki{
Z = {\cal N}\int \prod_j {d\phi_j \over \sqrt{2\pi}}\ e^{-S},
}{pathint}
using the valley trajectory, where ${\cal N}$ is a suitable
normalization factor.
The evaluation for observables is essentially the same.\footnote{The 
only possible exceptions are the cases then we have 
``large observables''; if for example we have a large number
of fields, which contributions could override the dominations
by the exponentiated action.  
This requires further development of the theory at hand.} 
Substituting \eqa{fpdef} into \eqa{pathint}, we obtain
\shiki{Z = {\cal N}\int d\alpha \int\prod_j {d\phi_j \over \sqrt{2\pi}} 
\, \delta(\tilde\phi_i R_i )
\left|{d\phi_i \over d\alpha} R_i \right|
e^{-S(\phi)},}
{fppathint}
where we have already used the one-loop expression 
for the FD determinant (the Jacobian) $\Delta(\phi(\alpha))$.
We note that the above expression is apparently invariant
under the general coordinate transformation of $\alpha$.
The second (functional) integration, 
$\int\prod_j d\phi_j \delta(\tilde\phi_i R_i)$,
is the integration over the subspace denoted by
the dotted line in Fig.~\ref{fig:valillust}.
At the one-loop order, the $\phi_i$ integration 
yields $\det^\prime$, the determinant of $D_{ij}$ restricted to the subspace.
This subspace is orthogonal to the gradient vector, 
which is the eigenvector with
the eigenvalue $\lambda$ according to the proper valley equation
\eqa{nvmdef}.
Therefore this subspace is the whole space less the direction of
the eigenvalue $\lambda$ of $D_{ij}$.
Therefore the resulting determinant is simply the ordinary determinant less
the  eigenvalue $\lambda$; 
\shiki{
{\rm det}^\prime D = {\det D \over \lambda}_{\displaystyle ,}
} 
{detprime}
This can be shown by evaluating \eqa{fppathint} by exponentiating
the delta-function and integrating over the 
fluctuations $\tilde\phi$ as follows;
\begin{eqnarray}
Z &=& {\cal N}\int d\alpha \int\prod_j {d\tilde\phi_j \over \sqrt{2\pi}} 
\int_{-\infty}^\infty {dk \over 2\pi} 
e^{ik\tilde\phi_i R_i} \left|{d\phi_i \over d\alpha} R_i \right|
e^{-S(\phi)}\nonumber\\
&=& {\cal N}\int d\alpha \int_{-\infty}^\infty {dk \over 2\pi} 
{1 \over \sqrt{\det D}}
\exp \left(-{1 \over 2} R D^{-1} R k^2 \right) \nonumber\\ 
&=& {\cal N} \int d\alpha \left|{d\phi_i \over d\alpha} R_i  \right|
{1 \over \sqrt{2 \pi (R D^{-1}R) \det D}}
\ e^{-S(\phi(\alpha))} .
\end{eqnarray}
Since the proper valley equation \eqa{nvmdef} yields
$D^{-1} R = R/\lambda$ and the vector $R$ is normalized, we obtain
\begin{equation}
(R D^{-1}R) \det D = {1 \over \lambda} \det D (= {\rm det}^\prime D),
\label{eq:nicelambda}
\end{equation}
as in \eqa{detprime}.
We thus obtain the following expression at the one-loop order;
\shiki{
Z = {\cal N}
\int d \alpha \left|{d\phi_i \over d\alpha} R_i  \right|
{1\over\sqrt{2 \pi \det^\prime D}} \ e^{-S(\phi(\alpha))}.
}
{zreduct}
In this sense, the proper valley method converts the
eigenvalue to the collective coordinate, {\it exactly}.
As we briefly mentioned at the beginning of this section,
this exact conversion is quite ideal for the actual calculations:
In physical situation one often suffers from a negative,
zero, or positive but very small eigenvalue, which renders the
ordinary Gaussian integration meaningless or unreliable.
The proper valley method saves this situation by converting
the unwanted eigenvalue to the collective coordinate, for
the factor $\det^\prime D$ is exactly free from this eigenvalue.

When there are solutions of the equation of motion with a zero mode,
as in the models with ordinary instantons,
the proper valley method reduces to the usual collective coordinate method.
This is seen as follows.
Let us denote the solution of the equation of motion by
$\phi_{\rm sol}$;
\begin{equation}
\partial_i S(\phi_{\rm sol})=0.
\label{eq:phisol}
\end{equation}
This satisfies the proper valley equation \eqa{nvmdef} at the same time.
The normalized gradient vector $R_i$ is apparently singular 
due to \eqa{phisol}. 
However, $R_i$ is well-defined by a limiting procedure
$\lambda \rightarrow 0$.
In order to show this, we denote the original action by $S^{(0)}$
and we add a term $\lambda S^{(1)}$ to the
original action, so that $\phi_{\rm sol}$ is no longer
a solution of equation of motion for the theory defined by
the total action $S = S^{(0)} + \lambda S^{(1)}$.
(The parameter $\lambda$ is an arbitrary small quantity and
$S^{(1)}$ can be any function of $\phi$ that satisfy the above 
requirement.)
The solution of the proper valley equation for $S$ 
can be expanded in terms of $\lambda$ around $\phi_{\rm sol}$.
For this purpose, We introduce a set of auxiliary variables $F_i$ and
rewrite the proper valley equation \eqa{nvmdef} as follows;
\begin{eqnarray}
D_{ij} F_j &=& \lambda F_i , \label{eq:nvfeq}\\
\partial_i S &=& F_i .
\label{eq:nvfdef}
\end{eqnarray}
The equation \eqa{phisol} guarantees that at $\lambda=0$, $F_i=0$.
Thus we use the following expansion;
\begin{eqnarray}
\phi &=& \phi_{\rm sol} + \lambda \phi^{(1)} + \cdots ,\nonumber \\
F &=& \lambda F^{(1)} + \cdots \nonumber .
\end{eqnarray}
Substituting the above expansion to \eqa{nvfeq} and \eqa{nvfdef}, we find
that the zeroth order equation in $\lambda$ 
is simply \eqa{phisol},
and that the first order equations are as the follows;
\begin{eqnarray}
D_{ij}^{(0)} F^{(1)}_j &=& 0, \label{eq:fiszeromode}\\
\partial_i S^{(1)}(\phi_{\rm sol}) + D_{ij}^{(0)} \phi_j^{(1)}
&=&  F^{(1)}_i ,\label{eq:fnormalized}
\end{eqnarray}
where $D_{ij}^{(0)} \equiv \partial_i \partial_j S^{(0)}(\phi_{\rm sol})$.
From the equation \eqa{fiszeromode}, we find that $F^{(1)}$
is proportional to the zero-mode of the original solution $\phi_{\rm sol}$.
The proportionality constant is determined by multiplying
$F^{(1)}_i$ on \eqa{fnormalized};
\begin{equation}
F^{(1)}_i \partial_i S^{(1)}(\phi_{\rm sol}) =  F^{(1)}_i F^{(1)}_i,
\end{equation}
where we have used the fact that $D_{ij}$ is symmetric
and \eqa{fiszeromode}. The quantity
$\phi_j^{(1)}$ is determined by \eqa{fnormalized} under the above condition.
Since the normalized gradient vector $R_i$ in question 
is the normalized $F$ vector, {\it i.e.}, $R_i = F_i / |F|$, 
it is, in the limit $\lambda=0$, simply the 
the normalized zero-mode of the solution of the equation of motion.
Therefore, we conclude that
the FP procedure \eqa{dfres} is equivalent to the 
usual collective coordinate method, as promised. 

The above analysis shows that the proper valley method can
be considered as an generalization of the usual
collective coordinate method.
An added bonus of this property is that the resulting det$^\prime D$
is quite easy to calculate; we simply calculate the whole determinant
and divide it by the smallest eigenvalue.
If several unwanted eigenvalues exist, 
the proper valley method can be extended 
straightforwardly.  The equation should then specify that
the gradient vector $\partial_i S$ lie in the subspace
spanned by the unwanted eigenvalues. This leads to a multi-dimensional
valley with the advantages noted above.

There is another valley method, called ``streamline 
method"\cite{rf:balitsky},
which has been extensively used in the literatures.
It proposes to trace the steepest descent line starting
from a region of larger action.
Namely, it uses the equation
\begin{equation}
{d \phi_i \over d \alpha} = f(\alpha) \partial_i S,
\end{equation}
where $f(\alpha)$ is an arbitrary function of $\alpha$
(there is a general reparametrization invariance of $\alpha$
in this formalism).
This valley line coordinate can be introduced in the 
same manner as before using the FP technique.
This makes the Jacobian trivial at the one-loop order.
This streamline method, however, suffers from many difficulties 
unknown to the proper valley method:
The most serious one is the fact that 
its subspace for the Gaussian integration
has no relation to the eigenvalues of $D_{ij}$ whatsoever.
The relation \eqa{nicelambda} does not hold;
the determinant is not 
guaranteed to be free from the unwanted eigenvalue.
Another problem is that it is a flow equation in 
the functional space:
Once given a point in the functional space, it
defines other valley configurations in the infinitesimally
small neighborhood of that point.  
In other words, it is not a local definition in the functional space.
As such, it does not define a solution by itself. 
One has to define the starting point to start the valley.  
Therefore the construction of the configurations
on the valley trajectory is quite difficult.
When the starting configuration is known (like a pair of instanton
and anti-instanton separated by infinite distance), this might
be said to be a technical difficulty.  In this method, however, 
the valley can be traced only from the above;
it suffers from instability if the valley is traced 
from the {\it bottom} of the valley.
An elementary analysis shows that the streamline most often
climbs up the side-walls of the valley, not the bottom line.
This problem becomes fatal when one is dealing with
the theories with zero-size instantons, like the standard model
noted in the previous section, in which the
higher starting point is not known at all.
For more detailed comparison of these methods 
and other features of the proper valley method,
we refer the readers to Ref.~\citen{rf:pvalley}.

There is yet one another formalism that
deals with this kind of situation.
It is the ``constrained (instanton) method'',\cite{rf:Aff} explained briefly
in the context of the standard model in Section 2.
This method is actually a valley method,
in the sense that we obtain a line of configuration parametrized
by the constraint parameter.
In the case mentioned above,  
this method allows the definition of configurations,
are very close to the zero-size instanton.
It has a problem that there is no 
prescription for the choice of the constraint that guarantees 
the effective evaluation of the path-integral.
In general one knows some way to choose the constraint
so that it allows desirable configurations. But this
is only a prescription for the constraint not to
be meaningless. Among all the meaningful constraints
which allows effective evaluation is the question unanswered.

Actually there is a subtle relation between the proper valley method
and the constraint method. Let us elaborate upon this point for
clarification.  The equation \eqa{lagra} can be interpreted as
a constrained equation in a different manner: By dividing
with $\lambda$, it can be viewed as extremizing the action $S$
under the constraint functional;
\begin{equation}
S_{\rm const.} = -{1 \over 2 \lambda} (\partial_i S)^2.
\label{eq:sconstraint}
\end{equation}
Therefore, the proper valley method can be interpreted
as a constraint method, but {\it with the specific choice
of the constraint} \eqa{sconstraint}.
However, with all the geometrical properties and the special features,
especially in relation with the Gaussian integrations, 
which are all unique to the proper valley method,
we consider it is best to distinguish it from 
the generic constraint method without any special consideration
to the constraint. Hereafter in this paper we simply call the latter 
the constraint method.

\begin{wrapfigure}{c}{14cm}
\centerline{\epsfxsize=8cm \epsfbox{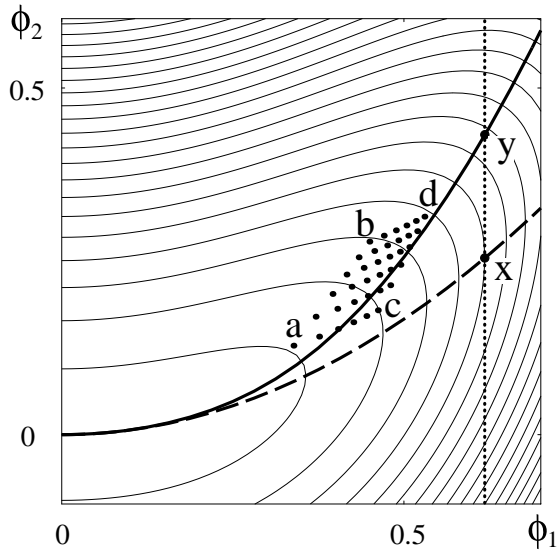}}
\caption{The valley instanton and the constrained instanton in the toy
model. The solid line denotes the valley instanton and the dashed line
shows the constrained instanton. The dots show the positions of the 
saddle points of the integrand of $O(p,q)$.}
\label{fig:toymodel}
\end{wrapfigure}

While the detailed comparison of the configurations obtained 
by the constraint method and the ones by the proper valley method 
in specific field theoretical models are done in Section 4,
here we discuss a toy two-dimensional functional space
to demonstrate the power of the proper valley method.
The two degrees of freedom in the model is denoted
as $\phi_i$ ($i = 1,2$), 
and the action is given by the following,
\shiki{
S(\phi_1, \phi_2) = {1 \over g^2}
\left(
(\phi_1^2 + \phi_2)^2 + 5(\phi_1^2 - \phi_2)^2
\right)_{\displaystyle .}
}{twoaction}
This is constructed by distorting a simple parabolic
potential so that the valley trajectory is not trivial.
Its contours ($S =$ constant lines) are plotted 
as thin lines in Fig.~\ref{fig:toymodel}.

The proper valley equation is now a simple algebraic equation,
which can be solved numerically. 
Alternatively, we could reparametrize the functional space
by $(r, \theta)$ defined by the following;
\shiki{
\phi_1^2 + \phi_2 = r \cos\theta, \hskip 3mm
\phi_1^2 - \phi_2 = {1 \over \sqrt{5}}r \sin\theta.
}{repra}
Under this parametrization $S = r^2/ g^2$.
Therefore following the variational interpretation of 
the proper valley equation, we can minimize the square of the norm of the 
gradient vector, $(\partial_i S)^2$, as a function of $\theta$.
In Fig.~\ref{fig:toymodel}, the thick solid line denotes 
the proper valley trajectory.

As an analogue of the constrained instanton formalism 
in this toy model, we introduce a simple constraint.
Near the origin the valley trajectory extends to the $\phi_1$ direction.
Any reasonable constraint has to reproduce this property.
Therefore, as a simple example for the constraint, 
we choose the following;
\shiki{\phi_1 = {\rm const.}}
{toyconst}
The solution of the constraint method is plotted in 
Fig.~\ref{fig:toymodel} by the dashed line.

In Fig.~\ref{fig:toymodel}, it is apparent that the proper valley trajectory
goes through the region of importance, while the constrained
trajectory does not.
This property can be displayed more explicitly 
by considering the analogue of
the physical observables $O(p,q)$ defined by the following;
\shiki{
O(p,q) = \int_0^\infty d\phi_1 \int_{-\infty}^\infty d\phi_2
\ \phi_1^p \phi_2^q \ e^{- S(\phi_1, \phi_2)}.
}{opq}
Following the prescription given above, we have carried out
the numerical evaluation of $O(p,q)$ 
exactly and by the Gaussian (``one-loop") 
approximation around the proper valley trajectory and 
the constrained trajectory.
Table.\ref{tab:toyratio} gives the ratio of
the one-loop values over the exact value.

As is seen in the table, the proper valley method gives
a better approximation than the constraint method
in this range of $(p, q)$ consistently.
This could be explained in the following manner;
one could calculate the position of the maximum of 
the integrand in \eqa{opq}, or the following 
``effective action";
\shiki{\tilde S^{(p, q)}(\phi) 
= S(\phi) - p \log \phi_1 - q \log \phi_2.}
{stilde}
The peak positions calculated in this manner 
are denoted in Fig.~\ref{fig:toymodel} by dots.
The point $a$ in the left bottom is for 
$(p, q) = (2,2)$, $b$ for $(2,12)$, $c$ for $(12,2)$,
and $d$ for $(12,12)$.
The dots are distributed around the proper valley trajectory.
This explains the fact that the proper valley trajectory
gives the better approximation over the constrained
trajectory.
Only when the power of one of the observable gets high
enough as in the point $c$, 
the constraint method starts to give a little better values.
However, this is exactly the range in which
the ordinary instanton calculation starts to fail;
For the $n$-point function with $n > O(1/g^2)$,
it is well known that one needs to take into account
the effect of the external particles on the instanton itself. 
The power of the valley method lies in the fact that
even in this range the calculation is done with good
accuracy for generic type of observables, like the (12,12)
case. Only when the observable is very special, like
the (12,2) case, the constraint method tuned to 
that operator may give reasonable estimates.

\begin{wraptable}{c}{14cm}
\begin{center}
\doublerulesep=0pt
\def\arraystretch{1.3}
\begin{tabular}{@{\vrule width 1.0pt \quad}c @{\quad} |
c|c|c|c|c|c @{\ \vrule width 1.0pt}}
\noalign{\hrule height 1.0pt}
$p\backslash q$ &2& 4 & $6$ & 8 &
10 & 12 \\
\hline
2  & 0.827  & 0.679  & 0.585  & 0.520  & 0.474  & 0.439 \\
   & 0.545  & 0.248  & 0.110  & 0.049  & 0.022  & 0.010 \\
\hline
4  & 0.997  & 0.870  & 0.767  & 0.689  & 0.629  & 0.581 \\
   & 0.667  & 0.324  & 0.149  & 0.068  & 0.030  & 0.014 \\
\hline
6  & 1.061  & 0.986  & 0.900  & 0.826  & 0.764  & 0.712 \\
   & 0.737  & 0.384  & 0.185  & 0.086  & 0.040  & 0.018 \\
\hline
8  & 1.064  & 1.046  & 0.991  & 0.931  & 0.876  & 0.826 \\
   & 0.783  & 0.432  & 0.218  & 0.105  & 0.050  & 0.023 \\
\hline
10 & 1.029  & 1.063  & 1.041  & 1.003  & 0.960  & 0.918 \\
   & 0.815  & 0.473  & 0.248  & 0.123  & 0.060  & 0.028 \\
\hline
12 & 0.971  & 1.046  & 1.058  & 1.043  & 1.017  & 0.985 \\
   & 0.839  & 0.508  & 0.276  & 0.141  & 0.070  & 0.034 \\
\noalign{\hrule height 1pt}
\end{tabular}
\end{center}
\caption{The ratio of the $O(p,q)$ estimated by 
proper valley method (the upper column) and the constraint method
(the lower column) over the exact value for $g=0.2$. }
\label{tab:toyratio}
\end{wraptable}

Let us note a tricky point in comparing the valley instanton
and the constrained instanton.
If we compare them with the same parameter values,
the constrained instanton has smaller action than the
valley instanton.  This is not a contradiction, nor it means
the constrained instanton has a larger contribution:
This becomes clear by considering the dotted vertical line
in Fig.~\ref{fig:toymodel}.
The point $x$ and $y$ are the valley and the constrained configurations,
respectively, which have the same value of $\phi_1$.
The configuration $y$ has a larger action than $x$,
by the definition of the constraint method.
However, as we have seen above, this does not mean
that the constrained trajectory gives a better approximation.
(One way to illustrate this point is that if we compare
the configurations at the same distance, $\int d\alpha
|(d\phi_i / d\alpha) R_i |$, from the origin, the valley configuration has
the smaller action.)
The same situation will be seen in the results of the following
sections; when we compare the instantons at the same value of the
constraint parameter, the constrained instanton has a smaller action than
that of the valley instanton.  
This is a red herring and has nothing to do with
the relevancy of the valley instanton.

In this analysis we took a very simple 
constraint \eqa{toyconst}.
Alternatively we could take 
some ad-hoc constraint as long as any of these
yield ``finite-size" instantons, {\it i.e.},
points away from the origin.
Most of these constraints yield essentially similar
results.
Only when the trajectory goes through the dotted area
in the Fig.~\ref{fig:valillust}, one can obtain good quantitative
results.  Such a trajectory, however, is destined to be
very close to the proper valley trajectory.
Therefore the proper valley method, defined without any
room for adjustment and guaranteed to give good results
is superior to the constrained method.

Finally, we note that in the actual calculation of the solutions in the 
continuum space-time 
the proper valley equation with auxiliary variable $F$, 
the equations \eqa{nvfeq} and \eqa{nvfdef}, is most convenient. 
Let us denote all the real bosonic fields in the theory by  $\phi_\alpha(x)$,
where we denote all the space-time coordinates by $x$
and the rest of the indices by $\alpha$.
The proper valley equation is as follows;
\begin{eqnarray}
\begin{array}{cc}
& \displaystyle \sum_\beta \int d^4 y
{\delta^2 S \over \delta\phi_{\alpha}(x) \delta\phi_{\beta}(y)}
F_{\beta}(y) = \lambda F_{\alpha}(x), 
\\[0.5cm]\label{eq:gennv}
&\displaystyle F_\alpha (x) = {\delta S \over \delta\phi_{\alpha}(x)}_{\displaystyle .}
\end{array}
\end{eqnarray}

This is a set of second-order differential equation, 
which can be analyzed by the conventional methods.
The solution of the ordinary field equation of motion 
is a solution of the above equation with $F_\alpha (x) =0$.
In other words, $F_\alpha (x)$ specifies where and 
how much the valley configuration
deviates from the solution of equation of motion.
This property is useful for the 
qualitative discussion of the properties of the valley 
configurations.
The analysis in the following sections will be carried out
in this auxiliary field formalism.

\section{Application of the proper valley method}

\subsection{Valley instanton in the 
asymmetric double-well potential} 

\begin{figure}[htb]
  \parbox{\halftext}{
       \centerline{\epsfxsize=6cm\epsfbox{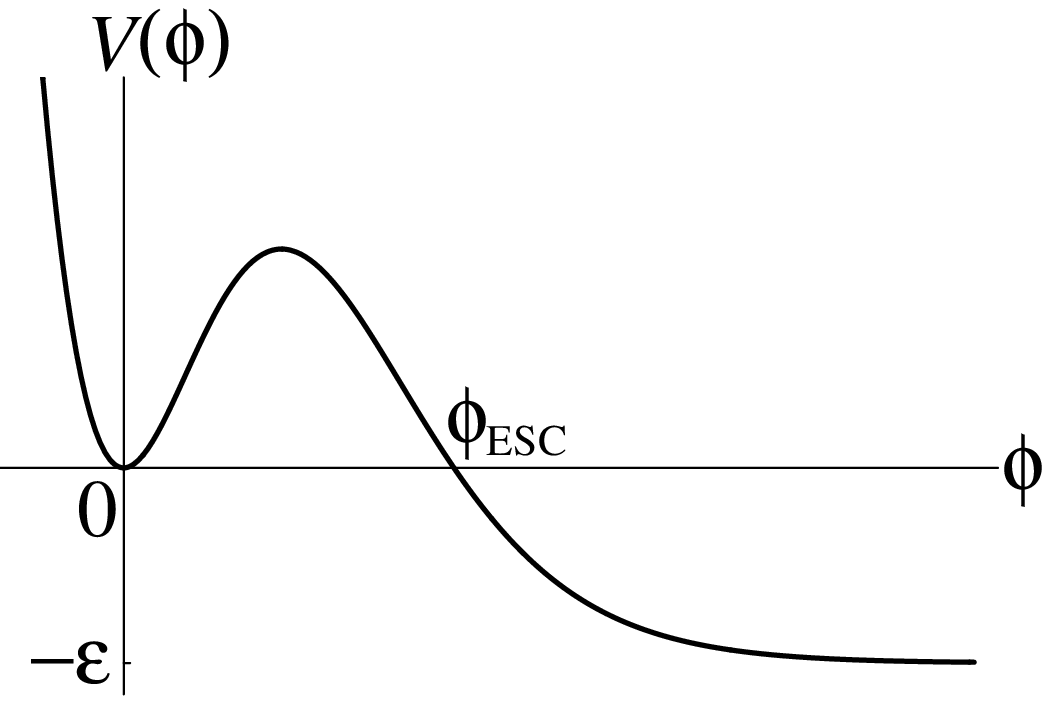}}
       \caption{A potential that is flat in the asymptotic direction,
$V(\infty) = -\epsilon$.}
\label{fig:s41potflat}}
\hspace{4mm}
  \parbox{\halftext}{
      \centerline{\epsfxsize=6cm\epsfbox{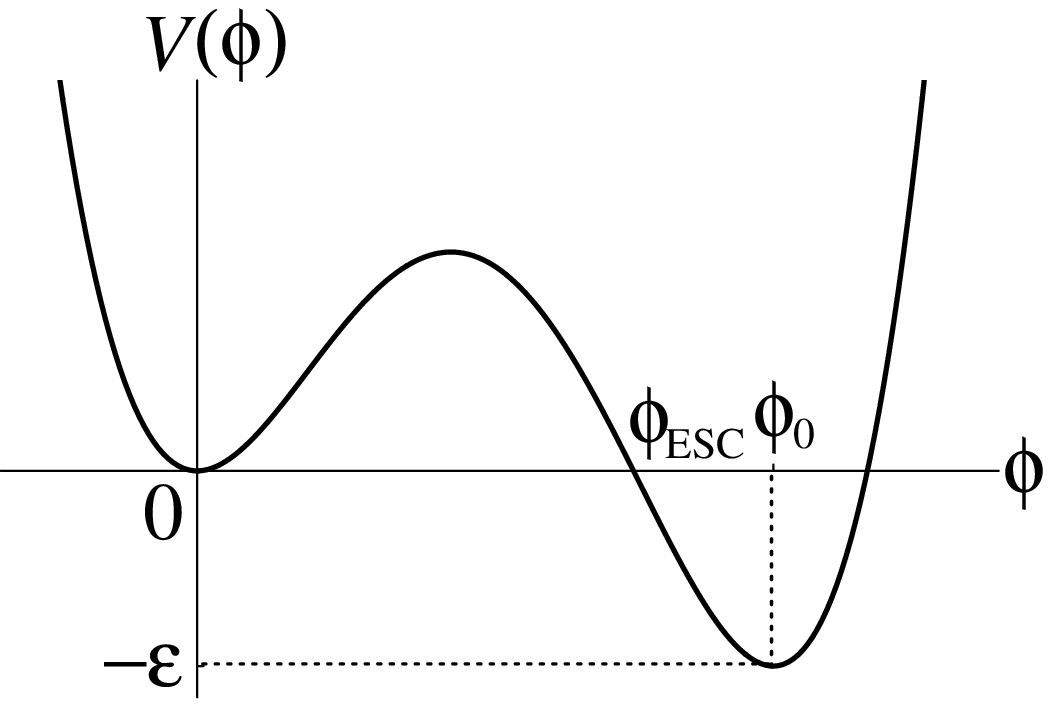}}
      \caption{An asymmetric double-well potential, in which $\phi=0$ is
only metastable.}
\label{fig:s41potdw}}
\end{figure}

In this subsection we apply the proper valley method to an
one-dimensional quantum mechanical model.
Our aim is to resolve a paradox\cite{rf:boy} that arises if we naively apply 
the semi-classical
approximation to the potentials, \(V(\phi)\),  depicted in 
Fig.~\ref{fig:s41potflat} and \ref{fig:s41potdw}.
Let us start with  reviewing the path-integral formalism in the quantum
mechanics and then clarify what is the paradox we are concerned with.

For a given Hamiltonian with the dynamical variable \(\phi\) and its
conjugate momentum \(p\),
\be H = {1\over 2}\; p^2 + V(\phi), \ee
the transition amplitude from the initial value \(\phi = q_1\) at
\(\tau = 0\) to the final one \( q_2\) at \(T\) in imaginary time
formalism is
evaluated by the path-integral
\be \langle q_2| e^{- HT/\hbar } | q_1 \rangle =
{\cal N} \int {\cal D} \phi \, e^{ - S[\phi]/ \hbar }
\label{eq:s41pathintegral}\ee
where 
\be S[\phi] = \int_0^T d\tau \left[ 
{1\over 2}\left({d \phi\over d\tau}\right)^2
+ V(\phi) \right]. \label{eq:s41action}\ee
In Eq.~(\ref{eq:s41pathintegral}), 
\(\phi(\tau)\)  is parameterized as
\be \phi(\tau) = \phi_{\rm B}(\tau) + \sum_n a_n \varphi^{(n)}(\tau), \ee
where \(\phi_{\rm B}\) is a background that satisfies
\be \phi_{\rm B}(0) = q_1 ,\quad \phi_{\rm B}(T) = q_2, \ee
and \(\varphi^{(n)}\) are elements of  a 
normalized complete set in the functional space for the 
interval \([0, T]\);
they satisfy
\be \left( \varphi^{(n)}\cdot \varphi^{(m)} \right) 
\equiv \int_0^T d\tau \varphi^{(n)}{}(\tau) 
\varphi^{(m)}(\tau) = \delta_{nm}
\ee
and
\be \varphi^{(n)} (0)  =  \varphi^{(n)}(T) = 0. \ee
The integral measure is then defined in terms of the coefficients 
\( a_n\) by
\be
{\cal D}\phi = \prod_n {d a_n\over \sqrt{2\pi \hbar}}, 
\quad {\cal N} = \left[ {1\over 2\pi \hbar T } \prod_{m=1}^\infty 
\left(\pi m \over T\right)^2\right]^{1/2}.
\ee
The infinite products over \(m\) and \(n\)
are understood to be combined properly in actual calculations
so that the final expression becomes a convergent product.
The information on the model will be given by calculating
Eq.~(\ref{eq:s41pathintegral}) for the various values of \(q_1\), \(q_2\),
and \(T\).
We are interested in the energy eigenvalues \(E\) 
of the lowest-lying states;
we calculate Eq.~(\ref{eq:s41pathintegral}) with the boundary condition
\(\phi(0) = \phi(T) = 0\) and with sufficiently large \(T\)  
(compared with the inverse of the 
excitation energies of the other states.)
The energy eigenvalues are extracted from the \(T\) dependence of 
the transition amplitude,
\(\langle 0 | e^{- HT/\hbar } | 0 \rangle \sim e^{ - E T/\hbar }\).

The semi-classical approximation for the path-integral 
Eq.~(\ref{eq:s41pathintegral}) is 
carried out by the following: use the classical solutions 
\(\phi_{\rm cl}\) 
\be {\delta S\over \delta \phi_{\rm cl} } = 0 \ee
as \(\phi_{\rm B}\) and 
expand the action around it,
\be S[\phi] = S[\phi_{\rm cl}] + {1\over 2} \int\int 
d\tau d\tau' D(\tau, \tau') \varphi(\tau)
\varphi(\tau'), \ee
where
\(D(\tau, \tau') \equiv \delta^2 S/ 
\delta \phi(\tau) \delta \phi(\tau')\) and
\(\varphi \equiv \phi - \phi_{\rm cl}\);
use the eigenfunctions of  \(D\) for the complete set \(\{\varphi^{(n)}\}\)
and approximate the  functional integral by Gaussian,
\be\langle 0 | e^{ - H T } |0 \rangle \simeq \sum_{\phi_{\rm cl}}
e^{ - S[\phi_{\rm cl}]}
{{\cal N } \over \sqrt{\mbox{det}D[\phi_{\rm cl}]}}\ee
where ``det" denote the product of all the eigenvalues of  \(D\)
(we will set \(\hbar = 1\) hereafter.)
The zero modes will be treated by the collective 
coordinate method (see for example Ref.~\citen{rf:clec})
if \(D\) has some.

For the potentials in 
Fig.~\ref{fig:s41potflat} and \ref{fig:s41potdw} 
we have the bounce solution,
that starts at \(\phi = 0\) and bounce at \(\phi_{\rm ESC}\) back 
to \(\phi = 0\) for the large time interval,
\(T \gg \omega_+^{-1}\) (\(\omega_+ \) is the frequency at \(\phi=0\).)
There is a subtlety  in the semi-classical approximation at
the bounce:  it has one negative eigenvalue  mode and
the Gaussian integral along this direction becomes 
ill-defined.
The way we usually adopt to circumvent this is to
assume an appropriate analytic continuation of the
contour for the ill-defined Gaussian integral;
The bounce then gives an  imaginary part for 
the energy eigenvalues\cite{rf:clec,rf:coleman,rf:callancoleman}
for both potentials in Fig.~\ref{fig:s41potflat} and \ref{fig:s41potdw}.

Although the exact form of the analytic continuation
is not so clear, the complex energy for the potential 
in Fig.~\ref{fig:s41potflat} has a sound interpretation:
if one restricts the wave functions to have only the outgoing component
at $\phi \gg \phi_{\rm ESC}$,  the hermiticity of the Hamiltonian
is violated and the energy eigenvalues become complex. 
This imaginary part reflects the instability of the localized wave packet at
$\phi \sim 0$.

For the potential in Fig.~\ref{fig:s41potdw}, however,
the imaginary part is obviously a wrong answer.\cite{rf:boy} \ 
We can only take wave functions with decaying exponential
at $\phi \gg \phi_0$
and the hermiticity  of the Hamiltonian cannot be violated.
There is no room for the energy eigenvalue to be complex.
This is the paradox we intend to solve.

We use the proper valley method to
see how the action behaves for the path-integral related to
the negative mode at the bounce and to make the corresponding
path-integral  well-defined. 
We construct ``valley instanton'', which should replace the bounce solution.
Interestingly, it has a zero mode and this expedites the calculation
of its determinant and Jacobian as was the case of the instanton.
It is a well-localized configuration with respect to the imaginary
time and their dilute-gas sum generates the reasonable
energy shift instead of the decay rate.
We will also show that it converges analytically to the instanton
in the limit of \(\epsilon \rightarrow 0\) and all the results
reproduce the well-known instanton results.

Specifically we parameterize the potential in Fig.~\ref{fig:s41potdw} as
\begin{equation}
V (\phi) = {1 \over 2} \phi^2 \left(1 - g\phi\right)^2
- \epsilon (4 g^3 \phi^3 - 3 g^4 \phi^4) ,
\label{eq:s41potone}
\end{equation}
where the coupling constants $g$ and $\epsilon$ are positive.
The potential (\ref{eq:s41potone}) has a local minimum, $V(0) = 0$ and
a global minimum at $\phi_0 = 1/g$, where $V(1/g) = -\epsilon$.
(Fig.~\ref{fig:s41potdw} is plotted for $g=0.3$ and $\epsilon=0.25$.)
The potential $V(\phi)$ in Eq.~(\ref{eq:s41potone}) is a
canonical form of quartic potentials with two minima, since
any such potential can be cast into this form by suitable linear
transformations on $\phi$ and $\tau$. In this sense the following
analysis is quite a general one.
(Especially, the above form is related to that of
Ref.~\citen{rf:bl} by a simple reparametrization.)
In the following, we consider the cases with small coupling $g \ll 1$,
but not necessarily small $\epsilon$.

The proper valley configurations are given by the equations
\begin{eqnarray}
-\partial_\tau^2 \phi + V'(\phi) &= F, \label{eq:s41nvone}\\
\left( -\partial_\tau^2  + V''(\phi)\right) F &= \lambda F.
\label{eq:s41nvtwo}
\end{eqnarray}
as has been explained in the section 3.
The bounce satisfies this equation trivially since it has
$F=0$. 
We take \(\lambda\) in the infinitesimal neighborhood of the bounce
as its negative eigenvalue. 
Then the valley trajectory extends to the direction of the negative mode
from the bounce.
By changing the parameter \(\lambda\), we will obtain a
trajectory in the functional space.
The general solutions can be parameterized by the eigenvalue $\lambda$,
or any arbitrary function $\alpha$ of $\lambda$.
We  denote the valley trajectory by $\phi (\alpha)$.
We consider the contribution of such a configuration to 
the  transition amplitude
\(\langle 0 | e^{ - H T} |0 \rangle\).
By the result of Eq.~(\ref{eq:zreduct}), this amplitude is given as
\begin{equation}
\langle 0 | e^{ - H T} |0 \rangle
= {\cal N} \int d\alpha {1 \over \sqrt{2 \pi \det^\prime D}} J_\alpha
e^{-S [\phi(\alpha) ]},
\label{eq:s41vvampalpha}
\end{equation}
to the leading order in $g$,
where $\det^\prime D$ is the usual determinant less the
eigenvalue $\lambda$ in the proper valley equation (\ref{eq:s41nvtwo}) and
\begin{equation}
J_\alpha \equiv {\displaystyle \int d\tau {d \phi(\alpha) \over d\alpha}F
\over \displaystyle\sqrt{\int d\tau F^2}}
= {\displaystyle{d S[\phi(\alpha)] \over d\alpha} \over
\displaystyle\sqrt{\int d\tau F^2}}.
\label{eq:s41jacob}
\end{equation}
The factor $J_\alpha$ is the Jacobian for this change of the integration
variable which is explained at the previous section.

\begin{figure}
\centerline{\epsfysize=10cm\epsfbox{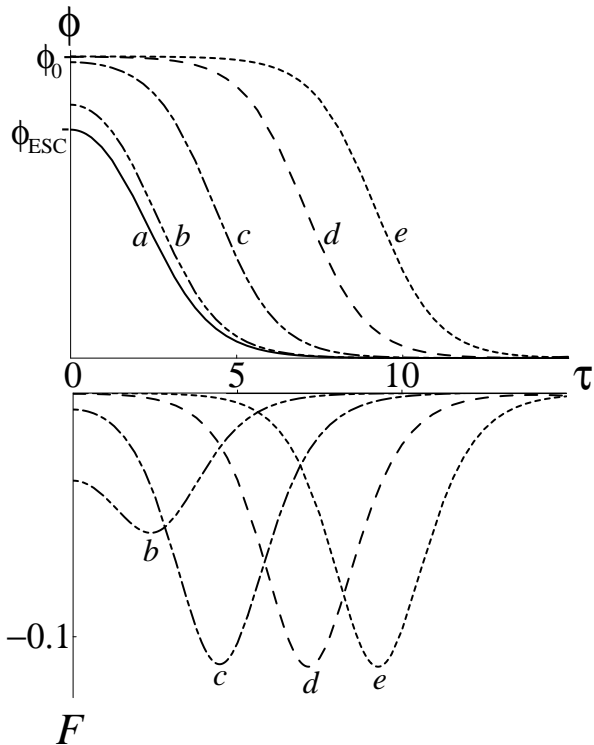}}
\vspace{3mm}
\caption{Valley bounce solutions $(\phi(\tau), F(\tau))$ of the proper
valley equations.
Center of all of the configurations are chosen to be at the origin, $\tau=0$,
around which the solutions are symmetric.
The solid line $a$ in the upper figure is the usual bounce solution, which has
$F(\tau)=0$.
The other lines, $b$--$e$, are unique to the proper valley equations.}
\label{fig:s41conf}
\end{figure}

\begin{figure}
\centerline{\epsfysize=9cm\epsfbox{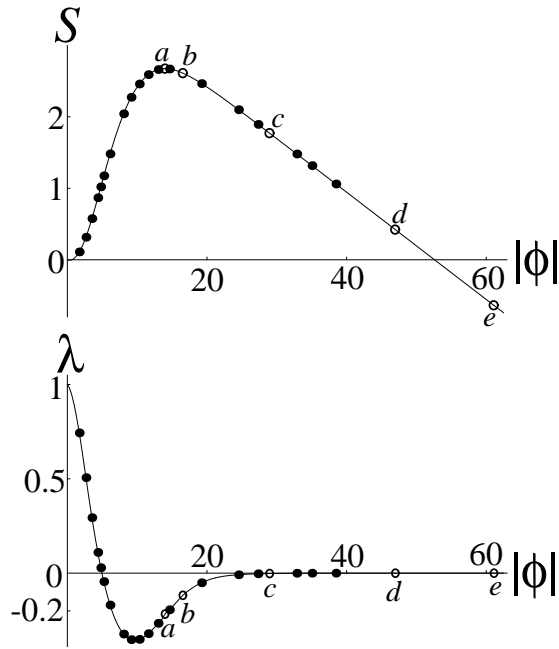}}
\vspace{3mm}
\caption{The values of the action, $S$, and the eigenvalue, $\lambda$, of the
valley bounces.
The peak of the action is given by the bounce solution, the line $a$ in
Fig.~\protect\ref{fig:s41conf}. 
The points corresponding to the valley bounces in
Fig.~\protect\ref{fig:s41conf} 
are plotted with circle.
The solid lines are drawn as the guide for eyes.
The valley parameter is chosen to be $|\phi| \equiv  \int d\tau \phi$.}
\label{fig:s41action}
\end{figure}

We have carried out the numerical analysis and obtained the solutions 
of the proper valley equation plotted in Fig.~\ref{fig:s41conf}.
The solid line, $a$, is the bounce solution of the equation of motion.
The rest do not satisfy the equation of motion.
We call these solutions (including the bounce solution) ``valley bounce".
The values of the action, $S$, and the eigenvalue, $\lambda$,
for the valley bounces are plotted in Fig.~\ref{fig:s41action}.
The bounce solution lies at the top of the line of the action
in Fig.~\ref{fig:s41action}, corresponding to the fact that it has
a negative eigenvalue.
This negative eigenvalue can be read from the corresponding
point of the plot of $\lambda$ in the lower half of Fig.~\ref{fig:s41action}.
The most notable feature is that the large size valley bounces
have a clean interior, where $\phi = \phi_0$ and $F = 0$.
(This property is shared by the higher dimensional configurations,
{\it i.e.}, the valley bubbles.)
The effect of this is apparent in the behavior of the action
in Fig.~\ref{fig:s41action};
the action decreases linearly with large $|\phi|$,
which is proportional to the size of the valley bounce.

In the large valley bounces, $c$, $d$, and $e$ in Fig.~\ref{fig:s41conf},
we notice that the shape of the wall, {\it i.e.,}
the transition region from $\phi \sim \phi_0$ to $\phi \sim 0$,
is almost identical to each other.
That is, they overlap with each other very well when translated in $\tau$.
Thus these large valley bounces can be approximated
by simply connecting the walls by a flat region, $\phi=\phi_0$, at
various separation.
The shape of the wall can be most readily identified
when the size ($|\phi| \equiv \int d\tau \phi $) of
the valley bounce becomes $\infty$.
In this limit, the wall is simply a localized transition from
$\phi = 0$ to $\phi = \phi_0$ (or vise-versa).
Such a solution is an analogue of the instanton (or anti-instanton).
The difference is that now it is not a solution of equation of motion,
but is a solution of the proper valley equation.
This kind of configuration is called the 
``valley (anti-)instanton." \cite{rf:ahsw}

The notable information from this numerical analysis is
that the path-integral for the negative mode direction at
the bounce can be  translated eventually 
into the collective coordinate
integral with respect to the relative distance between
the valley instanton and the valley anti-instanton.
This integral indeed gets the large contribution 
from the large separation and
is divergent for \(T\rightarrow \infty\),
 but we can handle it anyway;
we will be even able to carry out the dilute-gas summation
for the multi valley instantons.
Before doing so, we look into the details of the valley instanton
which is needed for the summation.

Since we define the valley instantons at the large-size limit,
$|\phi| \rightarrow \infty$, of the valley bounce,
the plot of the eigenvalue in Fig.~\ref{fig:s41conf} implies that
the eigenvalue $\lambda$ of the valley instanton is exactly zero.
This is not a trivial property.  A solution of equation of motion
is guaranteed to have zero modes corresponding to its symmetry
transformation, such as a time translation.
Arbitrary background configurations do not have this property in general.
However, we can prove the existence of the zero mode as follows:
Take a derivative of Eq.~(\ref{eq:s41nvone}) with respect to $\tau$,
multiply $F$, and integrate over $\tau$.
After partial integrations (which surface terms vanish), we then find that
\begin{equation}
\lambda \int_{-\infty}^\infty F \dot{\phi} d\tau
= \int_{-\infty}^\infty F \dot{F} d\tau \; .
\end{equation}
The integral in the left-hand side is generally non-zero
due to the boundary conditions of the valley instantons.
This can be seen from the behavior of $\phi$ and $F$ in the walls
in Fig.~\ref{fig:s41conf}.
Since the right-hand side is zero, we find that $\lambda = 0$.
(For the valley bounces, the integral in the left-hand side is zero
since $\phi(\tau)$ and $F(\tau)$ are even functions.
Therefore, $\lambda \ne 0$ is allowed.)

We have carried out the numerical investigation of the
valley instanton with $\lambda=0$ and have successfully obtained
the solutions, which have turned out to be almost identical
to the wall regions in Fig.~\ref{fig:s41conf}.

Although we know of no exact analytical expression of the valley instanton,
it can be constructed analytically for small $g^2$
in a manner used in the construction of the constrained
instanton \cite{rf:Aff}
as well as other types of the valley instantons.\cite{rf:ahsw} \ 
Consider the valley instanton in $\tau \in [-T/2, T/2]$ ($T \gg 1$).
We define its central coordinate to be at the origin,
$\tau = 0$, by $\phi(0)=1/2g$.
The naive perturbation in $\epsilon g^2$ yields the following perturbative
valley instanton solution;
\begin{eqnarray}
\phi &=& \phi_0^I + 3\epsilon g^2 \tau \dot{\phi}_0^{I}
+ O((\epsilon g^2)^2) ,\nonumber\\
&&{} \label{eq:s41pertsol} \\
F &=& - 6 \epsilon g^2 \dot{\phi}_0^I
+ 36 \epsilon^2 g^5 \tau \phi_0^I \dot{\phi}_0^I + O((\epsilon g^2)^3),
\nonumber
\end{eqnarray}
where $\phi_0^I$ is the instanton solution for $\epsilon=0$;
\begin{equation}
\phi_0^I = {1 \over g} {1 \over 1 + e^{- \tau}}.
\end{equation}
{}From Eq.(\ref{eq:s41pertsol}), it is apparent that this naive perturbation
is valid only in the region close to the instanton center,
$|\tau| \ll 1/(\epsilon g^2 )$.
On the other hand, in the asymptotic regions, $\tau \rightarrow \pm \infty$,
we linearize the proper valley equation and find the general solutions
valid for $|\tau| \gg 1$.
Coefficients of the general solutions are fixed by matching it with
the inner solution Eq.(\ref{eq:s41pertsol}) in the intermediate region,
$1 \ll |\tau| \ll 1/ (\epsilon g^2 )$.  We have found that
this procedure can be done consistently.
The resulting asymptotic behaviors are
for $\tau \rightarrow + \infty$;
\begin{eqnarray}
\phi &\simeq& {1 \over g} \left( 1 -
\left( 1 + {3 \epsilon g^2 \over \omega_-} \tau \right)
e^{- \omega_- \tau} \right) ,\nonumber\\
&& \label{eq:s41asymplus}\\
F &\simeq& -6 \epsilon g e^{-\omega_- \tau}, \nonumber
\end{eqnarray}
where $\omega_-^2 \equiv V^{\prime\prime}(\phi_0) = 1 + 12 \epsilon g^2 $,
and for $\tau \rightarrow -\infty$;
\begin{eqnarray}
\phi &\simeq& {1 \over g} \left( 1 + 3 \epsilon g^2 \tau \right)
e^{\tau} ,\nonumber\\
&& \label{eq:s41asymminus}\\
F &\simeq& -6 \epsilon g e^\tau. \nonumber
\end{eqnarray}
This way, the valley instanton is constructed in all regions of $\tau$
for small $\epsilon$.

The action of the valley instanton is given by
$S^I=1/6g^2 + \epsilon (-T/2  + 1/2) + O(\epsilon^2)$.
This action is divided to the volume part and the
remaining (proper) part as $S^I = -\epsilon T/2 + {\tilde S}^I$.
{}From the construction above, we find that
${\tilde S}^I = 1/6g^2 + \epsilon /2 + \cdots$.
However, there is a subtlety on this point:
In the following we integrate over the position coordinate of the
instantons and anti-instantons in the dilute-gas approximation.
These coordinates are originally the valley parameters ($\alpha$s) of the
valley bounces (and their central coordinates).
The $O(\epsilon)$ term in ${\tilde S}^I$ depends on the definition of
these valley parameters, since the definition of the volume ($T$)
is affected by it. Therefore, careful study of the small valley bounces
is needed to fix this term.  This term, however, has only the
non-leading contribution. Therefore, we will not pursue this
problem any further here.

The Jacobian for the instanton position is given by,
\begin{equation}
J^I={\epsilon \over \displaystyle\sqrt{\int d\tau F^2}} \; .
\label{eq:s41acja}
\end{equation}
Since the contribution to the integration is dominated by the
central region, the leading term of Eq.(\ref{eq:s41acja}) for small
$\epsilon$ is evaluated by the use of Eq.(\ref{eq:s41pertsol}).
The result is that $J^I = 1/\sqrt{6g^2}(1 + O(\epsilon g^2))$.
The first term is the instanton action for $\epsilon =0$.
Therefore, in the limit $\epsilon \rightarrow 0$, the Jacobian of the
valley instanton reduces to that of the ordinary instanton.

The determinant, $\det^\prime D$, can be calculated by extending
the Coleman's method,\cite{rf:clec} \  in spite of the fact that the valley
instanton
is not the solution of the equation of motion.
This is due to the fact that the valley instanton
possesses the exact zero mode $F(\tau)$.
We define the asymptotic coefficients $F_\pm$ by
$F(\tau) \simeq F_{\pm} e^{\mp\omega_{\pm}\tau}$,
where for the sake of notation we introduced
$\omega_+^2 = V^{\prime\prime}(0)=1$.
After some calculation, we find that the ratio of the determinants
for the valley instanton located at \(\tau_0 ( \in [ - T/2, T/2 ] )\)
is given by the following;
\begin{equation}
{\det^\prime (-\partial_\tau^2 + V^{\prime\prime}(\phi^I))
\over \det (-\partial_\tau^2 + \omega_+^2)}
= \kappa e^{ (\omega_- - \omega_+) (T/2 - \tau_0)},
\quad
\kappa \equiv {1 \over 2 \omega_+ \omega_- F_+ F_-} \int_{-\infty}^\infty
F^2.
\end{equation}
The exponential factor represents the contribution of the
difference between the zero point energies at the two local minima.
The factor $\kappa$ is the `proper' instanton contribution.
{}From Eq.(\ref{eq:s41asymplus}) and Eq.(\ref{eq:s41asymminus}), we find that
$\kappa$ reduces to the ordinary instanton determinant for
$\epsilon \rightarrow 0$.

We combine all factors  and calculate the amplitude 
using the ``dilute-gas" valley instanton approximation.
We find the transition
amplitude to be the following;
\begin{equation}
Z(T) \equiv
{\langle 0  | e^{ -H T} | 0\rangle
\over
\langle 0 | e^{ -H_0 T} |0\rangle}
= \sum_{n=0}^\infty \alpha^{2n} I_n ,
\label{eq:s41amplitude}
\end{equation}
where we have normalized the amplitude to that of the 
free Hamiltonian  
\(H_0 = (1/2) [p ^2 + \omega_+^2 \phi^2]\);
$n$ is the number of the valley instanton pairs,
and the factor $\alpha$ is the product of the proper contributions of
the action, the determinant ratio and the Jacobian;
$\alpha=(J^I / \sqrt{2 \pi \kappa}) e^{-{\tilde S}^I}$.
The actual integrations over the positions of the valley instantons are
in the factors $I_n$;
\begin{eqnarray}
I_n (T) \equiv  \cases{
1, & for $n=0$, \cr
&\cr
\displaystyle \int_0^T d\tau_{2n} \int_0^{\tau_{2n}} d\tau_{2n-1}
...
\int_0^{\tau_{2}} d\tau_{1} \,
e^{\tilde\epsilon (\tau_{2n}-\tau_{2n-1}+ ... +\tau_2 - \tau_1)},
& for $n \ge 1$, \cr}
\label{eq:s41idef}
\end{eqnarray}
where zero-energy contributions of the determinants are
absorbed in $\epsilon$ by
$\tilde \epsilon \equiv \epsilon - (\omega_- - \omega_+)/2$.

The infinite series can be summed by the use of the generating function
method:
\cite{rf:aq} \ 
{}From Eq.(\ref{eq:s41idef}), we find that the following differential equation
is satisfied by $Z(T)$;
\begin{equation}
Z(T)'' - \tilde \epsilon Z(T)' - \alpha^2 Z(T) =0.
\end{equation}
Also, $Z(0) = 1$, and $Z'(0) =0$.
Therefore, we find that
\begin{equation}
Z(T) = {k_+ e^{-k_- T} - k_- e^{-k_+ T} \over k_+ - k_-},
\label{eq:s41ztresult}
\end{equation}
where,
\begin{equation}
k_\pm \equiv - {\tilde\epsilon \over 2} \pm
\sqrt{{\tilde\epsilon^2 \over 4} + \alpha^2}.
\end{equation}
Thus we find that the energies of the two lowest states,
$E_\pm$ is given by
\begin{equation}
E_\pm  = {\omega_+ \over 2} + k_\pm =
{\omega_+ \over 2} - {\tilde\epsilon \over 2} \pm
\sqrt{{\tilde\epsilon^2 \over 4} + \alpha^2}.
\label{eq:s41enefi}
\end{equation}
Furthermore, from the coefficients of the respective exponents of
Eq.(\ref{eq:s41ztresult}), we find that
$|\langle \phi=0 | E_\pm\rangle|^2 = \pm k_\pm / (k_+ - k_-)$.
Note that there appear no fake imaginary parts in the energy spectrum.

We also examine the validity of the dilute-gas approximation.
The mean size of the bounce, which is the mean distance, $R$, between
the instanton and the anti-instanton located at the right of the instanton,
can be obtained as the expectation value,  $\langle\tau_2 - \tau_1 \rangle$
in the amplitude, Eq.(\ref{eq:s41amplitude}) and Eq.(\ref{eq:s41idef}).
(Any other $\langle \tau_{2m}-\tau_{2m-1}\rangle$ with integer $m$ would
yield the same result for $T \rightarrow \infty$.)
Using the same generating function method as above, we
obtain the following;
\begin{equation}
R = {1 \over \alpha^2} \left(
{\tilde\epsilon \over 2} + \sqrt{{\tilde\epsilon^2 \over 4} + \alpha^2}
\right) .
\end{equation}
Similarly, the mean distance, $d$, between
the anti-instanton and the instanton located at the right of the
anti-instanton $\langle \tau_3 - \tau_2 \rangle$ is given by,
\begin{equation}
d = {1 \over \alpha^2}
\left(-{\tilde\epsilon \over 2} + \sqrt{{\tilde\epsilon^2 \over 4} + \alpha^2}
\right).
\end{equation}
For $\epsilon \gg \alpha$, $d \sim 1/\tilde\epsilon$.
On the other hand, the thickness of the instanton is
O($1/\sqrt{\epsilon})$ for $\epsilon \gg 1$, which can be seen by a simple
scaling argument on the proper valley equations,
Eq.(\ref{eq:s41nvone}) and Eq.(\ref{eq:s41nvtwo}).
This means that the dilute-gas approximation is valid 
for $\tilde\epsilon < 1$.

The result (\ref{eq:s41enefi}) has now a simple explanation.
We observe that $E_\pm$ are equal to the eigenvalues of the
following matrix;
\begin{equation}
H = \pmatrix{ \displaystyle{\omega_+ \over 2} & \alpha \cr
              \alpha & \displaystyle -\epsilon + {\omega_- \over 2}\cr} .
\label{eq:s41hmatrix}
\end{equation}
Furthermore, the weights of the state localized in the left well agree with
Eq.(\ref{eq:s41ztresult}): Denoting the eigenvectors of Eq.(\ref{eq:s41hmatrix})
$V_\pm$ with eigenvalues $E_\pm$,
\begin{equation}
\pmatrix{1 \cr 0} = \sqrt{k_+ \over k_+ - k_-} V_+
+ \sqrt{-k_- \over k_+ - k_-} V_-.
\end{equation}
The energy spectrum we obtain is the same as  the two-level
system made of the perturbative ground state at $\phi=0$ and
$\phi = \phi_0$, with the tunneling matrix element $\alpha$.
This simple picture is correct if the higher excited 
states have large energy gap from the two, and the condition
is  exactly the one obtained from the validity of the dilute-gas summation, 
$\tilde\epsilon < 1$.

Finally let us check  the result (\ref{eq:s41enefi})
for a few cases.
For $\epsilon \ll \alpha$,
the energy spectrum Eq.(\ref{eq:s41enefi}) gives the following;
\begin{equation}
E_\pm  = {\omega_+ + \omega_- \over 4} \pm \alpha - {\epsilon \over 2}
+ O\left({\epsilon^2 \over \alpha}\right).
\end{equation}
In the limit $\epsilon \rightarrow 0$, this result reduces to
the well-known instanton result for the degenerate case.
In addition, we have the average of the perturbative zero-point energies
of the left-well $\omega_+ / 2$ and the right-well
$\omega_-/2 -\epsilon$.
Since the wave function is distributed evenly at the zeroth order of the
$\epsilon$ expansion, this is the correct formula for the two lowest energy
eigenvalues.  Since the perturbative contribution to the energy splitting,
$\delta E = E_+ - E_-$, is of order $\epsilon g^4$, even a purist\cite{rf:clec}
would retain our result, $\delta E = 2 \alpha$.
For $\alpha \ll \epsilon < 1$, Eq.(\ref{eq:s41enefi}) leads to,
\begin{equation}
E_+  = {\omega_+ \over 2} + {\alpha^2 \over \tilde\epsilon}
+ O\left({\alpha^4 \over \epsilon^3}\right), \quad\quad
E_-  = -\epsilon + {\omega_- \over 2} - {\alpha^2 \over \tilde\epsilon}
+ O\left({\alpha^4 \over \epsilon^3}\right).
\end{equation}
The lower energy, $E_-$ corresponds to the state almost localized in
the right well;
this is also an expected result
since the tunneling transition is small compared
with the energy difference.

In summary, we have developed the method to obtain
the correct energy shift in the asymmetric double-well
potential by applying the proper valley method.
In view of this development,
the ordinary calculation of the imaginary part of the
energy in the unstable cases (Fig.~\ref{fig:s41potflat})
needs to be examined under the new light.

\subsection{Valley bubble} 
\def\dprime{^{\prime\prime}}
\def\sp{^\prime}

In this subsection, we study another example of quantum metastability
problem, the false vacuum decay.
It also can be treated 
in the imaginary-time path-integral formalism
using the bounce solution.\cite{rf:coleman,rf:callancoleman}

Due to the existence of a negative-eigenvalue fluctuation-mode
around the bounce solution, 
the contour of the Gaussian integration through the bounce
has to be deformed to yield the imaginary part of the energy level.
Thus the decay rate of the false vacuum is obtained.

This bounce belongs to a valley of the
action in the whole functional space.\cite{rf:aw}
This situation is analogous to that of 
the tunneling processes via instanton, 
such as the baryon number violation process in the standard model 
explained in section 2.

In either case, when the initial condition is such that the 
state is the local minimum, the solutions of the equations of motion
(bounce or instanton) dominate the relevant imaginary-time
path-integral.
However, when the initial state is of higher energy,
different configurations could dominate.
For the quantum tunneling between degenerate vacua, 
the deformed instanton and anti-instanton pair is known to play that role.
\cite{rf:mattis} \ 
These configurations generally belong to the valley of action,
since they form a line of relatively small actions.
\cite{rf:pvalley} \ This is obtained by
separation of a collective coordinate that corresponds to small, zero or 
negative eigenvalue, which is dangerous for the Gaussian integration.
On the other hand, not much is known for quantum decay of metastable
states: In the false vacuum, decay occurs through a vacuum
bubble.\cite{rf:coleman} \ 
When the energy is high, we expect that modified vacuum bubbles
dominate the path-integral, just like instanton anti-instanton with
shorter separation dominates as energy increases.
In fact, one may estimate the $n$-point Green function
in a thin-wall bubble background as follows;
Far from the bubble at the origin, the scalar field $\phi(x)$ 
behaves as $\sim e^{R-\rho}$
for $\rho \equiv |x| \rightarrow \infty$, where we denote the radius
of the bubble by $R$.
Thus the $n$-point function has the integrand, $e^{nR-S} $.
Minimizing this expression with respect to $R$, 
we find that the radius shifts by 
$ - O(\epsilon n / S_1^2)$, where $S_1$ is the surface energy, 
and $\epsilon$ is the energy density difference.
This leads to a smaller bubble, whose mid-section is
a three-dimensional bubble with energy of $O(n)$.
This argument demonstrates that in fact
the induced decay of the false vacuum is mediated by smaller bubbles
under the presence of low energy incoming particles.
However, this is quite a crude estimate.
Accurate analysis should be carried out on the basis of
the imaginary-time path-integral formalism. 
The starting point is then to clarify the
structure of those modified bubbles.
This series of bubbles forms a valley, on which the bounce
is the saddle point.
In this subsection, we carry out the analysis of this valley,
employing the proper valley method.
We consider a quantum field theory of a scalar field $\phi(x)$ in 
3+1 dimensional space-time with an action $S$, which we shall specify later.
The proper valley equation (\ref{eq:gennv}) can be obtained by varying the 
following total action, $S_{\rm tot}$,
\begin{equation}
S_{\rm tot} = S_{\rm pv} +S_{\rm F}, 
\label{eq:s42snewvalley}
\end{equation}
where
\begin{eqnarray}
S_{\rm pv} &=& S +S_\lambda, \quad
S_\lambda = -\frac{1}{2 \lambda} \int d^4x 
\left( {\delta S \over \delta\phi(x)} \right)^2_, \\
S_{\rm F} &=& {1 \over 2 \lambda} \int d^4x 
\left(F(x) - {\delta S \over \delta\phi(x)} \right)^2_. 
\label{eq:s42snewvalley2}
\end{eqnarray}

We consider the following action, 
\begin{equation}
S = \int d^4 x \ 
        \left[ {1 \over 2} 
        \left( \partial_\mu \phi\right)^2 + V(\phi)
        \right]_,
\quad
V(\phi) = {1 \over 2} \phi^2 (1- \phi)^2
        -\epsilon (4 \phi^3 - 3 \phi^4).
\label{eq:s42action}
\end{equation}
Since the above potential is same as Eq.(\ref{eq:s41potone}) 
of $g=1$, it has the false minima at $\phi =0$ and 
the true minima at $\phi = 1$, regardless of the value of the
parameter $\epsilon ( \ge 0)$.
The energy density of the true vacuum is $-\epsilon$,
even for large $\epsilon$.
In this sense, this action defines a convenient model for study
of thick-wall bubbles as well as thin-wall ones.
\begin{figure}
\centerline{\epsfxsize=10cm\epsfbox{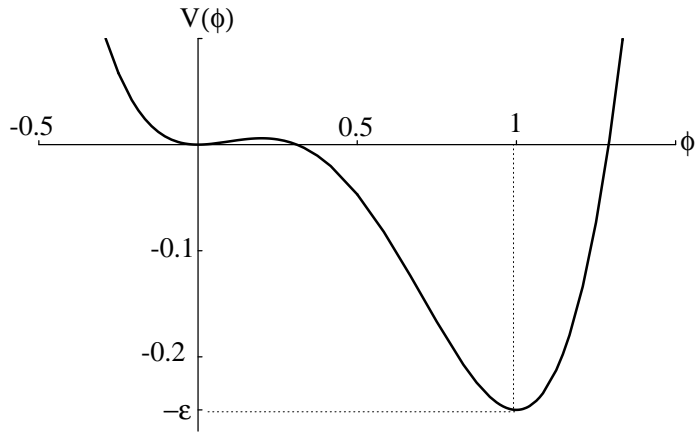}}
\vspace{3mm}
\caption{Potential \protect $V(\phi )$ for \protect $\epsilon =0.25$.}
\label{fig:s42pot}
\end{figure}

As the bounce solution is spherically symmetric, 
we shall confine ourselves to the study of spherically symmetric
configurations, 
$\phi(x) = \phi(\rho)$ and $F(x) = F(\rho)$, 
where $\rho =\sqrt{{x_\mu}^2}$.
The proper valley equations lead to the following;
\begin{eqnarray}
        \phi\dprime + {3 \over \rho} \phi\sp
                - {d V\over d\phi} + F &= 0, 
\label{eq:s42actualeq1}
\end{eqnarray}
\begin{eqnarray}
        F\dprime + {3 \over \rho} F\sp
                - {d^2 V \over d\phi^2} F + \lambda F &= 0, 
\label{eq:s42actualeq2}
\end{eqnarray}
where we denote the derivatives with respect to $\rho$ by primes.
Just as in Coleman's treatise of the bounce equation,
the above equations can be thought as the set of Minkowskian equations
of motion of a particle in two dimensional space $(\phi, F)$ 
at ``time" $\rho$.
The linear differential terms of $\phi$ and $F$ act as friction terms.
(Note that there are no friction term in Eq.(\ref{eq:s41nvone}) and 
Eq.(\ref{eq:s41nvtwo}).)
The other terms are space-dependent forces.
The big difference is that now this force is not conservative.
Therefore, no simple energy argument is possible.

The proper valley equations Eq.(\ref{eq:s42actualeq1}) and 
Eq.(\ref{eq:s42actualeq2}) require four 
boundary conditions. 
They are as follows:
For the solution to be regular at the origin $\rho=0$,
we require boundary conditions $\phi\sp(0)=F\sp(0)=0$.
The outside of the bubble has to be the false vacuum, so
$\phi(\infty)=F(\infty)=0$ have to be satisfied.

In solving Eq.(\ref{eq:s42actualeq1}) and Eq.(\ref{eq:s42actualeq2}) 
numerically, we have chosen to start at the origin
with the boundary conditions $\phi\sp(0)=F\sp(0)=0$ and
adjust $\phi(0)$ and $F(0)$ so that $\phi(\infty)=F(\infty)=0$
are (approximately) satisfied.
We have done this calculation for $\epsilon = 0.25$.
The potential for this value of $\epsilon$ is given in Fig.~\ref{fig:s42pot}.
Since the depth of the true vacuum is much more than the height
of the potential barrier, we expect that the bounce solution
is a bubble with a thick wall.

\begin{figure}
\centerline{\epsfxsize=10cm\epsfbox{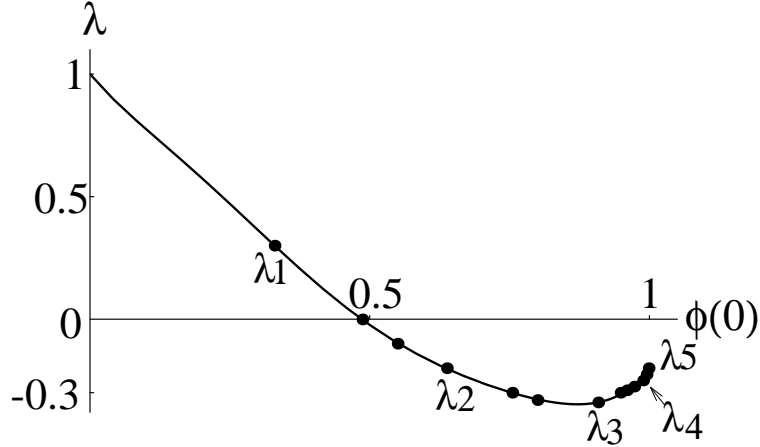}}
\caption{The value of the smallest eigenvalue \protect $\lambda$ of $D$
as a function of \protect $\phi(0)$.}
\label{fig:s42lam}
\end{figure}

In Fig.~\ref{fig:s42solutions}, we show the numerical solutions 
and Fig.~\ref{fig:s42lam} is the relation
between smallest eigenvalue $\lambda$ of $D$ and the starting point
$\phi(0)$.
We observe the following in these figures:
\begin{figure}
\centerline{\epsfxsize=10cm\epsfbox{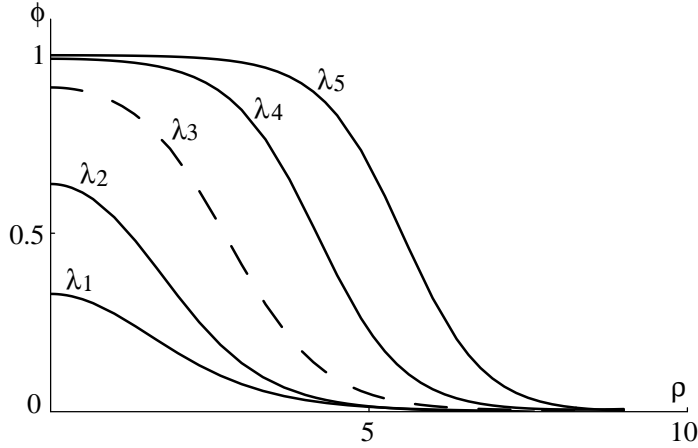}}
\vspace{3mm}
\caption{Shapes of the solutions of the proper valley 
equation Eq.(\protect \ref{eq:s42actualeq1}) and 
Eq.(\protect \ref{eq:s42actualeq2}).
The eigenvalues \protect $\lambda_{1\sim5}$ of each lines are 
\protect $\lambda_1 = 0.3$, \protect $\lambda_2 = -0.2$, 
\protect $\lambda_3 = -0.34$, 
\protect $\lambda_4 = -0.25$, \protect $\lambda_5 = -0.2$.
The broken line shows the bounce solution.}
\label{fig:s42solutions}
\end{figure}

(1) The proper valley contains the bounce solution 
(broken line in Fig.~\ref{fig:s42solutions}).
This solution has a thick wall, as expected.
It has a negative eigenvalue $\lambda_3$, 
which causes the instability.
Another solutions that construct the valley are not the classical
solutions.
We call these series of the thick-wall (including the bounce) to 
the thin-wall bubbles ''valley bubbles''.
(2) In Fig.~\ref{fig:s42lam}, when $\phi(0)$ goes to zero, $\lambda$ goes to one,
because the minimum eigenvalue in the false vacuum is equal to the
mass, which is equal to one.
The eigenvalue $\lambda_3$ of the bounce solution does not occupy any
particular position in Fig.~\ref{fig:s42lam}.
As $\phi(0)$ approaches to one, $\lambda$ approaches to zero.
This is the region where we have large thin-wall bubbles, as can be
seen in the $\lambda_5$-line of Fig.~\ref{fig:s42solutions}.
(There is another zero point near $\phi(0)\sim 0.5$, but this point does
not play any special role.)
The interior of these bubbles is the true vacuum, $\phi=1$. 
The latter property is especially notable: Even though the bounce solution
is a thick-wall bubble, we find that the valley contains large, thin-wall,
clean bubbles in the outskirts.

These thin-wall bubbles can be analyzed by extending the original
Coleman's argument:
The solution of the proper valley equation Eq.(\ref{eq:s42actualeq1})
and Eq.(\ref{eq:s42actualeq2}) extremizes the
action $S_{\rm PV}+S_{\rm F}$. 
Using the valley equation Eq.(\ref{eq:gennv}), we rewrite it as follows
for negative $\lambda$;
\begin{equation}
S_{\rm PV}+S_{\rm F}=S + S_\lambda, \quad
        S_\lambda = {1\over 2 |\lambda|}\int d^4x F^2. 
\label{eq:s42react}
\end{equation}
Now consider a fictitious particle in a one-dimensional space
$\phi$ at time $\rho$.
In the equation Eq.(\ref{eq:s42actualeq1}),
the auxiliary field $-F(\rho)$ 
acts as an ``external force" for this particle.
If the initial value $\phi(0)$ is sufficiently close 
to $1$ and $F$ is sufficiently small, $\phi$ remains close
to $1$ for a long time, until the friction term dies away.
When it finally rolls down the hill, it does so under the external force 
$-F(\rho)$.
If $-F(\rho)$ is just right, the particle approaches to
the top of the lesser hill $\phi=0$ asymptotically. 
Note that there is a major difference with Coleman's
argument here: When the roll-down occurs, it does so under
the external force, which is the sole source of the 
energy reduction, while in the bounce solution the timing of the
roll-down has to be such that the friction term is just right 
to take care of the extra energy.
[This does not prove the existence of the thin-wall solution, but 
Fig.~\ref{fig:s42solutions} shows that this in fact happens.]
Since $-F(\rho)$ is the stopping force, it deviates from zero only at the wall.
Therefore the second term in Eq.(\ref{eq:s42react}) contributes positively {\sl only} 
at the wall.
The actions $S$ and $S_{\lambda}$ can be approximately written in
terms of the radius of the bubble $R$ as the following;
\begin{eqnarray}
S &=& -\epsilon {\pi^2 \over 2} R^4 + S_1 2\pi^2 R^3, \\
        S_\lambda &=& {W_F\over 2 |\lambda|} 2\pi^2 R^3,
\label{eq:s42actrad}
\end{eqnarray}
where $S_1$ and $W_F$ are the numbers of $O(1)$.
Taking the derivative of $S + S_\lambda$ in the above 
with respect to $R$, we find the radius of the
solution of the proper valley equation to be
\begin{equation}
R_{\rm PV} = {3\over \epsilon}
        \left(S_1 + {W_F \over 2 |\lambda|} \right). 
\label{eq:s42rad}
\end{equation}
Therefore, even when $\epsilon$ is not small enough to guarantee
the large radius, $|\lambda|$ can be small enough to do so.
This is what is causing the thin-wall bubble to be a solution
of the proper valley equation.

The shape of the large bubble and its thin wall can be
examined in detail by looking at the 0+1 dimensional model,
since the friction term can be neglected.
We have thus also analyzed the 0+1 dimensional model.
(See the previous subsection.)
We have obtained the numerical value $W_F \simeq 0.2104$ from this analysis.
The radius seen in Fig.~\ref{fig:s42solutions} for $\lambda_{4,5}$ is in agreement
with Eq.(\ref{eq:s42rad}) for this value of $W_F$.

\begin{figure}
\centerline{\epsfxsize=10cm\epsfbox{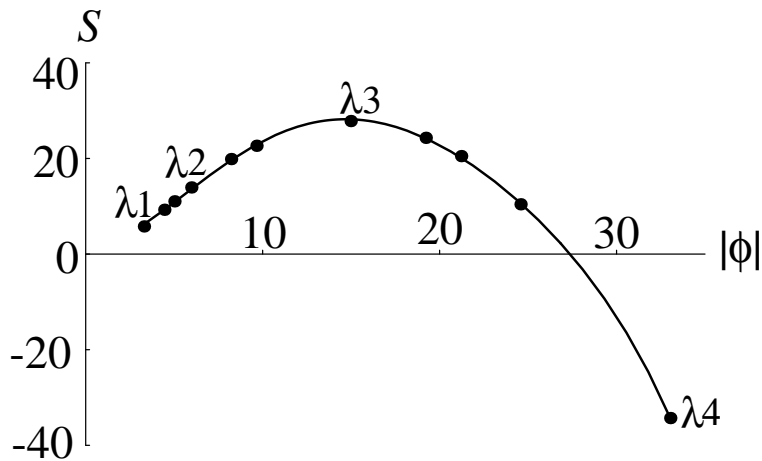}}
\caption{The action of the solutions of the proper valley equation
as a function of the norm \protect $|\phi|$.
The points with eigenvalues \protect $\lambda_{1\sim4}$ correspond to the lines in 
Fig.~\protect\ref{fig:s42solutions}.}
\label{fig:s42value}
\end{figure}

Numerical values of the action $S$ are plotted in Fig.~\ref{fig:s42value},
where the horizontal coordinate is the ``norm" of the
solution, $|\phi| \equiv \sqrt{\int d^4x \ \phi(x)^2}$,
which for large $R$ should be $\sim \sqrt{\pi^2/2}R^2$.
We see in this figure that the bounce solution denoted by 
its smallest eigenvalue, $\lambda_3$, is in fact at the maximum point of
the valley.
We have also compared the asymptotic expression of the action $S$
in Eq.(\ref{eq:s42actrad}) with numerical value of $S$ for large $|\phi|$ 
and have confirmed that
the action is in fact dominated by the volume and the surface terms.

\begin{figure}
\centerline{\epsfxsize=10cm\epsfbox{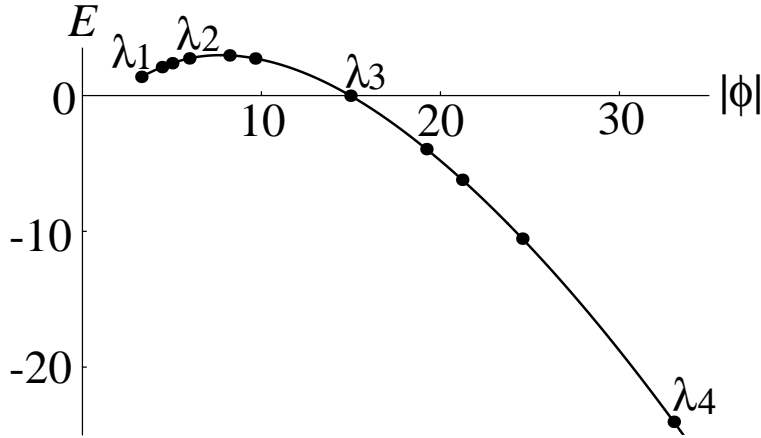}}
\caption{The energy of the bubbles at the mid-section,
where \protect $\partial \phi / \partial \tau = 0$.}
\label{fig:s42energy}
\end{figure}

On the other side of the valley are small bubbles.
In order to see their role, we have calculated the energy $E$
of the mid-section of bubbles and plotted the
result in Fig.~\ref{fig:s42energy}.
As is well known, the bounce solution ($\lambda_3$) has $E=0$, so that
the energy conservation allows it to contribute to 
the tunneling from the false vacuum.
We find that the bubbles smaller than the bounce have $E > 0$, while
the larger ones have $E < 0$.
This shows that the smaller bubble contribute to the
tunneling from the states with higher energy than the false vacuum.

We have numerically examined the valley that contains the bounce solution.
We have obtained a series of deformed bounce solutions, 
valley bubbles, on the valley.
These configurations play the roles similar to instanton
anti-instanton pairs with various distances.
We have found that even when the bounce solution is a thick-wall bubble,
the valley contains thin-wall large bubbles, whose interior is the true vacuum.
Smaller bubbles are also identified. We expect them to contribute
to the decay of the higher energy states.
This is quite an interesting possibility, which should be explored further.
We note that we confined ourselves to spherical configurations
in this letter. However, non-spherical configurations 
may also be interesting, as they could also contribute to the induced decay.

\subsection{Valley instanton in the scalar $\phi^4$ field theory} 
In the following two subsections, using the proper valley method, we
examine the \linebreak
instanton-like configurations in the unstable scalar field
theory and the gauge-Higgs system.\cite{rf:ahsw} 
In both systems, the instanton-like configuration has an important
role in the quantum tunneling, but a simple scaling argument shows
that no classical solution exists except a zero size one.

To construct finite size instanton-like configurations, it was often
used the so-called the constrained instanton method,\cite{rf:Aff}
which was explained briefly in section 2 and 3. 
In this method, one introduces a constraint to
define a sub-functional space of finite radius configurations.
The field equation is solved in this subspace
and the finite-radius configuration similar to instantons
is defined. After doing the one-loop (and possibly higher order)
integral in this subspace,
the constraint parameter is integrated over.
The configuration constructed this way is called ``constrained
instanton''.
One problem about this method is that its validity
depends on the choice of the constraint:
Since in practice one does the Gaussian integration around the
solution under the constraint,
the degree of approximation depends on the way constraint is introduced.
Unfortunately, no known criterion guarantees the
effectiveness of the approximation.

This situation could be remedied once one realizes that
what we have near the point-like (true) instanton is the
valley.\cite{rf:ahsw}
That is, although the zero-size instanton may be the dominating
configurations, it is expected to be followed by a series of
configurations that makes the valley of the action.
Therefore, instead of trying to cover all the neighborhoods of
the zero-radius instanton, one may cover the valley region,
which is expected to dominate the path-integral.
The trajectory along the valley bottom should correspond to
the scaling parameter, or the radius parameter of the instanton.
As such, the finite-size instanton can be defined as
configurations along the valley trajectory.
This is similar to a calculation in the electroweak theory in which
one evaluates the contribution of the instanton-anti-instanton
valley.\cite{rf:balitsky,rf:yung,rf:khozering}
Thus treating a single instanton as a configuration on the valley
provides a means of unifying the approximation schemes.
These configurations are named ``valley instanton''.

One convenient way to define the valley trajectory
is to use the proper valley method.
Since the constrained instanton has been used extensively in
the existing literature, our main purpose here is to establish
the existence of the valley instanton on a firm basis and
compare its properties with that of the constrained instanton.
This should serve as a starting point for the reanalysis of
the existing theories and results under the new light.

First, in this subsection, we consider the scalar $\phi^{4}$ theory in
the 4 dimensional space-time 
with Euclidean Lagrangian, 
\begin{eqnarray}
{\cal L}=\frac{1}{g^2}\left[ \frac{1}{2}(\partial_{\mu} \phi)^2 +\frac{1}{2}
\mu^2 \phi^2 -\frac{1}{4!}\phi^4 \right].
\label{eq:s43scalarlagrangian}
\end{eqnarray}
The negative sign for the $\phi^4$ term is useful for
estimating the asymptotic behavior of the perturbative expansion in
the theory with a positive $\phi^4$
term.\cite{rf:Aff,rf:asym,rf:brezin,rf:parisi,rf:fy} 
Namely, the instanton-like configurations in the negative $\phi^4$
theory reveal the large order behavior of the positive $\phi^4$ theory.
This model allows finite-size instantons only for $\mu=0$;
otherwise, a scaling argument shows  
that the action of any finite-size configuration
can be reduced by reducing its size.
As is shown in the following, when the size of the configuration is
small, the valley instanton can be constructed analytically.
The differences from the constrained instanton are rather small in
this case.
For a configuration with a large radius, we construct both
instanton numerically.
It will be found that remarkable differences between them appear in this
case.
While the action of the valley instanton increases monotonously and
remains positive, the one of the constrained instanton becomes
negative when the radius becomes large.

\subsubsection{Valley instanton}
For this model, the proper valley equation introduced in the previous
section is 
\begin{eqnarray}
\begin{array}{cc}
&\displaystyle -\partial_{\mu}\partial_{\mu} \phi + \mu^2 \phi
- \frac{1}{3!}\phi^2 =F,
\\[0.5cm]
&\displaystyle \left( -\partial_{\mu}\partial_{\mu}+\mu^2 -\frac{1}{2}
\phi^2\right)F=\lambda F.
\end{array}
\label{eq:s43scalarge:gennv}
\end{eqnarray}
Now we introduce the scale parameter $\rho$ 
defined by $\phi(0) \equiv 4\sqrt{3}/\rho$, in order to
fix the radius of the instanton solution to be unity.
We rescale fields and variables as in the following;
\begin{equation}
r=\frac{\sqrt{x^2}}{\rho},\quad
\lambda=\mu^2\nu,\quad
\phi(x)=\frac{h(r)}{\rho},\quad
F(x)=\frac{\mu^2}{\rho} f(r).
\label{eq:s43scalarrescale}
\end{equation}
The proper valley equation (\ref{eq:s43scalarge:gennv}) is given by
the following under this rescaling;
\begin{eqnarray}
&&-\frac{1}{r^3}\frac{d}{dr}\left(r^3\frac{dh}{dr}\right)
+(\rho\mu)^2h-\frac{1}{3!}h^3=(\rho\mu)^2f,
\label{eq:s43scalarnewvalleyeq1}\\
&&-\frac{1}{r^3}\frac{d}{dr}\left(r^3\frac{df}{dr}\right)
+(\rho\mu)^2f-\frac{1}{2}h^2f=(\rho\mu)^2\nu f.
\label{eq:s43scalarnewvalleyeq2}
\end{eqnarray}

This system has an instanton solution in the massless limit, $\rho \mu \to
0$.\cite{rf:asym,rf:brezin}
In this limit, (\ref{eq:s43scalarnewvalleyeq1}) reduces to the equation of
motion and (\ref{eq:s43scalarnewvalleyeq2}) 
the equation for the zero-mode fluctuation, $f$, around the instanton solution.
The solution is the following;
\begin{eqnarray}
\begin{array}{cc}
\displaystyle h_{0}=\frac{4\sqrt{3}}{1+r^{2}},
&\displaystyle f_{0}
=C\left(\frac{4\sqrt{3}}{1+r^2}-\frac{8\sqrt{3}}{(1+r^2)^2}\right),
\end{array}
\label{eq:s43scalarinstanton}
\end{eqnarray}
where $C$ is an arbitrary constant.
Note that 
solution $f_0$ is obtained from $\partial \phi_0 (x) /\partial \rho$,
$\phi_0=h_0/\rho$.

Let us construct the valley instanton in the scalar $\phi^4$ theory 
analytically.
When $\rho \mu$ is very small but finite, the valley instanton
is expected to have $\rho \mu$ corrections to (\ref{eq:s43scalarinstanton}).
On the other hand, at large distance from the core region of the 
valley instanton, since the term of $O(h^2)$ is negligible, the 
valley equation can be linearized.
This linearized equation can be solved easily.
By matching the solution near the core and the solution in the asymptotic 
region in the overlapping intermediate region, we can construct 
approximate solution analytically.
We will carry out this procedure in the following.

In the asymptotic region,  $(\rho\mu)^2\gg h^2$, 
the linearized valley equation is the following;
\begin{eqnarray}
\begin{array}{cc}
&\displaystyle -\frac{1}{r^3}\frac{d}{dr}\left(r^3\frac{dh}{dr}\right)
+(\rho\mu)^2h=(\rho\mu)^2f,
\\[0.5cm]
&\displaystyle -\frac{1}{r^3}\frac{d}{dr}\left(r^3\frac{df}{dr}\right)
+(\rho\mu)^2f=(\rho\mu)^2\nu f.
\end{array}
\label{eq:s43scalarlinernveq}
\end{eqnarray}
The solution of these equations is
\begin{equation}
h(r)=C_1G_{\rho\mu}(r)+\frac{f}{\nu},\quad
f(r)=C_2G_{\rho\mu\sqrt{1-\nu}}(r),
\label{eq:s43scalarasymptoticsol}
\end{equation}
where $C_1$ and $C_2$ are arbitrary functions of $\rho \mu$.
The function $G_{m}(r)$ is
\begin{equation}
G_{m}(r)=\frac{m K_{1}(m r)}{(2\pi)^{2}r},
\end{equation}
where $K_{1}$ is a modified Bessel function.
The functions $f$ and $h$ decay exponentially at large $r$.
In the region of $r \ll (\rho \mu)^{-1}$, $r \ll (\rho \mu \sqrt{1-\nu})^
{-1}$, $f$ and $h$ can be expanded in series as the following;
\begin{eqnarray}
\begin{array}{lll}
&\displaystyle h=\frac{C_1}{(2\pi)^2}
\left[\frac{1}{r^2}+\frac{1}{2}(\rho\mu)^2\ln(\rho\mu rc)
+\cdots\right]\\[0.5cm]
&\displaystyle \hspace{1.7ex}+\frac{C_2}{(2\pi)^2\nu}
\left[\frac{1}{r^2}+\frac{1}{2}(\rho\mu)^2(1-\nu)
\ln(\rho\mu\sqrt{1-\nu}rc)+\cdots\right],
\\[0.5cm]
&\displaystyle f=\frac{C_2}{(2\pi)^2}
\left[\frac{1}{r^2}+\frac{1}{2}(\rho\mu)^2(1-\nu)
\ln(\rho\mu\sqrt{1-\nu}rc)+\cdots\right].
\end{array}
\label{eq:s43scalarexpand}
\end{eqnarray}
In the above, $c= e^{\gamma-1/2}/2$, where $\gamma$ is
the Euler's constant.

Near the origin, we expect that the valley 
instanton is similar to the ordinary instanton.
It is convenient to define $\hat{h}$ and $\hat{f}$ as the following;
\begin{equation}
h=h_0+(\rho\mu)^2\hat{h},\qquad
f=f_0+(\rho\mu)^2\hat{f}, 
\label{eq:s43scalarperturbation}
\end{equation}
where $C$ in $f_0$ is the function of $\rho \mu$ and is decided in 
the following.
The ``core region'' is defined as 
\begin{equation}
h_0\gg (\rho\mu)^2\hat{h},\qquad
f_0\gg (\rho\mu)^2\hat{f}.
\label{eq:s43scalarassumption2}
\end{equation}
The valley equation for perturbation field $\hat{h}$, $\hat{f}$ becomes 
\begin{eqnarray}
\begin{array}{ll}
&\displaystyle -\frac{1}{r^3}\frac{d}{dr}\left(r^3\frac{d\hat{h}}{dr}\right)
-\frac{1}{2}h_0^2\hat{h}=f_0-h_0,
\\[0.5cm]
&\displaystyle -\frac{1}{r^3}\frac{d}{dr}\left(r^3\frac{d\hat{f}}{dr}\right)
-\frac{1}{2}h_0^2\hat{f}=(\nu-1)f_0+h_0f_0\hat{h}.
\end{array}
\label{eq:s43scalarasymptoticveq}
\end{eqnarray}
The left-hand side of (\ref{eq:s43scalarasymptoticveq}) has a zero mode, 
$\varphi$. It satisfies the equation,
\begin{eqnarray}
-\frac{1}{r^3} \frac{d}{dr}\left(r^3 \frac{d \varphi}{d r}\right) 
- \frac{1}{2} h_0^2 \, \varphi =0,
\label{eq:s43scalarzeromodeeq}
\end{eqnarray}
and is given by the following;
\begin{equation}
\varphi=\frac{4\sqrt{3}}{1+r^2}-\frac{8\sqrt{3}}{(1+r^2)^2}.
\label{eq:s43scalarzeromode}
\end{equation}
We multiply $r^3 \varphi$ to both sides of (\ref{eq:s43scalarasymptoticveq}), 
and integrate them from $0$ to $r$.
The existence of zero mode $\varphi$ makes it possible 
to integrate the left-hand side of (\ref{eq:s43scalarasymptoticveq}).
As a result of the integration by parts, only the surface terms remain 
and we obtain,
\begin{eqnarray}
& \displaystyle -r^3\varphi\frac{d\hat{h}}{dr}+r^3\frac{d\varphi}{dr}\hat{h}
=\int_0^{r}dr'r'^3\varphi\left[f_0-h_0\right],
\label{eq:s43valleyeqpart21}
\\[0.5cm]
& \displaystyle -r^3\varphi\frac{d\hat{f}}{dr}+r^3\frac{d\varphi}{dr}\hat{f}
=\int_0^{r}dr'r'^3\varphi\left[(\nu-1)f_0+h_0f_0\hat{h}\right].
\label{eq:s43valleyeqpart22}
\end{eqnarray}
First, using (\ref{eq:s43scalarinstanton}) and (\ref{eq:s43scalarzeromode}), 
we can find that the right-hand side of (\ref{eq:s43valleyeqpart21}) is 
proportional to $\ln r$ at $r \gg 1$.
Thus (\ref{eq:s43valleyeqpart21}) becomes
\begin{eqnarray}
&\displaystyle r \frac{d \hat{h}}{dr}+2 \hat{h}
 =4\sqrt{3}\,(1-C) \ln r.
\label{eq:s43valleyeqpart31}
\end{eqnarray}
This equation can be solved easily at $r\gg 1$.
Using the solution of (\ref{eq:s43valleyeqpart31}), we can find the 
right-hand side of (\ref{eq:s43valleyeqpart21}) also proportional to $\ln r$ as 
the following;
\begin{eqnarray}
&\displaystyle  r \frac{d \hat{f}}{dr}+2 \hat{f}
 =4\sqrt{3}\,C (1-\nu) \ln r.
\label{eq:s43valleyeqpart32}
\end{eqnarray}
Finally, we obtain $\hat{h}$ and $\hat{f}$ at $r\gg 1$,
\begin{equation}
 \hat{h}=2\sqrt{3}\, (1-C)\ln r+\cdots, \quad
 \hat{f}=(1-\nu)2\sqrt{3}\, C\ln r+\cdots.
\end{equation}
For these solutions to meet to (\ref{eq:s43scalarexpand}), 
the parameters need to be the following;
\begin{equation}
C_1=0,\quad
C_2=4\sqrt{3}\,(2\pi)^2,\quad
\nu=1,\quad
C=1,
\label{eq:s43scalarmatchingpara}
\end{equation}
as $\rho \mu =0$.
Now we have obtained the solution of the proper valley equation;
\begin{equation}
h(r)=\left\{
    \begin{array}{ll}
\displaystyle      \frac{4\sqrt{3}}{1+r^2},& 
\quad  \mbox{if} \quad r \ll (\rho \mu)^{-1/2};  \\[0.4cm]
\displaystyle      \frac{4\sqrt{3}}{r^2} + o\left((\rho \mu)^2 \right),& 
\quad \mbox{if} \quad (\rho \mu)^{-1/2} \ll r \ll (\rho \mu)^{-1}; \\[0.4cm]
\displaystyle      4 \sqrt{3}\, (2 \pi^2) \,G_{\rho \mu \sqrt{1-\nu}}\,(r),& 
\quad \mbox{if} \quad (\rho \mu)^{-1/2} \ll r. 
     \end{array} \right.
\label{eq:s43scalarinstanalyeq}
\end{equation}

Let us discuss the consistency of our analysis.
In the construction of the analytical solution, especially 
in the argument of the matching of the core and asymptotic region solution, 
we have implicitly assumed that there exists an overlapping region 
where both (\ref{eq:s43scalarlinernveq}) and 
(\ref{eq:s43scalarasymptoticveq}) are valid.
Using the solution (\ref{eq:s43scalarinstanalyeq}), it is found that 
(\ref{eq:s43scalarlinernveq}) is valid in the 
region of $r \gg (\rho \mu)^{-1/2}$, and 
(\ref{eq:s43scalarasymptoticveq}) is valid 
in the region of $r \ll (\rho \mu)^{-1}$.
Therefore in the above analysis we have limited 
our calculation in  
the overlapping region $(\rho \mu)^{-1/2} \ll r \ll (\rho \mu)^{-1}$.

We calculate the action of the valley instanton using the above solution.
Rewriting the action in terms of $h(r)$, we find
\begin{equation}
S=\frac{\pi^2}{g^2}\int_0^\infty dr\, r^3 \left[
\left(\frac{dh}{dr}\right)^2 + (\rho \mu)^2 h^2 - \frac{1}{12} h^4 \right].
\end{equation}
Substituting the analytic solution for $S$, we obtain
\begin{equation}
S=\frac{16 \pi^2}{g^2} + O\left((\rho \mu)^2\right).
\label{eq:s43scalarresultactionnv}
\end{equation}
The leading contribution term comes from the ordinary instanton 
solution, and the correction term comes from the distortion of 
the instanton solution.

\subsubsection{Constrained instanton}
Here, we consider the constrained instanton in the scalar 
$\phi^4$ theory, following the construction in Ref.~\citen{rf:Aff}.
We require the constraint in the path-integral.
The field equation under the constraint is
\begin{eqnarray}
\frac{\delta S}{\delta \phi} + \sigma \frac{\delta O}{\delta \phi}=0,
\end{eqnarray}
where $\sigma$ is a Lagrange multiplier.
The functional $O$ had to satisfy the certain scaling properties 
that guarantee the existence of the solution.\cite{rf:Aff}
We choose it as follows;
\begin{eqnarray}
O=\int d^4 x \,\frac{\phi^6}{6}.
\end{eqnarray}
This choice is one of the simplest for constructing the constrained 
instanton in the scalar $\phi^4$ theory.
Again, adopting the rescaling (\ref{eq:s43scalarrescale}), 
the equation of motion under this constraint becomes
\begin{eqnarray}
-\frac{1}{r^3}\frac{d}{dr}\left(r^3\frac{dh}{dr}\right)
+(\rho\mu)^2h-\frac{1}{3!}h^3+(\rho\mu)^2\tilde{\sigma}h^5=0,
\label{eq:s43scalarconstrainedeq}
\end{eqnarray}
where we rescale the parameter $\sigma$ as $\sigma=(\rho\mu)^2\widetilde
{\sigma}$.
The solution of this equation can be constructed in a manner similar to the 
previous subsection.
To carry out the perturbation calculation in the core region, 
we replace the field variable as $h=h_0 + (\rho \mu)^2 \hat{h}$, where 
$h_0 \gg (\rho \mu)^2 \hat{h}$.
The field equation of $\hat{h}$ becomes
\begin{equation}
-\frac{1}{r^3}\frac{d}{dr}\left(r^3\frac{d\hat{h}}{dr}\right)
-\frac{1}{2}h_0^2\hat{h}=-h_0-\tilde{\sigma}h_0^5.
\label{eq:s43scalarconsthhateq}
\end{equation}
We multiply this equation by the zero mode (\ref{eq:s43scalarzeromode}) and 
integrate this from $0$ to $r$.
Then we obtain 
\begin{equation}
-r^3\varphi\frac{d\hat{h}}{dr}+r^3\frac{d\varphi}{dr}\hat{h}=
-\int_0^{r}dr'r'^3\varphi\left[h_0+\tilde{\sigma}h_0^5\right].
\end{equation}
The solution of this equation in the region where $r \gg 1$ is
\begin{equation}
\hat{h}=2\sqrt{3}\ln r -\frac{192\sqrt{3}}{7}\tilde{\sigma}
+\cdots.
\label{eq:s43scalarconstcoresol}
\end{equation}
In the asymptotic region, we consider the field variable and the 
parameter as $h^2 \ll (\rho \mu)^2$, $\widetilde{\sigma}h^4\ll 1$.
Under this condition, the field equation of the asymptotic region becomes 
\begin{equation}
-\frac{1}{r^3}\frac{d}{dr}\left(r^3\frac{dh}{dr}\right)
+(\rho\mu)^2h=0.
\label{eq:s43scalarconstfieldeqatasym}
\end{equation}
The solution is 
\begin{equation}
h(r)=C_1G_{\rho\mu}(r),
\end{equation}
where $C_1$ is arbitrary constant.
In the region of $r \ll (\rho \mu)^{-1}$, this solution can be expanded as
 the following;
\begin{equation}
h=\frac{C_1}{(2\pi)^2}\left[\frac{1}{r^2}
+\frac{1}{2}(\rho\mu)^2\ln(\rho\mu rc)+\cdots\right].
\label{eq:s43scalarconstasymsol}
\end{equation}
Matching this solution and the core region solution 
(\ref{eq:s43scalarconstcoresol}), parameters are determined as the following;
\begin{equation}
C_1=4\sqrt{3}(2\pi)^2,\quad \tilde{\sigma}=-\frac{7}{96}\ln(\rho\mu).
\label{eq:s43scalarconstparameter}
\end{equation}

To summarize, the analytical solution of the constrained instanton
we have obtained is the following,
\begin{equation}
h(r)=\left\{
    \begin{array}{ll}
\displaystyle \frac{4\sqrt{3}}{1+r^2},& 
\quad \mbox{if} \quad r \ll (\rho \mu)^{-1/2}; \\[0.4cm]
\displaystyle \frac{4\sqrt{3}}{r^2} + 2 \sqrt{3} (\rho \mu)^2 \ln(\rho \mu rc) 
+ o\left((\rho \mu)^2 \right), & 
\quad \mbox{if} \quad (\rho \mu)^{-1/2} \ll r \ll (\rho \mu)^{-1}; \\[0.4cm]
\displaystyle 4 \sqrt{3}\, (2 \pi^2) \,G_{\rho \mu }\,(r),& 
\quad \mbox{if} \quad (\rho \mu)^{-1/2} \ll r. 
     \end{array} \right.
\label{eq:s43scalarconstanalyeq}
\end{equation}

The action of the constrained instanton is given by the following;
\begin{equation}
S=\frac{16 \pi^2}{g^2} - 96 \pi^2 (\rho \mu)^2 \ln (\rho \mu) 
+ O\left((\rho \mu)^2\right).
\label{eq:s43scalarresultaction}
\end{equation}
This differs from the action of the valley instanton at the next-to-leading 
order.
This correction term shows that the constrained instanton is more distorted 
from the ordinary instanton than the 
valley instanton.

\subsubsection{Numerical analysis}
In this subsection, we calculate the valley equation 
(\ref{eq:s43scalarnewvalleyeq1}), (\ref{eq:s43scalarnewvalleyeq2}) and 
the constrained equation (\ref{eq:s43scalarconstrainedeq}) numerically.
Then we compare the valley and the constrained instanton.

Each of the equations is the second order differential equation, so 
we require two boundary conditions for each field variable to 
decide the solution.
We require all the field variables are regular at the origin.
The finiteness of the action requires $h, f\to 0$ faster than $1/r^2$ at 
infinity.
In solving (\ref{eq:s43scalarnewvalleyeq1}) and (\ref{eq:s43scalarnewvalleyeq2}), 
we adjust the parameter $\nu$ and $f(0)$ so that $h,f \to 0$ at infinity 
for the fixed $\rho \mu$.
In the similar way, in case of the constrained instanton, the parameter 
$\sigma$ is determined so that $h\to 0$ at infinity.

\begin{figure}
\epsfxsize=6.5cm
\centerline{
\epsfbox{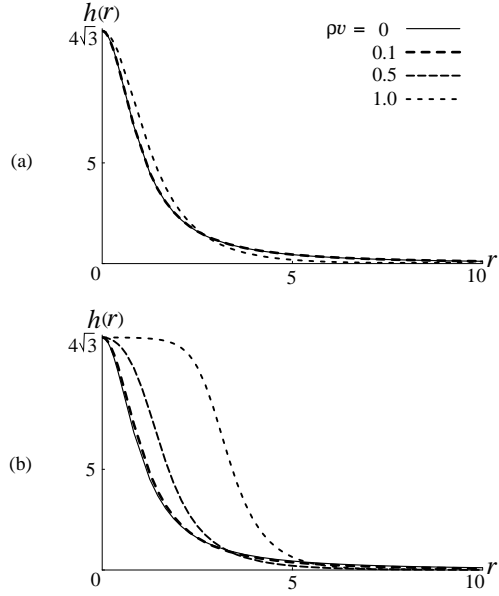}
}
\caption{(a) Shapes of the numerical solution of the valley
instanton,
 $h(r)$, for $\rho \mu=$$0,$ $0.1,$ and $1.0$ near the origin.
The solid line denotes the original instanton, $h_0$.
(b) Shapes of the constrained instanton for $\rho \mu=$$0,$ $0.1,$ $0.5,$ 
and $1.0$.}
\label{fig:s43scalarconfig}
\end{figure}

Numerical solutions of the valley equation near the origin are plotted 
in Fig.~\ref{fig:s43scalarconfig} (a) for $\rho \mu=0.1,\, 1.0$.
The solid line shows the instanton solutions (\ref{eq:s43scalarinstanton}), 
which corresponds to $\rho \mu=0$. 
The numerical solutions of the constrained instanton are also plotted in 
Fig.~\ref{fig:s43scalarconfig} (b) for $\rho \mu=0.1, \,0.5,\,1.0$.
Both the valley and the constrained solution for $\rho \mu=0.1$ agree 
with the analytical result.
As $\rho \mu$ becomes large, both solutions are deformed from the original 
instanton solution.
We find that the distortion of the constrained instanton 
is much larger than that of the valley instanton.
This also agrees with the analytic result.
In the analytical solution (\ref{eq:s43scalarconstanalyeq}), the correction 
term $2 \sqrt{3} (\rho \mu)^2 \ln(\rho \mu rc)$ contributes to this 
distortion.
On the other hand, the correction term of the valley instanton 
(\ref{eq:s43scalarconstanalyeq}) is $o\left((\rho \mu)^2\right)$, which is 
smaller than the previous one.
In addition, we find that the exponentially damping behavior 
of the analytical solution in the asymptotic region, where 
$r \gg (\rho \mu)^{-1}$ agrees with the result of numerical analysis.

\begin{figure}
\epsfxsize=6.5cm
\centerline{
\epsfbox{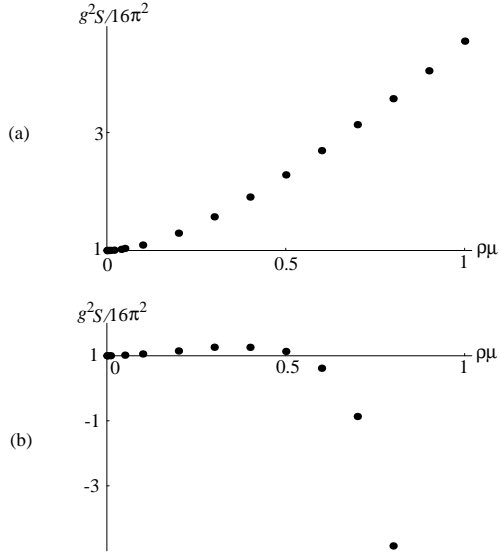}
}
\caption{(a) The action $S$ (in units of $g^2/16 \pi^2$) of the
numerical solution 
of the valley equation as a function of the parameter $\rho \mu$.
(b) The action $S$ of the constrained instanton.}
\label{fig:s43scalaraction}
\end{figure}

The values of the action of the valley instanton for $\rho \mu=$$0.0001 \sim 
1.0$ are plotted in Fig.~\ref{fig:s43scalaraction} (a).
If $\rho \mu$ is very small, this result is consistent with 
(\ref{eq:s43scalarresultactionnv}).
The values of the action of the constrained instanton for $\rho \mu=$ 
$0.0001 \sim 0.8$ are plotted in Fig.~\ref{fig:s43scalaraction} (b).
This figure shows that the behavior of the action is similar to that of  
the valley instanton when $\rho \mu$ is very small.
When  $\rho \mu$ becomes large, the behavior of the action 
is different from the valley instanton case.

\subsection{Valley instanton in the gauge-Higgs system} 

\newcommand{\lsim}{\,\lower0.9ex\hbox{$\stackrel{\displaystyle >}{\sim}$}\,}
\newcommand{\rsimil}{\,\lower0.9ex\hbox{$\stackrel{\displaystyle <}{\sim}$}\,}

In this subsection, we consider the SU(2) gauge theory with one scalar
Higgs doublet, which has the following action $S=S_g+S_h$;
\begin{eqnarray}
&&S_g=\frac{1}{2 g^2} \int d^4 x\ {\rm tr} F_{\mu \nu}  F_{\mu \nu},
\label{eq:s44gaugeaction}
\\
&&S_h=\frac{1}{ \lambda}\int d^4 x\left\{\left(D_{\mu} H\right)^{\dagger}
\left(D_{\mu} H\right)+
\frac{1}{8}\left(H^{\dagger}H-v^2\right)^2\right\},
\label{eq:s44higgsaction}
\end{eqnarray}
where $F_{\mu\nu}=\partial_{\mu} A_{\nu}-\partial_{\nu}
A_{\mu}-i\left[A_{\mu},A_{\nu}\right]$ and
$D_{\mu}=\partial_{\mu}-iA_{\mu}$.
The masses of the gauge boson and the Higgs boson are given by,
\begin{eqnarray}
&&m_{_W}=\sqrt{\frac{g^2}{2\lambda}} v,\quad m_{_H}=\frac{1}{\sqrt{2}} v.
\end{eqnarray}
When $v$ is not zero, this theory is known not to have any finite-size
instanton solutions, in spite of its importance.
In the following, we will construct instanton-like configurations 
for this theory, relevant for tunneling phenomena, including
the baryon and lepton number violation processes.
As well as the unstable scalar theory, the valley instanton can be
constructed analytically when the size of the configuration is small. 
For a configuration with a large radius, numerically constructions are 
done.
There is differences between the valley instanton and the constrained
instanton in this case. 
Therefore, when the large size configuration dominate the path
integral, we must use the valley instanton instead of the constrained
instanton. 

\subsubsection{Valley instanton}\label{sec:s441}
The valley equation for this system is given by,
\begin{eqnarray}
&&\frac{\delta^2 S}{\delta A_{\mu}\delta A_{\nu}}F^A_{\nu}
+\frac{\delta^2 S}{\delta A_{\mu}\delta H^{\dagger}}F^H
+\frac{\delta^2 S}{\delta A_{\mu}\delta H}F^{H\dagger}
=\lambda_e F^A_{\mu}\nonumber,
\\
&&\frac{\delta^2 S}{\delta H^{\dagger}\delta A_{\mu}}
F^A_{\mu}+
\frac{\delta^2 S}{\delta H^{\dagger}\delta H}F^{H\dagger}
+\frac{\delta^2 S}{\delta H^{\dagger}\delta H^{\dagger}}F^H
=\lambda_e F^{H\dagger},
\label{eq:s44valley}
\\
&&F^A_{\mu}=\frac{\delta S}{\delta A_{\mu}},
\quad F^H =\frac{\delta S}{\delta H},\nonumber
\end{eqnarray}
where the integration over the space-time is implicitly assumed.
The valley is parameterized by the eigenvalue $\lambda_e$ that
is identified with the zero mode corresponding to the scale invariance
in the massless limit, $v\rightarrow 0$.

To simplify the equation, we adopt the following ansatz;
\begin{eqnarray}
A_{\mu}(x)=\frac{x_{\nu}\bar{\sigma}_{\mu\nu}}{x^2}\cdot 2a(r),
\quad H(x)=v\left( 1-h(r)\right)\eta,
\label{eq:s44ansatz}
\end{eqnarray}
where $\eta$ is a constant isospinor, and $a$ and
$h$ are real dimensionless functions of dimensionless variable
$r$, which is defined by $r=\sqrt{x^2}/\rho$. The matrix 
$\bar{\sigma}_{\mu\nu}$ is defined, according to the conventions of
Ref.~\citen{rf:espinosa}, as
$\bar{\sigma}_{\mu\nu}=\bar{\eta}_{a\mu\nu}\sigma^a/2$. 
We have introduced the scaling parameter $\rho$ so that we adjust the
radius of valley instanton as we will see later in 
subsection 4.4.3. The tensor structure
in (\ref{eq:s44ansatz}) is the same as that of the instanton in the singular
gauge.\cite{rf:thooft}

Inserting this ansatz to (\ref{eq:s44valley}), the structure
of $F^A_{\mu}$ and $F^{H\dagger}$ is
determined as the following;
\begin{eqnarray}
F^A_{\mu}(x)=\frac{x_{\nu}\bar{\sigma}_{\mu\nu}}{x^2}\cdot
\frac{2v^2}{\lambda}f^a(r),
\quad F^{H\dagger}(x)=-\frac{v^3}{\lambda}f^h(r)\eta.
\label{eq:s44fansatz}
\end{eqnarray}
By using this ansatz, (\ref{eq:s44ansatz}) and (\ref{eq:s44fansatz}), the
valley equation (\ref{eq:s44valley}) is reduced to the following;
\begin{eqnarray}
&&-\frac{1}{r}\frac{d}{dr}\left(r\frac{da}{dr}\right)
+\frac{4}{r^2}a(a-1)(2a-1)
+\frac{g^2}{2\lambda}(\rho v)^2 a(1-h)^2=\frac{g^2}{\lambda}(\rho
v)^2 f^a,
\label{eq:s44eqn:va}\\
&&-\frac{1}{r^3}\frac{d}{dr}\left(r^3\frac{dh}{dr}\right)
+\frac{3}{r^2}(h-1)a^2
+\frac{1}{4}(\rho v)^2 h(h-1)(h-2)=(\rho v)^2 f^h,
\label{eq:s44eqn:vh}\\
&&-\frac{1}{r}\frac{d}{dr}\left(r\frac{df^a}{dr}\right)
+\frac{4}{r^2}(6a^2-6a+1)f^a+\frac{g^2}{2\lambda}(\rho v)^2 (h-1)^2f^a
\nonumber
\\
&&\hspace{30ex}+\frac{g^2}{\lambda}(\rho v)^2 a(h-1)f^h=\frac{g^2}{\lambda}(\rho v)^2 \nu f^a,
\label{eq:s44eqn:vfa}\\
&&-\frac{1}{r^3}\frac{d}{dr}\left(r^3\frac{df^h}{dr}\right)
+\frac{3a^2}{r^2}f^h+\frac{1}{4}(\rho v)^2 (3h^2-6h+2)f^h
\nonumber
\\
&&\hspace{30ex}+\frac{6a}{r^2}(h-1)f^a=(\rho v)^2 \nu f^h,
\label{eq:s44eqn:vfh}
\end{eqnarray}
where $\nu$ is defined as $\lambda_e=v^2\nu/\lambda$.

In the massless limit, $\rho v \rightarrow 0$, (\ref{eq:s44eqn:va}) and
(\ref{eq:s44eqn:vh}) reduce to the equation of motion and (\ref{eq:s44eqn:vfa})
and (\ref{eq:s44eqn:vfh}) to the equation for the zero-mode fluctuation around the
instanton solution. The solution of this set of equations is the following;
\begin{eqnarray}
\begin{array}{cc}
\displaystyle a_{0}=\frac{1}{1+r^{2}},
&\displaystyle h_{0}=1-\left(\frac{r^{2}}{1+r^{2}}\right)^{1/2},
\\[4mm]
\displaystyle f^a_{0}=\frac{2Cr^{2}}{(1+r^{2})^{2}},
&\displaystyle f^h_{0}=\frac{Cr}{(1+r^{2})^{3/2}},
\end{array}
\label{eq:s44eq:a0}
\end{eqnarray}
where $C$ is an arbitrary function of $\rho v$.
Note that $a_{0}$ is an instanton solution in the singular gauge and
$h_{0}$ is a Higgs configuration in the instanton background.\cite{rf:thooft}
We have adjusted the scaling parameter $\rho$ so that the radius of the
instanton solution is unity. The mode solutions
$f^a_{0}$ and $f^h_{0}$ are obtained from $\partial a_0/\partial \rho$ and
$\partial h_0/\partial \rho$, respectively.

Now we will construct the valley instanton analytically.
When $\rho v=0$, it is given by the ordinary instanton configuration $a_{0}$,
$h_{0}$, $f^{a}_{0}$ and $f^{h}_{0}$.
When $\rho v$ is small but not zero, it is expected that small $\rho v$
corrections appear in the solution.
On the other hand, at large distance from the core of the valley
instanton, this solution is expected to decay exponentially, because
gauge boson and Higgs boson are massive.
Therefore, the solution is similar to the
instanton near the origin and decays exponentially in the asymptotic region.
In the following, we will solve the valley equation in both regions
and analyze the connection in the intermediate region.
In this manner we will find the solution.

In the asymptotic region, $a$, $h$, $f^{a}$
and $f^{h}$ become small and the valley equation can be linearized;
\begin{eqnarray}
&&-\frac{1}{r}\frac{d}{dr}\left(r\frac{da}{dr}\right)
+\frac{4}{r^{2}}a+\frac{g^{2}}{2\lambda}(\rho v)^{2}a
=\frac{g^{2}}{\lambda}(\rho v)^{2}f^{a},
\label{eq:s44eqn:la}\\
&&-\frac{1}{r^{3}}\frac{d}{dr}\left(r^{3}\frac{dh}{dr}\right)
+\frac{1}{2}(\rho v)^{2}h=(\rho v)^{2}f^{h},
\label{eq:s44eqn:lh}\\
&&-\frac{1}{r}\frac{d}{dr}\left(r\frac{df^{a}}{dr}\right)
+\frac{4}{r^{2}}f^{a}+\frac{g^{2}}{2\lambda}(\rho v)^{2}f^{a}
=\frac{g^{2}}{\lambda}(\rho v)^{2}\nu f^{h},
\label{eq:s44eqn:lfa}\\
&&-\frac{1}{r^{3}}\frac{d}{dr}\left(r^{3}\frac{df^{h}}{dr}\right)
+\frac{1}{2}(\rho v)^{2}f^{h}=(\rho v)^{2}\nu f^{h}.
\label{eq:s44eqn:lfh}
\end{eqnarray}
The solution of this set of equations is
\begin{eqnarray}
&&a(r)=C_{1}\,r\frac{d}{dr}G_{\rho m_{_W}}(r)+\frac{1}{\nu}f^{a}(r),\\
&&h(r)=C_{2}\,G_{\rho m_{_H}}(r)+\frac{1}{\nu}f^{h}(r),\\
&&f^{a}(r)=C_{3}\,r\frac{d}{dr}G_{\rho \mu_{_W}}(r),\\
&&f^{h}(r)=C_{4}\,G_{\rho \mu_{_H}}(r),
\end{eqnarray}
where $C_{i}$ are arbitrary functions of $\rho v$ and $\mu_{_{W,\,H}}$
are defined as
$\mu_{_{W,\,H}}=m_{_{W,\,H}}\sqrt{1-2\nu}$.
As was expected above, these solutions decay exponentially at
infinity and
when $r\ll(\rho
v)^{-1}$ they have the series expansions;
\begin{eqnarray}
&&a(r)=\frac{C_{1}}{(2\pi)^{2}}
\left[-\frac{2}{r^{2}}+\frac{1}{2}(\rho m_{_W})^{2}+\cdots\right]
+\frac{C_{3}}{\nu(2\pi)^{2}}
\left[-\frac{2}{r^{2}}+\frac{1}{2}(\rho \mu_{_W})^{2}+\cdots\right],
\label{eq:s44eqn:aa}\\
&&h(r)=\frac{C_{2}}{(2\pi)^{2}}
\left[\frac{1}{r^{2}}+\frac{1}{2}(\rho m_{_H})^{2}\ln(\rho m_{_H}rc)
+\cdots\right]\nonumber\\
&&\hspace{30ex}+\frac{C_{4}}{\nu(2\pi)^{2}}
\left[\frac{1}{r^{2}}+\frac{1}{2}(\rho \mu_{_H})^{2}\ln(\rho \mu_{_H}rc)
+\cdots\right],
\label{eq:s44eqn:ah}\\
&&f^{a}(r)=\frac{C_{3}}{(2\pi)^{2}}
\left[-\frac{2}{r^{2}}+\frac{1}{2}(\rho \mu_{_W})^{2}+\cdots\right],
\label{eq:s44eqn:afa}\\
&&f^{h}(r)=\frac{C_{4}}{(2\pi)^{2}}
\left[\frac{1}{r^{2}}+\frac{1}{2}(\rho \mu_{_H})^{2}\ln(\rho \mu_{_H}rc)
+\cdots\right],
\label{eq:s44eqn:afh}
\end{eqnarray}
$c$ being a numerical constant $e^{\gamma-1/2}/2$, where $\gamma$ is
the Euler's constant.

Near the origin, we expect that the valley instanton is similar to the
ordinary instanton.
Then the following replacement of the field variables is convenient;
$a=a_{0}+(\rho v)^{2}\hat{a}$,
$h=h_{0}+(\rho v)^{2}\hat{h}$,
$f^{a}=f^{a}_{0}+(\rho v)^{2}\hat{f^{a}}$,
$f^{h}=f^{h}_{0}+(\rho v)^{2}\hat{f^{h}}$.
If we assume
$a_{0}\gg(\rho v)^{2}\hat{a}$,
$h_{0}\gg(\rho v)^{2}\hat{h}$,
$f^{a}_{0}\gg(\rho v)^{2}\hat{f^{a}}$
and $f^{h}_{0}\gg(\rho v)^{2}\hat{f^{h}}$,
the valley equation becomes
\begin{eqnarray}
&&-\frac{1}{r}\frac{d}{dr}\left(r\frac{d\hat{a}}{dr}\right)
+\frac{4}{r^{2}}(6a_{0}^{2}-6a_{0}+1)\hat{a}
+\frac{g^{2}}{2\lambda}a_{0}(h_{0}-1)^{2}
=\frac{g^{2}}{\lambda}f^{a}_{0},
\label{eq:s44eqn:ha}\\
&&-\frac{1}{r^{3}}\frac{d}{dr}\left(r^{3}\frac{d\hat{h}}{dr}\right)
+\frac{3}{r^{2}}a_{0}^{2}\hat{h}+\frac{6}{r^{2}}(h_{0}-1)a_{0}\hat{a}
+\frac{1}{4} h_{0}(h_{0}-1)(h_{0}-2)=f^{h}_{0},
\label{eq:s44eqn:hh}\\
&&-\frac{1}{r}\frac{d}{dr}\left(r\frac{d\hat{f^{a}}}{dr}\right)
+\frac{4}{r^{2}}(6a_{0}^{2}-6a_{0}+1)\hat{f^{a}}
+\frac{24}{r^{2}}(2a_{0}-1)f^{a}_{0}\hat{a}\nonumber\\
&&\hspace{28ex}+\frac{g^{2}}{2\lambda}(h_{0}-1)^{2}f^{a}_{0}
+\frac{g^{2}}{\lambda}a_{0}(h_{0}-1)f^{h}_{0}
=\frac{g^{2}}{\lambda}\nu f^{a}_{0},
\label{eq:s44eqn:hfa}\\
&&-\frac{1}{r^{3}}\frac{d}{dr}\left(r^{3}\frac{d\hat{f^{h}}}{dr}\right)
+\frac{3}{r^{2}}a_{0}^{2}\hat{f^{h}}
+\frac{6}{r^{2}}a_{0}f^{h}_{0}\hat{a}
+\frac{1}{4}(3h_{0}^{2}-6h_{0}+2)f^{h}_{0}\nonumber
\\
&&\hspace{20ex}+\frac{6}{r^{2}}a_{0}(h_{0}-1)\hat{f^{a}}
+\frac{6}{r^{2}}(h_{0}-1)f^{a}_{0}\hat{a}
+\frac{6}{r^{2}}a_{0}f^{a}_{0}\hat{h}
=\nu f^{h}_{0}.
\label{eq:s44eqn:hfh}
\end{eqnarray}
To solve this equation, we introduce solutions of the following
equations;
\begin{equation}
\begin{array}{l}
\displaystyle
-\frac{1}{r}\frac{d}{dr}\left(r\frac{d\varphi_{a}}{dr}\right)
+\frac{4}{r^{2}}(6a_{0}^{2}-6a_{0}+1)\varphi_{a}=0,
\\
\\
\displaystyle
-\frac{1}{r^{3}}\frac{d}{dr}\left(r^{3}\frac{d\varphi_{h}}{dr}\right)
+\frac{3}{r^{2}}a_{0}^{2}\varphi_{h}=0.
\label{eq:s44eqn:zh}
\end{array}
\end{equation}
They are given as,
\begin{equation}
\varphi_{a}=\frac{r^{2}}{(1+r^{2})^{2}},
\quad\varphi_{h}=\left(\frac{r^{2}}{1+r^{2}}\right)^{1/2}.
\end{equation}
Using these solutions, we will integrate the valley equation.
We multiply (\ref{eq:s44eqn:ha}) and (\ref{eq:s44eqn:hfa}) by $r\varphi_{a}$, and
multiply (\ref{eq:s44eqn:hh}) and (\ref{eq:s44eqn:hfh}) by
$r^{3}\varphi_{h}$ then integrate them from $0$ to $r$.
Integrating by parts and using (\ref{eq:s44eqn:zh}), we obtain
\begin{eqnarray}
&&-\varphi_{a}r\frac{d\hat{a}}{dr}+\frac{d\varphi_{a}}{dr}r\hat{a}
=\frac{g^{2}}{\lambda}\int_{0}^{r}dr' r'\varphi_{a}
\left[f^{a}_{0}-\frac{1}{2}a_{0}(h_{0}-1)^{2}\right],
\label{eq:s44eqn:dha}\\
&&-\varphi_{h}r^{3}\frac{d\hat{h}}{dr}+\hat{h}r^{3}\frac{d\varphi_{h}}{dr}
=\int_{0}^{r}dr' r'^{3}\varphi_{h}
\left[f^{h}_{0}-\frac{1}{4}h_{0}(h_{0}-1)(h_{0}-2)
-\frac{6}{r'^{2}}(h_{0}-1)a_{0}\hat{a}\right],
\nonumber\\
\label{eq:s44eqn:dhh}\\
&&-\varphi_{a}r\frac{d\hat{f^{a}}}{dr}
+\frac{d\varphi_{a}}{dr}r\hat{f^{a}}\nonumber\\
&&\hspace{2ex}=\int_{0}^{r}dr' r'\varphi_{a}
\left[\frac{g^{2}}{\lambda}\nu f^{a}_{0}
-\frac{24}{r'^{2}}(2a_{0}-1)f^{a}_{0}\hat{a}
-\frac{g^{2}}{2\lambda}(h_{0}-1)^{2}f^{a}_{0}
-\frac{g^{2}}{\lambda}a_{0}(h_{0}-1)f^{h}_{0}\right],
\label{eq:s44eqn:dhfa}\nonumber\\
\\
&&-\varphi_{h}r^{3}\frac{d\hat{f^{h}}}{dr}
+\hat{h}r^{3}\frac{d\varphi_{h}}{dr}\nonumber\\
&&\hspace{2ex}=\int_{0}^{r}dr' r'^{3}\varphi_{h}
\left[\nu f^{h}_{0}-\frac{6}{r'^{2}}a_{0}f^{a}_{0}\hat{a}
-\frac{1}{4}(3h_{0}^{2}-6h_{0}+2)f^{h}_{0}\right.\nonumber\\
&&\hspace{29ex}\left.-\frac{6}{r'^{2}}a_{0}(h_{0}-1)\hat{f^{a}}
-\frac{6}{r'^{2}}(h_{0}-1)f^{a}_{0}\hat{a}
-\frac{6}{r'^{2}}a_{0}f^{a}_{0}\hat{h}\right].
\label{eq:s44eqn:dhfh}
\end{eqnarray}

First we will find $\hat{a}$.
The right-hand side of (\ref{eq:s44eqn:dha}) is proportional to $(C-1/4)$
and when $r$ goes to infinity this approaches a constant.
At $r\gg1$, (\ref{eq:s44eqn:dha}) becomes
\begin{equation}
-\frac{1}{r}\frac{d\hat{a}}{dr}-\frac{2}{r^{2}}\hat{a}
=\frac{1}{3}\frac{g^{2}}{\lambda}\left(C-\frac{1}{4}\right).
\end{equation}
Then at $r\gg1$, $\hat{a}(r)$ is proportional to
$(C-1/4)r^{2}$ and $a(r)$ becomes
\begin{equation}
a=\frac{1}{r^{2}}
-\frac{(\rho v)^{2}g^{2}}{12\lambda}\left(C-\frac{1}{4}\right)r^{2}+\cdots.
\end{equation}
To match this with (\ref{eq:s44eqn:aa}), it must be hold that $C=1/4$ when
$\rho v=0$.
When $C=1/4$, the right-hand sides of (\ref{eq:s44eqn:dha}) vanishes and
$\hat{a}$ satisfy $-\varphi_{a}d\hat{a}/dr+\hat{a}d\varphi_{a}/dr=0$.
Hence $\hat{a}$ is $\hat{a}=D\,\varphi_{a}$, where $D$ is a constant.
Identifying $a_{0}+(\rho v)^{2}\hat{a}$ with (\ref{eq:s44eqn:aa}) again at $r\gg1$,
we find that
$C_{1}+C_{3}/\nu=-2\pi^{2}$ and $C_{3}=-\pi^{2}$ at $\rho v=0$.
In the same manner, $\hat{h}$,
$\hat{f^{a}}$ and $\hat{f^{h}}$ are obtained.
At $r\gg1$, we find
\begin{eqnarray}
&&\hat{h}={\rm const.}+\cdots,
\nonumber\\
&&\hat{f^{a}}=\frac{g^{2}}{48\lambda}\left(\frac{1}{4}-\nu\right)r^{2}
-\frac{g^{2}}{16\lambda}(1-2\nu)+\cdots,
\\
&&\hat{f^{h}}=\frac{1}{16}(1-2\nu)\ln r+\cdots.
\nonumber
\end{eqnarray}
Here ${\rm const.}$ is a constant of integration.
Comparing (\ref{eq:s44eqn:ah})-(\ref{eq:s44eqn:afh}) with
them, we find that $C_{2}+C_{4}/\nu=2\pi^{2}$,
$C_{4}=\pi^{2}$ and $\nu=1/4$ at $\rho v=0$.

Now we have obtained the solution of the proper valley equation. Near the
origin of the valley instanton, 
$r \ll (\rho m_{_{W,H}})^{-1/2}$, 
it is given by, 
\begin{eqnarray}
\begin{array}{ll}
\displaystyle a(r)=\frac{1}{1+r^{2}},
&\displaystyle \hspace{8ex}h(r)=1-\left(\frac{r^{2}}{1+r^{2}}\right)^{1/2},
\\[4mm]
\displaystyle f^a(r)=\frac{r^{2}}{2(1+r^{2})^{2}},
&\displaystyle \hspace{8ex}f^h(r)=\frac{r}{4(1+r^{2})^{3/2}},
\end{array}
\label{eq:s44eq:vnear}
\end{eqnarray}
where we ignore the correction terms that go to zero as $\rho v
\rightarrow 0$, since they are too small comparing with the leading terms.
As $r$ becomes larger, the leading terms are getting smaller 
and so the correction terms become more important;
\begin{equation}
\begin{array}{ll}
\displaystyle a(r)=\frac{1}{r^{2}}+o((\rho v)^2),
&\displaystyle h(r)=\frac{1}{2r^{2}}-\frac{(\rho v)^{2}}{16}\ln2+\cdots,
\\[4mm]
\displaystyle f^{a}(r)=\frac{1}{2r^{2}}
-\frac{g^{2}(\rho v)^{2}}{32\lambda}+\cdots,
&\displaystyle f^{h}(r)=\frac{1}{4r^{2}}-\frac{(\rho v)^{2}}{32}\ln r+\cdots,
\end{array}
\label{eq:s44eq:vcorr}
\end{equation}
for $(\rho m_{_{W,H}})^{-1/2} \ll r \ll (\rho m_{_{W,H}})^{-1}$.
Finally, far from the origin, $r \gg (\rho m_{_{W,H}})^{-1/2}$, 
the solution is given by the following:
\begin{equation}
\begin{array}{l}
\displaystyle a(r)=2\pi^2\, r\frac{d}{dr}G_{\rho m_{_W}}(r)
+\frac{1}{\nu}f^{a}(r),\\
\displaystyle h(r)=-2\pi^2\,G_{\rho m_{_H}}(r)+\frac{1}{\nu}f^{h}(r),\\
\displaystyle f^{a}(r)=-\pi^2\,r\frac{d}{dr}G_{\rho \mu_{_W}}(r),\\
\displaystyle f^{h}(r)=\pi^2\,G_{\rho \mu_{_H}}(r),\\
\end{array}
\label{eq:s44eq:vfar}
\end{equation}
where $\nu = 1/4$ for $\rho v = 0$. In (\ref{eq:s44eq:vfar}), we ignore 
correction terms, since they are too small.

Let us make a brief comment about the consistency of our analysis.
Until now, we have implicitly assumed that there exists an overlapping
region where both (\ref{eq:s44eqn:la})-(\ref{eq:s44eqn:lfh}) and
(\ref{eq:s44eqn:ha})-(\ref{eq:s44eqn:hfh}) are valid.
Using the above solution, it is found that (\ref{eq:s44eqn:la})-(\ref{eq:s44eqn:lfh})
are valid when $r\gg (\rho m_{_{W,H}})^{-1/2}$ and (\ref{eq:s44eqn:ha})-(\ref{eq:s44eqn:hfh})
are valid when $r\ll (\rho m_{_{W,H}})^{-1}$.
If $\rho m_{_{W,H}}$ is small enough,
there exists the overlapping region $(\rho m_{_{W,H}})^{-1/2}\ll r\ll (\rho
m_{_{W,H}})^{-1}$.
Then our analysis is consistent.

The action of the valley instanton can be calculated using the above
solution.
Rewriting the action in terms of  $a$ and $h$, we find
\begin{eqnarray}
&&S_{g}=\frac{12\pi^{2}}{g^{2}}
\int_{0}^{\infty}\frac{dr}{r}\left\{
\left( r\frac{da}{dr}\right)^{2}+4a^{2}(a-1)^{2}
\right\},
\\
&&S_{h}=\frac{2\pi^{2}}{\lambda}(\rho v)^{2}\int_{0}^{\infty}
r^{3}dr\left\{
\left(\frac{dh}{dr}\right)^{3}+\frac{3}{r^{2}}(h-1)^{2}a^{2}
+\frac{1}{8}(\rho v)^{2}h^{2}(h-2)^{2}\right\}.
\end{eqnarray}
Substituting the above solution for $S$, we obtain
\begin{equation}
\label{eq:s44eqn:aaction}
S=\frac{8\pi^{2}}{g^{2}}
    +\frac{2\pi^{2}}{\lambda}(\rho v)^{2}-\frac{\pi^2}{4\lambda}(\rho
v)^{4}\ln(\rho v)+O((\rho v)^4).
\end{equation}
The leading contribution $8\pi^{2}/g^{2}$ comes from $S_{g}$ for $a_{0}$,
which is the action of the instanton, and the next-to-leading and the third
contributions come from $S_{h}$ for $a_{0}$ and $h_{0}$.

\subsubsection{Constrained instanton}\label{sec:s442}
Here, we consider the constrained instanton. 
According to Affleck's analysis,\cite{rf:Aff} the constrained
instanton satisfies the following equation: 
\begin{eqnarray} 
&&\frac{\delta S}{\delta A_{\mu}}+\sigma\frac{\delta O_A}{\delta A_{\mu}}=0,
\\
&&\frac{\delta S}{\delta H}+\sigma\frac{\delta O_H}{\delta H}=0,
\end{eqnarray}
where $\sigma$ is a Lagrange multiplier and depends on the constraint.
Both $O_A$ and $O_H$ are functionals of $A_{\mu}$ and $H$
respectively
that give a solution of the constrained equation, as the scalar theory
in the subsection 4.3.2. 
Here we adopt the ansatz (\ref{eq:s44ansatz}) again. By the similar analysis
as the valley instanton, it turns out that the behavior of the
constrained instanton is the following. Near the origin of the
instanton, the solution is given by $a_0$, $h_0$ in (\ref{eq:s44eq:a0}) as
well as the valley instanton. In the region 
where $(\rho m_{_{W,H}})^{-1/2} \ll r \ll (\rho m_{_{W,H}})^{-1}$, the solution is given
by 
\begin{equation}
a(r)=\frac{1}{r^{2}}-\frac{(\rho v)^2}{8}\frac{g^2}{\lambda}+\cdots,
\hspace{5ex}h(r)=\frac{1}{2r^{2}}+\frac{(\rho v)^{2}}{8}\ln(\rho v rc)+\cdots.
\label{eq:s44eq:ccorr}
\end{equation}
Let us compare (\ref{eq:s44eq:vcorr}) and (\ref{eq:s44eq:ccorr}). The correction
term of the valley instanton is smaller than one of the constrained instanton.
Finally, for $r \gg (\rho m_{_{W,H}})^{-1/2} $, the constrained instanton is given by
\begin{eqnarray}
a(r)=-2 \pi^2 r\frac{d}{dr}G_{\rho m_{_W}}(r),
\hspace{5ex}h(r)=2 \pi^2 G_{\rho m_{_H}}(r).
\end{eqnarray}
The action of the constrained instanton is the following:
\begin{equation}
S=\frac{8\pi^{2}}{g^{2}}
    +\frac{2\pi^{2}}{\lambda}(\rho v)^{2}+O((\rho v)^4 \ln(\rho v)).
\label{eq:s44eq:conaction}
\end{equation}
The leading contribution $8\pi^{2}/g^{2}$ comes from $S_{g}$ for $a_{0}$,
 which is the action of the instanton, and the next-to-leading
contributions comes from $S_{h}$ for $a_{0}$ and $h_{0}$. The
difference from the valley instanton is that we cannot determine the
term of $O((\rho v)^4\ln(\rho v))$ by the current analysis.  

Now we choose a constraint and analyze the constrained instanton
. We adopt the following functionals for the constraint:
$O_A=i g^2\int d^4 x {\rm tr}F_{\mu\nu}F_{\nu\rho}F_{\rho\mu},\ 
O_H=0$. This constraint is one of the simplest for giving the constrained instanton.
Then the constrained equation of motion is given by, 
\begin{eqnarray}
&&-\frac{1}{r}\frac{d}{dr}\left(r\frac{da}{dr}\right)
+\frac{4}{r^2}a(a-1)(2a-1)
+\frac{g^2}{2\lambda}(\rho v)^2 a(1-h)^2\nonumber\\
&&~\hspace{5ex}+6(\rho v)^2
\frac{\tilde{\sigma}}{r^2}\Biggl\{
(2a-1)\left(\frac{da}{dr}\right)^2 
+2a(a-1)\frac{d^2 a}{dr^2}
 -\frac{2}{r}a(a-1)\frac{da}{dr}\label{eq:s44eqn:ca}\\
&&~\hspace{20ex} -\frac{4}{r^2}a^2(a-1)^2(2a-1)\Biggr\}=0
,\nonumber\\
&&-\frac{1}{r^3}\frac{d}{dr}\left(r^3\frac{dh}{dr}\right)
+\frac{3}{r^2}(h-1)a^2
+\frac{1}{4}(\rho v)^2 h(h-1)(h-2)=0,
\label{eq:s44eqn:ch}
\end{eqnarray}
where $\sigma=\rho^4v^2\tilde{\sigma}$.
By solving these equations approximately, in fact, we obtain the
behavior of the solution that we have given previously, and  the
multiplier is determined by $\tilde{\sigma}=5g^2/48\lambda $ as $\rho
v \rightarrow 0$. 
  
\subsubsection{Numerical analysis}\label{sec:s443}
In this subsection, we solve the valley equation (\ref{eq:s44eqn:va})-(\ref{eq:s44eqn:vfh})
and the constrained equation (\ref{eq:s44eqn:ca}), (\ref{eq:s44eqn:ch})
numerically, and compare the valley instanton and constrained
instanton. 

We need a careful discussion for solving the valley equation
(\ref{eq:s44eqn:va})-(\ref{eq:s44eqn:vfh}):
Since the solution must be regular at the origin, we assume the following
expansions for $r \ll 1$;
\begin{eqnarray}
\begin{array}{cc}
\displaystyle a(r)=\sum_{n=0}^{\infty} a_{(n)} r^n,
&\displaystyle h(r)=\sum_{n=0}^{\infty} h_{(n)} r^n,
\\
\displaystyle f^a(r)=\sum_{n=0}^{\infty} f^a_{(n)} r^n,
&\displaystyle f^h(r)=\sum_{n=0}^{\infty} f^h_{(n)} r^n.
\end{array}
\label{eq:s44expand}
\end{eqnarray}
Inserting (\ref{eq:s44expand}) to (\ref{eq:s44eqn:va})-(\ref{eq:s44eqn:vfh}), we obtain
\begin{eqnarray}
&&a_{(0)}=1,h_{(0)}=1,f^a_{(0)}=0,f^h_{(0)}=0,
\nonumber
\\
&&a_{(1)}=0,f^a_{(1)}=0,
\\
&&h_{(2)}=0,f^h_{(2)}=0.
\nonumber
\end{eqnarray}
The coefficients $a_{(2)}$, $h_{(1)}$, $f^a_{(2)}$ and
$f^h_{(1)}$ are not determined and remain as free parameters.
The higher-order coefficients ($n \ge 3$) are determined
in terms of these parameters.
Four free parameters are determined by boundary conditions at infinity.
The finiteness of action requires
$a$, $h$ $\rightarrow 0$ faster than $1/r^2$ at infinity. This
condition also requires $f_a$, $f_h$ $\rightarrow 0$.

We have introduced $\rho$ as a free scale parameter.
We adjust this parameter $\rho$ so that $a_{(2)}=-2$
to make the radius of the valley instanton unity.
As a result we have four parameters $h_{(1)}$, $f^a_{(2)}$,
$f^h_{(1)}$ and $\rho v$ for a given $\nu$.
These four parameters are determined so that $a$, $h$, $f^a$,
and $f^h$ $\rightarrow 0$ at infinity. In the case
of the constrained instanton, the two parameters $h_{(1)}$ and
$\tilde{\sigma}$ are determined under $a_{(2)}=-2$ so that $a$, $h$
$\rightarrow 0$ at infinity.

\begin{figure}[htb]
  \parbox{\halftext}{
     \centerline{\epsfxsize=7cm\epsfbox{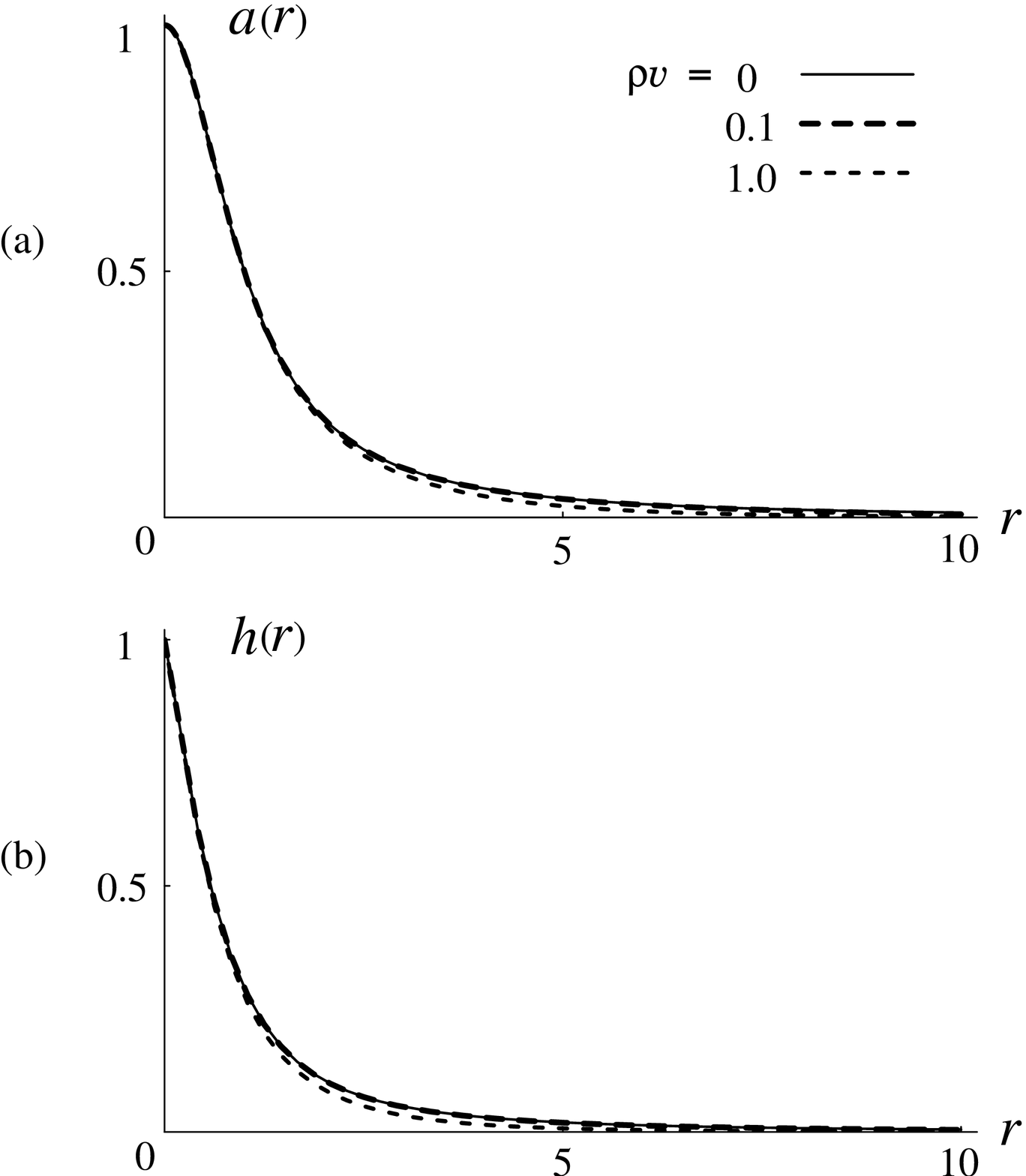}}
     \caption{Shapes of the numerical solution of the valley 
              equation, $a(r)$ and $h(r)$ for $\rho v =0.1, 
              1.0$ near the origin. The solid lines denote the 
              original instanton solution, $a_0$ and $h_0$.}
  \label{fig:s44shval}}
\hspace{4mm}
  \parbox{\halftext}{
     \centerline{\epsfxsize=7cm\epsfbox{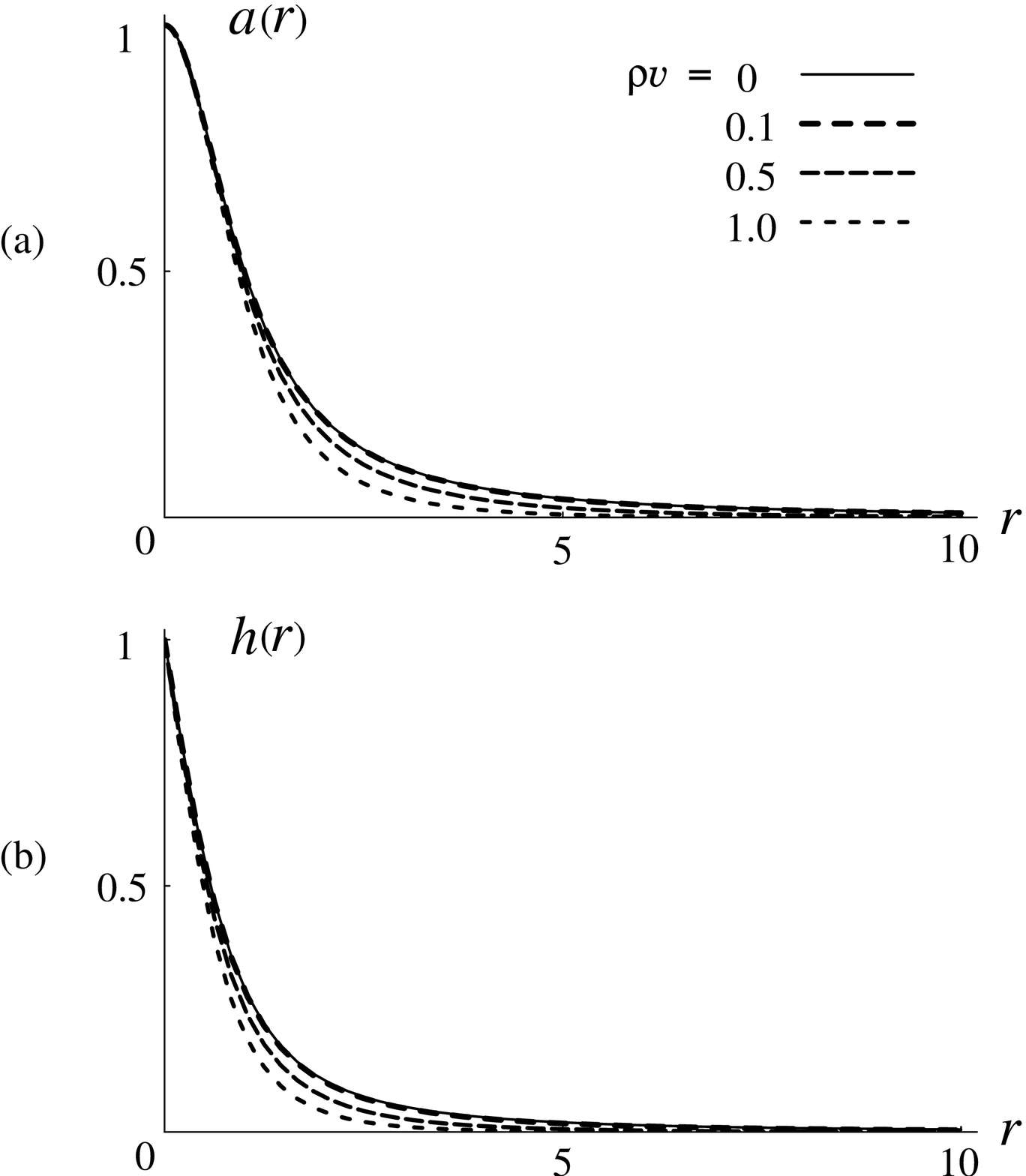}}
\caption{Shapes of the numerical solution of the constrained
instanton, $a(r)$ and $h(r)$ for $\rho v =0.1, 0.5, 1.0$ near the
origin. The solid lines denote the original instanton solution, $a_0$
and $h_0$.}
\label{fig:s44shcon}}
\end{figure}

A numerical solution of the valley equation near the origin is plotted 
in Fig.~\ref{fig:s44shval} for $\rho v=0.1,\ 1.0$ at $\lambda /g^2 =1$,
when $m_{_W}=m_{_H}$. 
We plot the instanton solution (\ref{eq:s44eq:a0}) by the solid line, 
which corresponds to the valley instanton for $\rho v=0$. 
This behavior of the numerical solution for $\rho v=0.1$ 
agrees with the result of the previous subsection, (\ref{eq:s44eq:vnear}). 
Moreover, even when
$\rho v=1.0$, the numerical solution is quite similar to the instanton
solution. We find that the behavior of $f^a(r)$ and $f^h(r)$
also agrees with the analytical result (\ref{eq:s44eq:vnear}) 
as well as $a(r)$ and $h(r)$. 
We also find that the numerical solution 
in the asymptotic region where $r \gg (\rho v)^{-1/2}$, is damping
exponentially and agrees with the analytical result (\ref{eq:s44eq:vfar}).

\begin{figure}[htb]
  \parbox{\halftext}{
\centerline{\epsfxsize=8.5cm \epsfbox{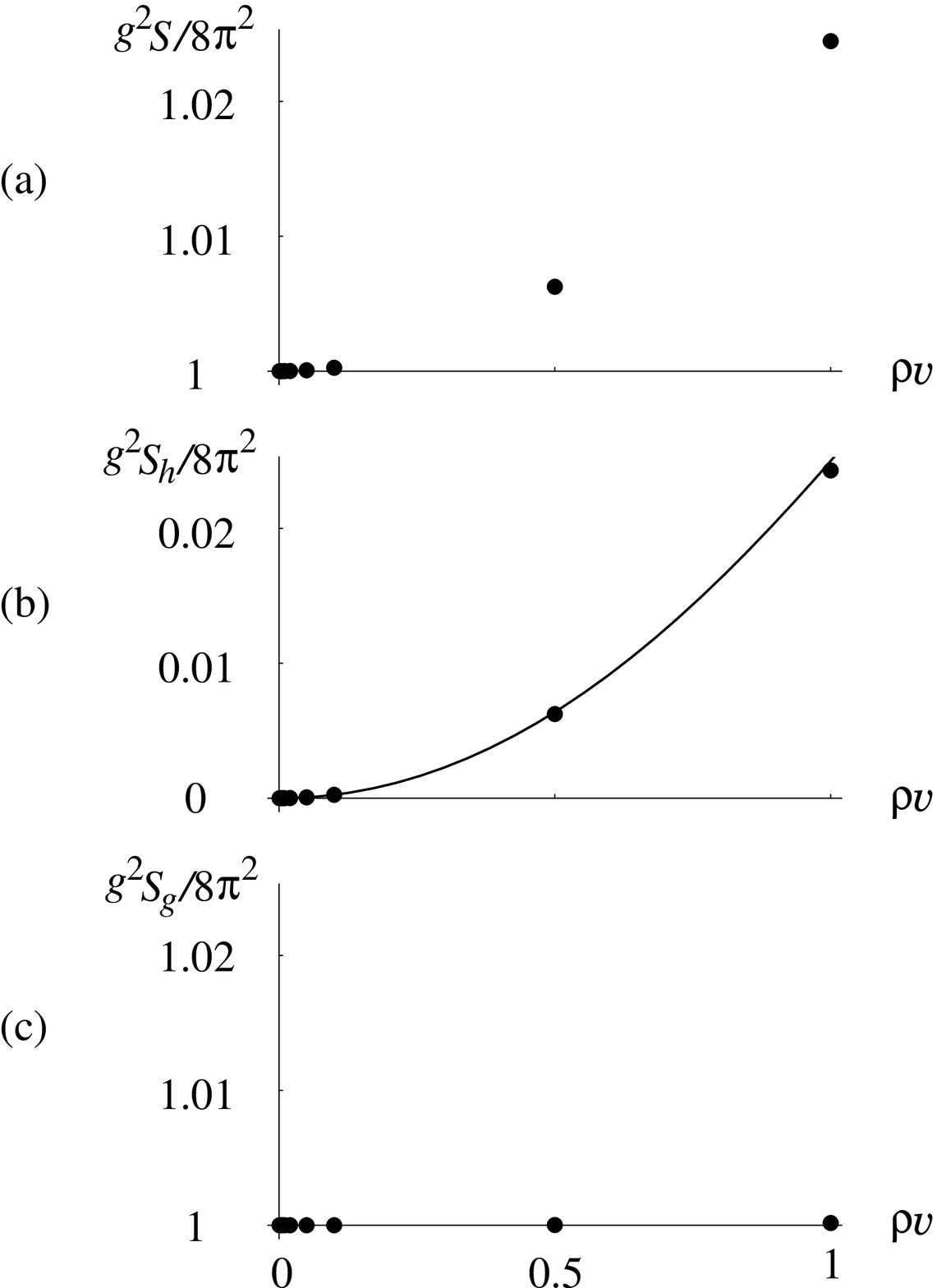}}
\caption{(a) The action $S$ (in unit of $g^2 /8 \pi^2$) of 
the numerical solution of the valley equation, at $\lambda /g^2 =1$, 
as a function of the parameter $\rho v$. 
(b) The contribution from the Higgs sector, $S_h$.
The solid line shows the behavior of the analytical result
that $g^2 S_h/8 \pi^2 = (\rho v)^2/4 -(\rho v)^4 \ln (\rho v)/32$.
(c) The contribution from the gauge sector, $S_g$.}
\label{fig:s44actval}}
\hspace{4mm}
  \parbox{\halftext}{
\centerline{\epsfxsize=8.5cm \epsfbox{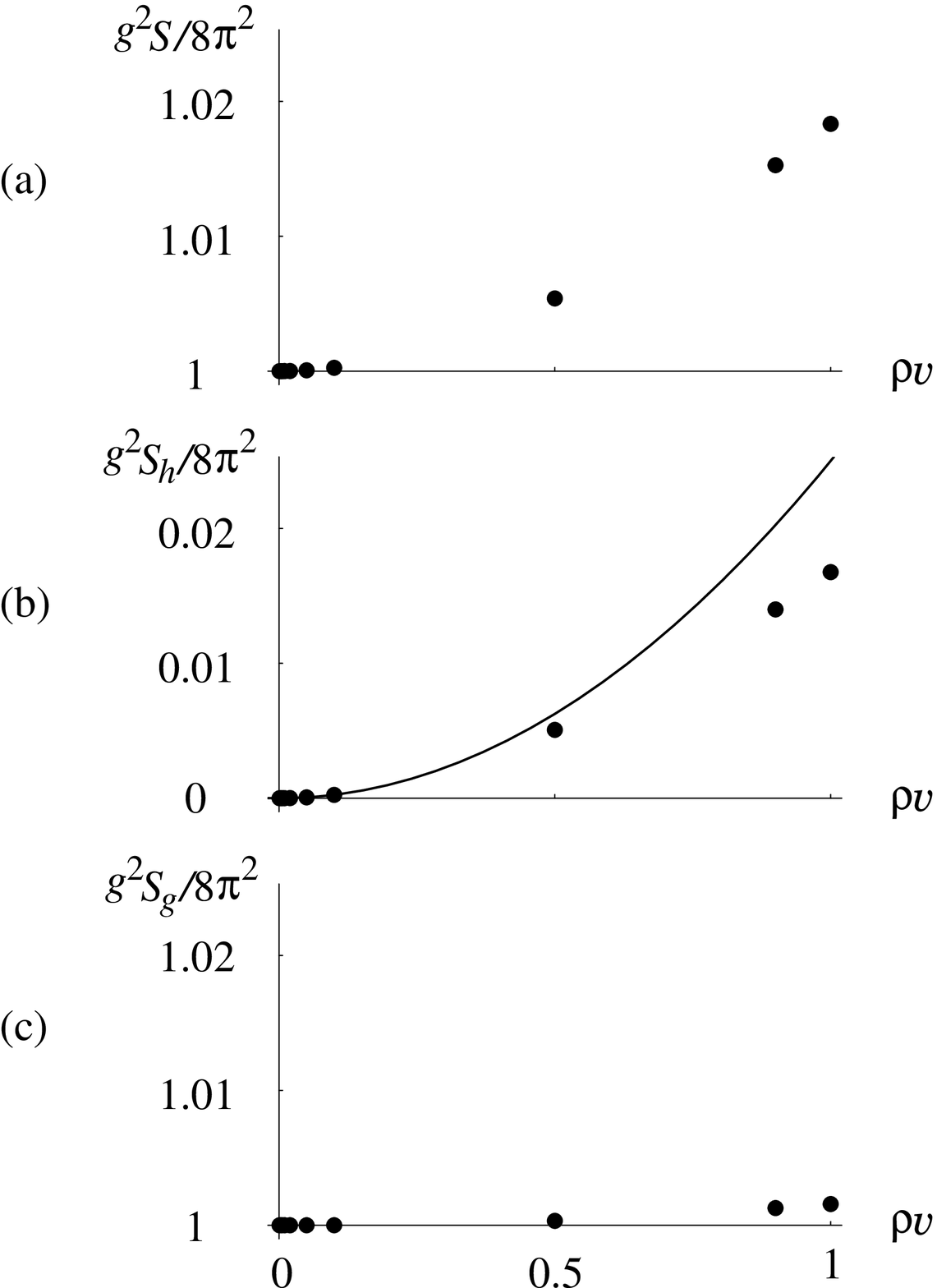}}
\caption{(a) The action $S$ (in units $g^2 /8 \pi^2$) of the numerical 
solution of the constrained instanton, at $\lambda /g^2 =1$, as a 
function of the parameter $\rho v$. 
(b) The contribution from the Higgs sector, $S_h$.
The solid line shows the behavior of the analytical result
that $g^2 S_h/8 \pi^2 = (\rho v)^2/4$.
(c) The contribution from the gauge sector, $S_g$.}
\label{fig:s44actcon}}
\end{figure}

On the other hand, a solution of the constrained equation is
plotted for $\rho v=0.1,\ 0.5,\ 1.0$ at $\lambda /g^2 =1$ 
in Fig.~\ref{fig:s44shcon}. 
The numerical solution for $\rho v=0.1$ also 
agrees with the analytical result. As $\rho v $ is larger, 
both the valley instanton and the constrained instanton 
are more deformed from the original instanton solution. 
Nevertheless, the
correction of the constrained instanton from the original instanton 
is much larger than one of the valley instanton, especially
when $\rho v \lsim 0.5$. 

The values of the action of the valley instanton for $\rho v=0.001\sim
1.0$ are plotted in Fig.~\ref{fig:s44actval}: 
Fig.~\ref{fig:s44actval}(a) depicts the behavior of the
total action $S$, while the contribution from the gauge part $S_g$
and from the Higgs part $S_h$ are in (b) and (c) respectively. 
The solid line shows the behavior of $S_h$
of the analytical result (\ref{eq:s44eqn:aaction}). From Fig.~\ref{fig:s44actval}
(c), it turns that $S_g$ is almost independent of $\rho v$.  
This figure shows that our numerical solutions are quite consistent
 with (\ref{eq:s44eqn:aaction}), even when $\rho v = 1.0$.

The values of the action of the constrained instanton for $\rho
v=0.001\sim 1.0$ are
plotted in Fig.~\ref{fig:s44actcon}. 
When $\rho v \lsim 0.5$, we can no
longer use the analytical result (\ref{eq:s44eq:conaction}). This is
attributed 
to the large deformation from the original instanton. We notice that 
the action of the constrained instanton is smaller than that of the
valley instanton for the same value of $\rho v$. This result is
natural at the point that the parameter $\rho v$ corresponds to the 
scale parameter. 
This point was elaborated upon in section 3, using
the toy model (\ref{eq:toyconst}).
To repeat, the fact that the constrained instanton 
has the smaller action than the valley instanton for the
same value of $\rho v$ does not mean that the constrained
instanton gives more important contribution to the functional integral
than the valley instanton. 

We summarize all the numerical data in Table \ref{tab:valresult}
and  \ref{tab:conresult}, for the valley instanton and for the
constrained instanton, respectively.

\begin{wraptable}{c}{14cm}
\begin{center}
\doublerulesep=0pt
\def\arraystretch{1.3}
\begin{tabular}{@{\vrule width 1pt \quad}c 
@{\quad \vrule width 1pt \, }c|c|c|c|c @{\ \vrule width 1pt}}
\noalign{\hrule height 1pt}
{$\rho v$ } & {$\nu$} & $h_{(1)}$ & $f^a_{(2)}$ &
$f^h_{(1)}$ & $g^{2}S/8\pi^2$ \\
\hline
0.001 & 0.2500 & -1.000 & 0.5001 & 0.2500 & 1.000000 \\
\hline
0.005 & 0.2500 & -1.000 & 0.5001 & 0.2500 & 1.000006 \\
\hline
0.01 & 0.2501  & -1.000 & 0.5001 & 0.2500 & 1.000025 \\
\hline
0.02 & 0.2501  & -1.000 & 0.5003 & 0.2501 & 1.000100 \\
\hline
0.05 & 0.2503  & -1.000 & 0.5014 & 0.2505 & 1.000625 \\
\hline
0.1  & 0.2509  & -1.000 & 0.5044 & 0.2515 & 1.002503 \\
\hline
0.5 & 0.2537   & -1.007 & 0.5478 & 0.2669 & 1.062550\\
\hline
1.0 & 0.2521   & -1.021 & 0.6226 & 0.2926 & 1.244646\\
\noalign{\hrule height 1pt}
\end{tabular}
\end{center}
\caption{The numerical data of the valley instanton in the gauge-Higgs
system.}
\label{tab:valresult}
\end{wraptable}

\begin{wraptable}{c}{14cm}
\begin{center}
\doublerulesep=0pt
\def\arraystretch{1.3}
\begin{tabular}{@{\vrule width 1pt \quad}c @{\quad \vrule width 1pt \, }c|c|c|c
@{\ \vrule width 1pt}}
\noalign{\hrule height 1pt}
{$\rho v$ } & {$\tilde{\sigma}$} & $h_{(1)}$ & {$\rho_0 v$} &
 $g^{2}S/8\pi^2$ \\
\hline
0.001 & 0.1042 & -1.000 & 0.000100   & 1.000000\\
\hline
0.005 & 0.1042 & -1.000 & 0.005000   & 1.000006\\
\hline
0.01  & 0.1041 & -1.000 & 0.010000   & 1.000025\\
\hline
0.02  & 0.1040 & -1.000 & 0.019998   & 1.000100 \\
\hline
0.05  & 0.1035 & -1.002 & 0.049972  & 1.000623 \\
\hline
0.1   & 0.1019 & -1.005 & 0.099779  & 1.002471 \\
\hline
0.5   & 0.0798 & -1.070 & 0.481738  & 1.053903 \\
\hline
0.9   & 0.0637 & -1.144 & 0.833000   & 1.152691 \\
\hline
1.0   & 0.0609 & -1.162 & 0.917406   & 1.183336 \\
\noalign{\hrule height 1pt}
\end{tabular}
\end{center}
\caption{The numerical data of the constrained instanton in the
gauge-Higgs system.}
\label{tab:conresult}
\end{wraptable}

In Table \ref{tab:conresult}, we
show the parameter $\rho_0$ used as ``$\rho$'' in Ref.~\citen{rf:Aff}.
The parameter $\rho_0$ is defined by,
\begin{eqnarray}
O = c_0 \rho_0^{4-d},
\end{eqnarray}
where $d$ is a dimension of the operator and $c_0$ is a conveniently
chosen constant. For $O_A=i g^2\int d^4 x {\rm
tr}F_{\mu\nu}F_{\nu\rho}F_{\rho\mu}$ and $O_H=0$, we choose 
that $c_0=96 \pi^2 /5g^2$ so that $\rho_0 =\rho$ as $\rho v \rightarrow 0$.
We find that $\rho_0$ nearly agrees with $\rho$.

\subsection{Valley calculation of the B and L violation} 
In this subsection, we calculate the cross section for the process of the
type\  $\bar{\rm q}+\bar{\rm q}\rightarrow(3n_f-2){\rm q}+n_f
l+n_w {\rm W}+n_h {\rm H}$, using the valley instanton.\cite{rf:hs} 
As explained in section 2, Ringwald and Espinosa studied this
process by using the constrained instanton.\cite{rf:ringwald,rf:espinosa}
They suggested that the amplitudes of the multi-boson process increase 
with the energy and reach 1pb for a center of mass energy of O(10) TeV.
Unfortunately, the high-energy behavior of the process with many
bosons, which is most interesting, is out of the validity of their
calculations.
One of the problem that they used the analytically constructed
configuration given in subsection \ref{sec:s442}. 
In the process with many bosons, the large radius instanton dominates
the path-integral.
As is shown in subsection \ref{sec:s443}, the large radius constrained 
instanton is not approximated well by the analytically constructed
configuration.
On the other hand, the valley instanton can be approximated by the
analytically constructed configuration even for the large radius one.
(See subsection \ref{sec:s443}.)
Thus the analytically constructed configuration can be used in the
calculation.
Although induced vertex by the valley instanton is similar to
that given by Ref.~\citen{rf:ringwald} and \citen{rf:espinosa}, the
latter can not be justified in the calculation of the
process with many bosons.  
 
Another problem of Ref.~\citen{rf:espinosa} is that the orientation
effects of the instanton are not taken into account.
The integration of the orientation in the Lorentz $SU(2)_R$ rotations
is not trivial but very important especially for the process with many 
bosons.
Moreover, in considering the multi-boson processes at high energies,
masses of the gauge boson, Higgs particle and top quark can not be
neglected. 
All the effects are incorporated in the
following calculations. 
\subsubsection{Valley instanton in the electroweak theory} 

We consider the standard model with $n_f$ generations of quarks
and leptons. 
For simplicity, we set ${\rm SU(3)}_c$ and 
${\rm U(1)}_Y$ coupling constant to be zero.  
Thus we treat the
simplified standard model that reduces to an SU(2) gauge-Higgs system
with fermions. As for quarks, we have the left-handed doublets
$q_L^{i \alpha}$ and the right-handed singlets $q_R^{i \alpha}$, where
$\alpha=1,2,3$ is the color index and $i=1,\cdots,n_f$ 
is the index of the generations. Similarly, we have the left-handed
doublets $l_L^i$ and the right-handed singlets $l_R^i$ for leptons. 
The bosonic part of the action is given by
\begin{eqnarray}
&&S_g=\frac{1}{2 g^2} \int d^4 x\ {\rm tr} F_{\mu \nu}  F_{\mu \nu},
\label{eq:s45gaugeaction}
\\
&&S_h=\frac{1}{ \lambda}\int d^4 x\left\{\left(D_{\mu} H\right)^{\dagger}
\left(D_{\mu} H\right)+
\frac{1}{8}\left(H^{\dagger}H-v^2\right)^2\right\},
\label{eq:s45higgsaction}
\end{eqnarray}
where $F_{\mu\nu}=\partial_{\mu} A_{\nu}-\partial_{\nu}
A_{\mu}-i\left[A_{\mu},A_{\nu}\right]$ and
$D_{\mu}=\partial_{\mu}-iA_{\mu}$.
The masses of the gauge boson and the Higgs boson are 
\begin{eqnarray}
&&m_{_W}=\sqrt{\frac{g^2}{2\lambda}} v,\quad m_{_H}=\frac{1}{\sqrt{2}} v.
\end{eqnarray}
The fermionic part of the action is 
\begin{eqnarray}
&&S_q=\int d^4 x (i q_L^{\dagger}\sigma_{\mu}D_{\mu}q_L
-i u_R^{\dagger}\bar{\sigma}_{\mu}\partial_{\mu} u_R
-i d_R^{\dagger}\bar{\sigma}_{\mu}\partial_{\mu} d_R\\
\nonumber
&& \hspace{10ex}+y_u q_L^{\dagger}\epsilon H^*
u_R-y_d q_L^{\dagger}H d_R +h.c.),\\
&&S_l=\int d^4 x (i l_L^{\dagger}\sigma_{\mu}D_{\mu}l_L
-i e_R^{\dagger}\bar{\sigma}_{\mu}\partial_{\mu} e_R
-y_l l_L^{\dagger} H e_R +h.c.),
\end{eqnarray}
where $\epsilon^{\alpha\beta}=-\epsilon_{\alpha\beta} = i \sigma^2$,
 $\sigma_{\mu}= (\sigma,i)$ and
$\bar{\sigma}_{\mu}= (\sigma,-i)$.
The masses of fermions are given by $m_{u,d,l}=y_{u,d,l}v$.
The total action is given by
\begin{eqnarray}
S=S_g+S_h+\sum_{i=1}^{n_f}\left(S_{l}+\sum_{\alpha=1}^{3}S_{q}\right). 
\end{eqnarray}

As was shown in the previous subsection, the dominating configuration in
the path-integral of the bosonic variables is the valley instanton. 
The valley instanton is given by the following;
\begin{eqnarray}
&&A_{\mu}(x)=\frac{x_{\nu}}{x^2}U\bar{\sigma}_{\mu\nu}U^{\dagger}\cdot
2a(r),
\\
&&H(x)=v(1-h(r))\eta,
\end{eqnarray}
where $U$ is a Lorentz $SU(2)_R$ matrix, $\eta$ is a constant
isospinor, and $a$ and $h$ are real dimensionless functions of
dimensionless variable $r$, which is defined by $r=\sqrt{x^2}/\rho$.  
We choose $\eta_r$ as (0,1).

Near the origin of the valley instanton,
$r \ll (\rho m_{_{W,H}})^{-1/2}$,
the valley instanton is given by,
\begin{eqnarray}
\begin{array}{ll}
\displaystyle a(r)=\frac{1}{1+r^{2}},
&\displaystyle \hspace{8ex}h(r)=1-\left(\frac{r^{2}}{1+r^{2}}\right)^{1/2},
\end{array}
\label{eq:s45eq:vnear}
\end{eqnarray}
where we ignore the correction terms that go to zero as $\rho v
\rightarrow 0$, since they are too small comparing with the leading terms.
As $r$ becomes larger, the leading terms are getting smaller
and so the correction terms become more important;
\begin{equation}
\begin{array}{ll}
\displaystyle a(r)=\frac{1}{r^{2}}+o((\rho v)^2),
&\displaystyle h(r)=\frac{1}{2r^{2}}-\frac{(\rho v)^{2}}{16}\ln2+\cdots,
\end{array}
\label{eq:s45eq:vcorr}
\end{equation}
for $(\rho m_{_{W,H}})^{-1/2} \ll r \ll (\rho m_{_{W,H}})^{-1}$.
Finally, far from the origin, $r \gg (\rho m_{_{W,H}})^{-1/2}$,
the solution is given by the following:
\begin{equation}
\begin{array}{l}
\displaystyle a(r)=\frac{\pi^2}{2\nu} r\frac{d}{dr}G_{\rho m_{_W}}(r)
+\frac{1}{\nu}f^{a}(r),\\
\displaystyle h(r)=-\frac{\pi^2}{2 \nu} G_{\rho m_{_H}}(r)+\frac{1}{\nu}f^{h}(r),\\
\end{array}
\label{eq:s45eq:vfar}
\end{equation}
where $\nu = 1/4$ for $\rho v = 0$, and by the numerical calculation,
we obtain $\nu\sim 1/4$ for $\rho v\rsimil 1$.  
In (\ref{eq:s45eq:vfar}), we ignore
correction terms, since they are too small.
The function $G_{m}(r)$ is
\begin{equation}
G_{m}(r)=\frac{m K_{1}(m r)}{(2\pi)^{2}r},
\end{equation}
where $K_{1}$ is a modified Bessel function.
The functions $f$ and $h$ decay exponentially at large $r$.

The fermionic zero mode around the valley instanton is given by;
\begin{eqnarray}
&&q_{Lr}^{\dot{\alpha}}(x)=-\frac{1}{\pi\rho^3}
x_{\nu}\bar{\sigma}_{\nu}^{\dot{\alpha} \alpha}
U^{\dagger \beta}_{\alpha}\left(-\epsilon_{rs}\eta^{\dagger
s}\eta_{\beta}u_L(r)+\eta_r \epsilon_{\beta
\gamma}\eta^{\dagger\gamma}d_L(r)\right),\\
&&u_{R\alpha}(x)=\frac{i}{2\pi}
\frac{m_u}{\rho}u_R(r)U^{\dagger\beta}_{\alpha}\eta_{\beta},\\
&&d_{R\alpha}(x)=\frac{i}{2
\pi}\frac{m_d}{\rho}d_R(r)U^{\dagger\beta}_{\alpha}\epsilon_{\beta\gamma}
\eta^{\dagger\gamma},
\end{eqnarray}
where the Greek and the Roman letters denote indices of spinor and isospinor,
respectively. 
For small $\rho v$, the solution is given  approximately, by the
following;
\begin{eqnarray}
&&u_L(r)=\left\{
    \begin{array}{ll}
\displaystyle \frac{1}{r(r^2+1)^{3/2}},&
\quad \mbox{if} \quad r \ll (\rho m_u)^{-1/2}\ ; \\[0.4cm]
\displaystyle  -2 \pi^2 \frac{1}{r}\frac{d}{dr}G_{\rho m_u}(r),&
\quad \mbox{if} \quad r \gg (\rho m_u)^{-1/2}\ ;
     \end{array} \right.\\
&&u_R(r)=\left\{
    \begin{array}{ll}
\displaystyle \frac{1}{r^2+1},&
\quad \mbox{if} \quad r \ll (\rho m_u)^{-1/2}\ ; \\[0.4cm]
\displaystyle  2 \pi^2 G_{\rho m_u}(r),&
\quad \mbox{if} \quad r \gg (\rho m_u)^{-1/2},
     \end{array} \right.
\end{eqnarray}
and $d_L$ and $d_R$ are obtained by replacing $m_u$ with $m_d$.
The above approximative behavior is correct even though $\rho v$ is
$O(1)$, since the valley instanton keep its shape unchanged as was
seen in section \ref{sec:s443}.

\subsubsection{Calculation of the cross section}
To evaluate
the multi-point Green function, we must introduce collective
coordinates $x_0$ and $U$ to restore the translational and Lorentz
invariance as the valley instanton breaks these symmetries.
In addition to them, the $\rho$ integral must be brought into, because 
along the valley direction the action varies gently and the
corresponding integration can not be approximated by the Gaussian
integration. 

Since the valley instanton is not a classical solution, a slight
complication arises to introduce the collective coordinates. 
For the classical solution, corresponding to the broken symmetries, there 
exist zero modes and the collective coordinates are introduced
naturally to extract these zero modes from the functional integral.
However, the existence of the zero modes is not guaranteed for the
valley instanton.
In Ref.~\citen{rf:hs}, another collective coordinate
method have been used. 
The point is that the
collective coordinates for the translations and the Lorentz $SU(2)_R$
rotations can be treated as transformation parameters of a kind of ``gauge
symmetries''.\cite{rf:hk} 
The orbits of the ``gauge symmetries'' can be extracted by the usual
Faddeev-Popov method and they becomes the collective coordinates for the
translations and the Lorentz $SU(2)_R$ rotations. 
As well as the usual gauge symmetry, 
the translation and the Lorentz invariance of the Green function is
automatically guaranteed.
For the collective coordinate of the valley direction, we bring into it
so as the quantum fluctuations around the valley instanton are
restricted to be orthogonal to the valley direction.\cite{rf:pvalley}
From this prescription, liner terms of the quantum fluctuation drop
naturally from the action while we do not treat the classical solution. 

\begin{figure}
\centerline{
\epsfxsize=8cm
\epsfbox{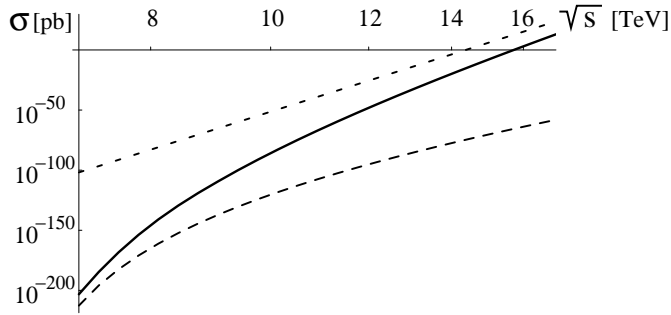}
}
\caption{The energy dependence of the cross section for $n_w=75$ and
$n_h=0$. The solid line denotes the result by the saddle point
approximation. The dashed line and the dotted line denote the
approximate result in the non-relativistic limit and extremely relativistic limit, respectively.}
\label{fig:s45crs75}
\end{figure}

\begin{figure}
\centerline{
\epsfxsize=8cm
\epsfbox{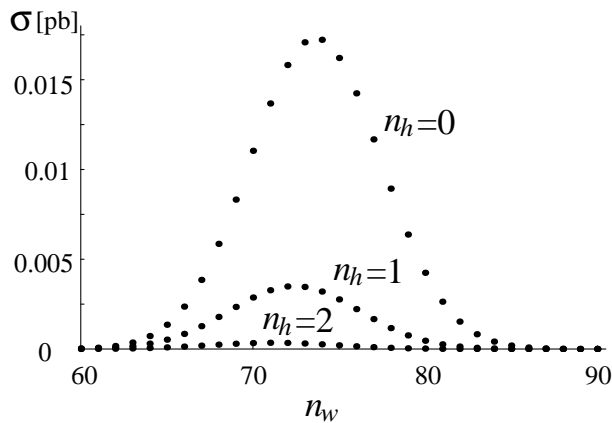}
}
\caption{The cross section for the various $n_w$ and $n_h$ at
$s^{1/2}=$ 15.55 TeV.
}
\label{fig:s45crs-nw}
\end{figure}

\begin{figure}
\centerline{
\epsfxsize=8cm
\epsfbox{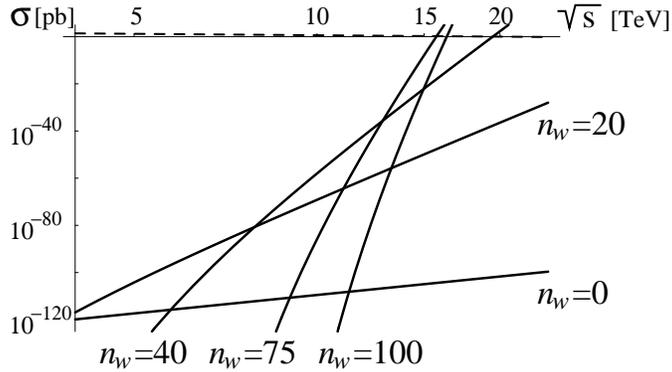}
}
\caption{The energy dependence of the cross section for $n_w$=0, 20, 40,
75 and 100, and $n_h$=0. A dashed line denotes the unitarity bound.}
\label{fig:s45crs-s}
\end{figure}

Since the valley instanton is very similar
to the original instanton, we approximate the determinant and
Jacobian of the valley instanton by one of the original
instanton\cite{rf:thooft}. 
Then the Fourier transformed Green function is given by,
\begin{eqnarray}
&&\tilde{G}({p},{q},{k})=e^{-8\pi^2/g^2} g^{-8}\int dU\int d\rho
e^{-2\pi^2(\rho v)^2/\lambda}\rho^{2 n_f-5}c(\rho \mu)^{(43-8 n_f)/6}
\nonumber\\
&&\hspace{15ex}\times \tilde{\psi}(p_1)\cdots
\tilde{A}_{\mu}(q_1)\cdots \tilde{\phi}(k_1)\cdots .
\end{eqnarray}
Here the Fourier transformations of the fermionic zero mode
$\tilde{\psi}$ is given by 
\begin{eqnarray}
  \label{psitilde}
  \tilde{\psi}=
\left(
  \begin{array}{c}
\tilde{\psi}_{R\alpha}\\
\tilde{\psi}_L^{\dot{\alpha}}
  \end{array}
\right)
=2\pi i \rho 
\left(
  \begin{array}{c}
-m_f U^{\dagger}\chi\\
\bar{\rlap/p}U^{\dagger}\chi
  \end{array}
\right)\frac{1}{p^2+m_f^2}+\cdots ,
\end{eqnarray}
where $\cdots$ includes the regular term in the limit:
$p^2+m_f^2\rightarrow 0$, and $\chi_{\alpha}$ is the constant spinor,
$(0,-1)$ for $\psi = u,\nu$ and (1,0) for $\psi = d,e$. The Fourier
transforms of the valley instanton, $\tilde{A}_{\mu}$ is given by
\begin{eqnarray}
  \tilde{A}_{\mu} = -\frac{1}{\nu}\pi^2 i \rho^2 \frac{q_{\nu}U \bar{\sigma}_{\mu\nu}U^{\dagger}}{q^2+m_w^2}+\cdots.
\end{eqnarray}
We define the shifted Higgs field $\phi$ by $H=(v+\phi)\eta$ and the
Fourier transform is given by,
\begin{eqnarray}
  \label{phitilde}
  \tilde{\phi}=\frac{1}{2\nu}\pi^2 \rho^2 \frac{v}{k^2+m_h^2}+\cdots .
\end{eqnarray}
The integration of the valley parameter $\rho v$ is given by
\begin{eqnarray}
\int_0^{\infty} d(\rho v) e^{-2 \pi^2(\rho v)^2/\lambda}(\rho v)^{2 t-1}
=\frac{1}{2}\left(\frac{\lambda}{2\pi^2}\right)^t \Gamma(t),
\end{eqnarray}
where $t=\frac{7}{3}n_f+n_w+n_h+\frac{19}{12}$. 
The integration is dominated by the contribution around the saddle
point; 
\begin{eqnarray}
\rho v_s=\left\{\frac{\lambda}{2\pi^2}(t-\frac{1}{2})\right\}^{1/2}
\end{eqnarray}
Therefore, we can check the consistency of the approximation by this
value of the saddle point. The saddle point $\rho v_s$ depends on the
number of bosons monotonously. The valley instanton is quite similar to
the original instanton even at $\rho v= 1$, and therefore the
approximation is valid for $n_w+n_h\sim 40$, if we assume
$m_{_H}=m_{_W}$ and use $g^2\sim 0.42$. On the other hand, the
constrained instanton is deviated from the original instanton at $\rho
v= 0.5$ and so the approximation breaks for $n_w +n_h\sim 4$.
Therefore we can calculate the amplitude of the multi-boson process,
using the valley instanton. Finally, we obtain the Green function;
\begin{eqnarray}
&&\tilde{G}({p},{q},{k})=(-i)^{n_w} c^{\prime}\cdot
2^{-{n_h}/2}\left(\frac{\mu}{m_{_W}}\right)^{(43-8 n_f)/6}
m_{_W}^{-6 n_f-2 n_w-n_h+4}e^{-8\pi^2/g^2}\nonumber\\
&&\hspace{15ex}\times g^{14n_f/3+n_w+n_h-29/6}
\int dU L(U;{p},{q},{k}),
\end{eqnarray}
where 
\begin{eqnarray}
&&c^{\prime}=c\cdot2^{-2n_f/3-25/6}\pi^{-2n_f/3-19/6}\Gamma(t),
\nonumber\\
&&\hspace{2ex}=1.26\times 10^6\cdot e^{6.516n_f}\cdot\Gamma(t),
\end{eqnarray}
and $L(U;p,q,k)$ is defined by,
\begin{eqnarray}
&&L(U;p,q,k)=\prod_{i=1}^{4n_f}
\left(
\begin{array}{c}
\displaystyle -m_f U^{\dagger}\chi 
\\
\displaystyle
\bar{\rlap/p}_i U^{\dagger}\chi
\end{array} 
\right)\frac{1}{p_i^2+m_f^2}
\prod^{n_w}_{j=1}\frac{q_{j\nu}U\bar{\sigma}_{\mu\nu}U^{\dagger}}{q_j^2+m_{_W}^2}
\prod^{n_h}_{l=1}\frac{1}{k_l^2+m_{_H}^2}.
\end{eqnarray}
We ignore the momentum dependence that goes to zero on mass-shell.
Performing the LSZ procedure, we obtain the invariant amplitude
$T_{n_f,n_w,n_h}$. Summing up the polarization and charge of the gauge
field, isospinor and spinor of the fermion field, we obtain the
amplitude; 
\begin{eqnarray}
&&\sum |T_{n_f,n_w,n_h}|^2=c^{\prime
2}2^{-n_w-n_h}\left(\frac{\mu}{m_{_H}}\right)^{(43-8n_f)/3}
(m^2_{_W})^{-6 n_f-2 n_w-n_h+4}e^{-16\pi^2/g^2}
\nonumber\\
&&\hspace{10ex}\times
(g^2)^{14n_f/3+n_w+n_h-29/6}
\int dU\prod_{i=1}^{4
n_f}{\rm tr} (\bar{\rlap/p}_i
U)\prod_{j=1}^{n_w}(-)g^{\mu\rho}q^{\nu}_jq^{\sigma}_j {\rm tr} (\sigma_{\mu\nu}U\bar{\sigma}_{\rho\sigma}U^{\dagger}).
\nonumber\\
\end{eqnarray}
The cross section is given by the following;
\begin{eqnarray}
&&\sigma=\frac{1}{n_w!n_h!}\frac{1}{2^4}\int 
\prod_{i=3}^{4n_f}\frac{d^3 p_i}{(2\pi)^3 2 E_{p_i}}
\prod_{j=1}^{n_w}\frac{d^3 q_j}{(2\pi)^3 2 E_{q_j}}
\prod_{l=1}^{n_h}\frac{d^3 k_l}{(2\pi)^3 2 E_{k_l}}
(2\pi)^4\delta^{(4)}(p_{\rm in}-p_{\rm out})
\nonumber\\
&&\hspace{15ex}\times \sum|T_{n_f,n_w,n_h}|^2\cdot \frac{1}{4\sqrt{(p_1\cdot
p_2)^2-m_1^2m_2^2} }
\\
&&\hspace{2ex}=\frac{c^{\prime 2}}{n_w!n_h!}2^{-n_h-5}(2\pi)^{-12
n_f-3n_w-3n_h+10}\left(\frac{\mu}{m_{_W}}\right)^{(43-8n_f)/3}(m_{_W})^{-6n_f-2n_w-n_h+4}
\nonumber\\
&&\hspace{15ex}\times e^{-16\pi^2/g^2}
(g^2)^{14n_f/3+n_w+n_h-29/6}\cdot I_{2n_f,n_w,n_h}(s)\cdot s^{-1},
\nonumber
\end{eqnarray}
where $s=p_{\rm in}^2=(p_1+p_2)^2$ and the masses of fermions of the initial state
are ignored. We denote the sum of the momentums of the final state by $p_{\rm out}$.
All the information about phase space and group integration is encoded
in the function $I_{2n_f,n_w,n_h}(s)$. We define $I_{l,m,n}(s)$
\begin{eqnarray}
&&I_{l,m,n}(s)=
\int\prod_{i=3}^{2l}\frac{d^3 p_i}{2 E_{p_i}}
\prod_{j=1}^{m}\frac{d^3 q_j}{2 E_{q_j}}
\prod_{l=1}^{n}\frac{d^3 k_l}{2 E_{k_l}}
\delta^{(4)}(p_{\rm in}-p_{\rm out})\nonumber\\
&&\hspace{20ex}\times
2^{m}\int dU\prod_{i=1}^{2l}{\rm tr} (\bar{\rlap/p}_i U)\prod_{j=1}^{m}(-)g^{\mu\rho}q^{\nu}_jq^{\sigma}_j {\rm tr} (\sigma_{\mu\nu}U\bar{\sigma}_{\rho\sigma}U^{\dagger})
\\
&&\hspace{13ex}=\int\prod_{i=3}^{2l}\frac{d^3 p_i}{2 E_{p_i}}
\prod_{j=1}^{m}\frac{d^3 q_j}{2 E_{q_j}}
\prod_{l=1}^{n}\frac{d^3 k_l}{2 E_{k_l}}
\delta^{(4)}(p_{\rm in}-p_{\rm out})
\nonumber\\
&&\hspace{20ex}\times
\int dU\prod_{i=1}^{2l}{\rm tr} (\bar{\rlap/p}_i U)\prod_{j=1}^{m}\left[\left\{{\rm
tr}(\bar{\rlap/q}_jU)\right\}^2-m_{_W}^2\right]. 
\nonumber
\end{eqnarray}
To evaluate the function $I_{l,m,n}(s)$, we consider the Laplace
transformation of this function. At first, we assume that $p_{\rm in}$, $p_1$
and $p_2$ are independent variables, and at last we input $p_{\rm
in}=p_1+p_2$.
The Laplace transform $\Phi_{l,m,n}(\alpha;p_1,p_2)$ is given by,
\begin{eqnarray}
&&\Phi_{l,m,n}(\alpha;p_1,p_2)=\int d^4 p_{\rm in}\ e^{-\alpha\cdot
p_{\rm in}}I_{l,m,n}(p_{\rm in};p_1,p_2) 
\nonumber\\
&&\hspace{3ex}=\int dU{\rm tr}(\bar{\rlap/p}_1U){\rm
tr}(\bar{\rlap/p}_2U)\phi_f(\alpha,U)^{2l-2}\phi_g(\alpha,U)^{m}\phi_h(\alpha)^{n},
\end{eqnarray}
where 
\begin{eqnarray}
&&\phi_f(\alpha,U)=\int\frac{d^3 p}{2 E_p}{\rm
tr}(\bar{\rlap/p}U)e^{-\alpha\cdot p}
=2\pi \frac{m_f^2}{\alpha^2} K_2(\alpha m_f){\rm tr}(\bar{\rlap/\alpha}U),
\\
&&\phi_g(\alpha,U)=\int\frac{d^3 q}{2 E_q}\left[\left\{{\rm
tr}(\bar{\rlap/q}U)\right\}^2-m_{_W}^2\right]e^{-\alpha\cdot q}
=\frac{2\pi m_{_W}^3}{\alpha^3}K_3(\alpha
m_{_W})\left[\left\{{\rm tr}(\bar{\rlap/\alpha}U)\right\}^2-\alpha^2\right],
\nonumber\\
\\
&&\phi_h(\alpha)=\int\frac{d^3 k}{2 E_k}e^{-\alpha\cdot k}=\frac{2\pi
m_{_H}}{\alpha} K_1(\alpha m_{_H}). 
\end{eqnarray}
The above functions  $K_{\nu}$  are the modified Bessel functions and
$\alpha=\sqrt{\alpha^2}$. Then we can perform the group
integration easily,
\begin{eqnarray}
&&\int dU {\rm tr}(\bar{\rlap/p}_1 U){\rm tr}(\bar{\rlap/p}_2 U)\left\{{\rm
tr}(\bar{\rlap/\alpha} U)\right\}^{2l-2}\left[\left\{{\rm
tr}(\bar{\rlap/\alpha}U)\right\}^2-\alpha^2\right]^{m} 
\nonumber\\
&&\hspace{5ex}=\sum_{i=0}^{m}\hspace{-1ex}\ _{m}C_i\int dU 
{\rm tr}(\bar{\rlap/p}_1 U){\rm tr}(\bar{\rlap/p}_2 U)
\left\{{\rm tr}(\bar{\rlap/\alpha}
U)\right\}^{2l-2+2i}(-\alpha^2)^{n-i}
\\
&&\hspace{5ex}=\left\{C_0(p_1\cdot\alpha)(p_2\cdot\alpha)+
C_1(p_1\cdot p_2)\alpha^2 \right\}\alpha^{2(l+m-2)},
\nonumber
\end{eqnarray}
where 
\begin{eqnarray}
&&C_0=4\sum_{i=0}^{m}(-)^{m-i}\hspace{-1ex}\ _{m}C_i
\frac{\left\{2(l+i-1)\right\}!}{(l+i+1)!(l+i-2)!},
\\
&&C_1=2\sum_{i=0}^{m}(-)^{m-i}\hspace{-1ex}\ _{m}C_i
\frac{\left\{2(l+i-1)\right\}!}{(l+i+1)!(l+i-1)!}.
\end{eqnarray}
Finally, using the inverse Laplace transformation, we obtain the
function $I_{l,m,n}$. We can evaluate $I_{\l,m,n}$ analytically in the
case of the relativistic limit and the non-relativistic limit.
In the relativistic limit, $I_{\l,m,n}$ is given by,
\begin{eqnarray}
I_{l,m,n}=C^{\rm ER}_{l,m,n}s^{3l+2m+n-4}
\end{eqnarray}
where
\begin{eqnarray}
&&C^{\rm
ER}_{l,m,n}=2^m\left(\frac{\pi}{2}\right)^{2l+m+n-3}\left\{(3l+2m+n-3)!(3l+2m+n-5)!\right\}^{-1}
\nonumber\\
&&\hspace{3ex}
\times\sum_{i=0}^{m}(-)^{m-i}\ _{m}C_i
\frac{\left\{2(l+i-1)\right\}!}{(l+i+1)!(l+i-1)!}
\left\{1+(l+i)(3l+2m+n-4)\right\}.
\end{eqnarray}
In the non-relativistic limit, we obtain 
\begin{eqnarray}
I_{l,m,n}=C^{\rm
NR}_{l,m,n}m_f^{3/2}m_{_W}^{5m/2}m_{_H}^{n/2}(\sqrt{s}-M)^{6l+3(m+n)/2-10}\cdot
s^{1/4},
\end{eqnarray}
where
\begin{eqnarray}
&&C^{\rm
NR}_{l,m,n}=\left\{\left(6l+\frac{3}{2}m+\frac{3}{2}n-10\right)!\right\}^{-1}\cdot
2^{4l+(m+n)/2-7}\pi^{2l+3(m+n)/2-3}
\nonumber\\
&&\hspace{10ex}\times \sum_{i=0}^m(-)^{m-i}\
_mC_i\frac{\left\{2(l+i-1)\right\}!}{(l+i+1)!(l+i-1)!}(l+i).
\end{eqnarray}
In the general cases, we can perform the inverse Laplace transformation
numerically, using the steepest descent approximation. We show the
result in Fig.~\ref{fig:s45crs75} for the case where $n_w=75$ and $n_h=0$. In
Fig.~\ref{fig:s45crs75}, the solid line denotes the numerical result by the
steepest descent method. The dashed line and the dotted line denote the
approximate result in the non-relativistic limit and extremely
relativistic limit, respectively. We show cross
sections for the various values of $n_w$ and $n_h$ in Fig.~\ref{fig:s45crs-nw} 
at $\sqrt{s}$=15.55 TeV. From this, we understand that the gauge boson
plays an important role rather than the Higgs particle. In
Fig.~\ref{fig:s45crs-s}, we show the energy dependence of the cross section
for $n_w$=0, 20, 40, 75 and 100, and $n_h$=0. We show the unitarity
bound as a dashed line in Fig.~\ref{fig:s45crs-s}. The approximation breaks in 
the region of the energy
where the cross section overcomes the unitarity bound. In the region,
the multi-instanton effect\cite{rf:ii}(see section 2) or interaction
between the gauge bosons in the final state\cite{rf:khozering} are
important, which are not evaluated in our analysis. 
Since the asymptotic behavior of the valley instanton is different
from the constrained instanton, a naive calculation yields that the
interaction between the valley instanton and the anti valley instanton
is different from that between the constrained instanton and the anti
constrained instanton. 
Therefore, the analysis given in section 2 must be reconsidered.
This is interesting problem, but remains as an open question for now. 


\section{Complex-time methods} 
\renewcommand{\baselinestretch}{1.2}
\newcommand{\bra}[1]{\left\langle #1 \right|}
\newcommand{\ket}[1]{\left| #1 \right\rangle}
\newcommand{\VEV}[1]{\left\langle #1 \right\rangle}
\newcommand{\braket}[2]{\VEV{#1 | #2}}
\newcommand{\calP}{{\cal P}}
\newcommand{\bx}[1]{\vbox{\hrule height1pt%
           \hbox{\vrule width1pt\hskip1pt%
           \vbox{\vskip1pt\hbox{#1}\vskip1pt}%
           \hskip1pt\vrule width1pt}}}%
\newcommand{\Half}{\frac{1}{2}}
\newcommand{\calD}{{\cal D}}
\newcommand{\Higgs}{{\rm \Phi}}
\newcommand{\Wboson}{{\rm W}}
\newcommand{\vev}{\VEV{\phi}}
\newcommand{\calO}{{\cal O}}
\newcommand{\cl}{{\rm cl}}
\newcommand{\psiwkb}{{\psi_{\rm WKB}}}
\newcommand{\twopi}{\frac{1}{\sqrt{2\pi}}}
\newcommand{\bl}{_\lambda}
\newcommand{\kb}{\mbox{\boldmath $k$}}
\newcommand{\xb}{\mbox{\boldmath $x$}}
\newcommand{\qb}{\mbox{\boldmath $q$}}
\newcommand{\etab}{\mbox{\boldmath $\eta$}}
\newcommand{\pib}{\mbox{\boldmath $\pi$}}
\newcommand{\xib}{\mbox{\boldmath $\xi$}}
\newcommand{\rbox}[1]{\vbox{\hrule height.8pt%
                \hbox{\vrule width.8pt\kern5pt
                \vbox{\kern5pt\hbox{#1}\kern5pt}\kern5pt
                \vrule width.8pt}
                \hrule height.8pt}}

As we have seen so far in this paper, the imaginary-time path-integral 
formalism combined with the proper valley method has been successful 
for the treatment of the quantum tunneling phenomena.
In spite of this, some basic as well as practical questions 
related to analytic continuation remain.
How does the boundary condition matches that of the valley configurations?
In estimating many-point Green functions, how does
the external fields enter (affect) the imaginary time configuration?
[This is crucial for the validity of the
path-deformations in the valley calculation, as well as bounce calculations.]
Some other points are elaborated by Boyanovsky, Willey
and Holman.\cite{rf:boy}

In quantum mechanics, the complex-time method has
been studied by various authors.
\cite{rf:mcl,rf:bender,rf:levi,rf:patra,rf:lap,rf:weiss,rf:carlitz}
It was argued that it allows semi-classical approximation
for tunneling phenomena and overcomes various problems
associated with the pure-imaginary-time method.
In it, one considers the analytic continuation of the
time-integral in Fourier transform of the Feynman kernel.
The existence of the tunneling leads to the
existence of the complex saddle-points in the time-plane.
It was claimed that semi-classical approximation is done by deforming the
time-integral so that it goes through all such saddle-points.\cite{rf:carlitz}

This gives one hope that calculations in field theory
may be improved by using the complex-time method.
Indeed, Son and Rubakov
\cite{rf:khleb,rf:son,rf:rubakoc}
adopted a complex-time method for the estimate of the cross section.
They used periodic instanton-anti-instanton solution
in imaginary time, and the initial Minkowski state
overlaps with the instanton as a coherent state.

Closer look, however, reveals many questions unanswered so far.\cite{rf:ah}
It is known that the complex-time plane is plagued
with singularities and infinite number of saddle-points, which form lattice
structure.
Among saddle-points, there are two kinds, those that are
solutions of the equation of motions (we shall call these
``physical saddle-points" in this paper),
and those that are not (``unphysical saddle-points").
How the path is deformed to avoid the singularities
and to go through only the physical saddle-points is unknown.
One simply assumes that all and only the physical saddle-points contribute.
Even so, the weight of each saddle-point is a riddle.
They are determined so that the result agrees with that of the WKB
approximation.
It is not known how it comes about from the path.

In this section, we give the solution of these problems,
based on a reduction formula in the number of the
turning points of the path.
In the first subsection, we give a brief overview
of the saddle-point calculations in complex-time.
There we elaborate some of the points briefly mentioned above.
In order to solve the riddles pointed in this subsection, 
we analyze the orthonormality of the WKB wavefunctions
and construct the Green function from them in the subsection 5.2.
This way, the arbitrariness in the path-integral formulation
is avoided and the relevant Green functions are constructed
from the first principle.
Based on this construction of the Green functions, 
we derive the connection formula for the WKB wavefunction
in the subsection 5.3, where it leads to the
reduction formula for Green function.
Expanding this formula, we obtain a series,
which can be interpreted as a sum over the physical saddle-points.  
The weights and the phases of saddle-points
are determined from this formula.


\subsection{Saddle-point method}
We consider one-dimensional quantum mechanics with action;
\begin{eqnarray}
S = \int dt \left[ \frac{1}{2} \dot{x}^2 - V(x) \right]_. 
\label{eq:s51-action}
\end{eqnarray}
The potential $V(x)$ is assumed to be smooth enough to 
allow WKB approximation in asymptotic regions I ($x \rightarrow -\infty$)
and II ($x \rightarrow \infty$).
The finite-time Green function (Feynman kernel) is defined by the 
following in the Heisenberg representation;
\begin{eqnarray}
G(x_{\rm i}, x_{\rm f}; T) = \bra{x_{\rm f}} e^{-iHT} \ket{x_{\rm i}},
\label{eq:s51-tgreen}
\end{eqnarray}
in terms of the Hamiltonian of the system, $H$.
The advanced and retarded resolvents are, respectively;
\begin{eqnarray}
G^A(x_{\rm i}, x_{\rm f}; E) &=& 
                i \int_{-\infty}^0 dT e^{i(E - i\delta)T} G(x_{\rm i}, x_{\rm f}; T)
                = \bra{x_{\rm f}} \frac{1}{E-i\delta -H} \ket{x_{\rm i}}, 
\label{eq:s51-advancedresolv}
\\      
G^R(x_{\rm i}, x_{\rm f}; E) &=& 
                -i \int_0^\infty dT e^{i(E + i\delta)T} G(x_{\rm i}, x_{\rm f}; T)
                = \bra{x_{\rm f}} \frac{1}{E+i\delta -H} \ket{x_{\rm i}},
\label{eq:s51-retardedresolv}
\end{eqnarray}
where real-positive $\delta$ is introduced to guarantee the
convergence of the integrals.
The poles of these Green functions come from the bound states,
and the cut from the continuous spectra. 

Let us first look at the saddle-point approximation for the path
integral,
\begin{eqnarray}
G(x_{\rm i}, x_{\rm f}; T)=\int_{x(0)=x_{\rm i}}^{x(T)=
x_{\rm f}} {\cal D} x e^{iS}. 
\end{eqnarray}
One obtains the following from the saddle-point method;
\begin{eqnarray}
G(x_{\rm i}, x_{\rm f}; T) &=& \sum_{x_{\cl}}\left[\frac{2 \pi}{i}\frac{\dot {x}_{\cl}(0) 
\dot{x}_{\cl}(T)}{-\partial^2 S_{\cl}/ \partial
T^2}\right]^{-1 / 2} e^{iS_{\rm cl}},
\label{eq:s51-grnt} 
\end{eqnarray}
where $x_{\rm cl}$ is the solution of the classical equation of motion and
$S_{\cl}$ is the action evaluated at the classical solution. This
approximation is valid as long as both $x_{\rm i}$ and  $x_{\rm f}$ are far from the
turning points where classical velocities vanish. 
Substituting (\ref{eq:s51-grnt}) into the retarded resolvent in
(\ref{eq:s51-advancedresolv}), (\ref{eq:s51-retardedresolv}), 
one arrives at the following expression; 
\begin{eqnarray}
G^R(x_{\rm i}, x_{\rm f}; E) = -i \int_0^\infty dT \sum_{x_{\cl}} \left[\frac{2\pi }{
i}\ \frac{\dot {x}_{\cl}(0) \dot {x}_{\cl}(T) }{ -\partial^2 S_{\cl} / \partial
T^2}\right]^{-{1/ 2}}e^{i\left( ET + S_{\cl}\right)}. 
\label{eq:s51-green}
\end{eqnarray}
In the above, we absorbed the infinitesimal imaginary convergence factor 
($i\delta$) in $E$.

In calculating (\ref{eq:s51-green}), one hopes to apply the saddle-point
approximation to the $T$-integral.
The saddle-point condition (the stable-phase condition, to be more
exact) is then,
\begin{eqnarray}
E=-\frac{\partial S_{\cl}}{ \partial T} = 
\frac{1}{2} \dot{x}_{\cl}^2 + V(x_{\cl}) \ .
\label{eq:s51-tcondition}
\end{eqnarray}
In other words, the solution of the equation of motion $x_{\rm \cl}(t)$
has to have the energy $E$.

\begin{figure}[b]
\centerline{
\epsfxsize=8cm
\epsfbox{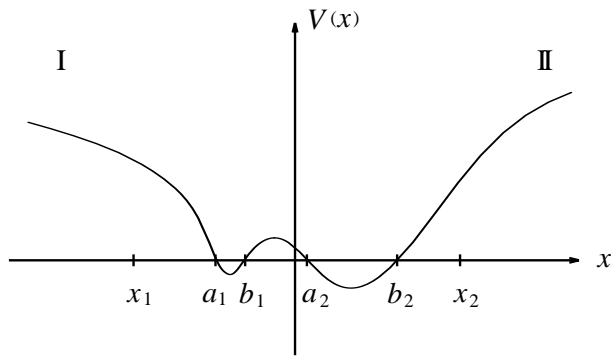}
}
\caption{A potential $V(X)$ with asymptotic regions I and II.}
\label{fig:s51-potshape}
\end{figure}

Let us consider a situation depicted in Fig.~\ref{fig:s51-potshape}.
Both the initial point $x_{\rm i}$ and the final point $x_{\rm f}$
lie in the forbidden region.
There is no real-time solution of the equation of motion.
Thus no semi-classical approximation is possible for
the real-time formalism.

If one assumes
that the expression (\ref{eq:s51-grnt}) gives correct analytic continuation
of the Green function in the complex $T$-plane when 
complex-$T$ solutions of the equation of motion exist,
one could go to the complex $T$-plane and apply the saddle-point method.
One then arrives at the following expression;  
\begin{eqnarray}
G(x_{\rm i}, x_{\rm f}; E) = -i\sum_{T_{\rm s}} \sum_{x_{\cl}} \vert \dot {x}_{\cl}(0) 
\dot {x}_{\cl}(T_{\rm s})
\vert^{-{1/2}}\ w(T_{\rm s})\ p(T_{\rm s})\  e^{i W_{\cl}(T_{\rm s})} ,
\label{eq:s51-grn} 
\end{eqnarray} 
where $T_{\rm s}$ is given by (\ref{eq:s51-tcondition}).
Here $W_{\cl}(T_{\rm s})$ is the WKB phase, $w(T_{\rm s})$ is weight and $p(T_{\rm s})$ is
phase for the
corresponding saddle-point. The original contour is deformed to pass
the saddle-points by a series of steepest descent paths 
from the origin $T=0$ to 
the positive infinity on the real axes (Re $T=\infty$, Im $T=0$). 
The weight of a saddle-point, $w(x_{\cl})$, 
is determined by how the contour crosses
the saddle-point: If the contour does not cross the saddle-point, the
corresponding weight is zero. If the contour crosses it along
the steepest descent direction, the weight is $1$. 
In case the contour reaches a sub-dominant saddle-point from the 
direction orthogonal to the steepest descent 
and leaves it along the steepest descent direction, the weight is $1/2$
(as in the calculation of the false vacuum decay).  
The phase of a saddle-point, $p(T_{\rm s})$ comes from the square root in
(\ref{eq:s51-green}). 

Carlitz and Nicole applied this method to 
linear potential, quadratic well and quadratic barrier, all of which
are exactly solvable.
Using the exact expression for the Green function (\ref{eq:s51-green}),
they found that the saddle-points explained above, namely
the ones associated with the complex-time solution, indeed
lead to saddle-points in the $T$-plain
[we shall call these ``physical saddle-points"].
They, however found more: there are saddle-points
that are {\sl not} associated with any solutions of
equations of motion (unphysical ones), and also singularities due to the
periodicities.
Deforming the $T$-integration path to avoid singularities,
they found that the path indeed goes through only the physical saddle-points.
As a result, the weights and phases in (\ref{eq:s51-grn}) are determined,
which leads to the correct WKB result.

In the case of a double-well potential, the deformation of the
integration contour is not specified,
which is understandable in view of the fact that
the analytic structure of (\ref{eq:s51-tgreen}) in complex $T$-plain
is fairly complicated.
The claim made in the literatures is simply that 
the contour passes all saddle-points that correspond to
classical trajectories with weights that are extracted
from the above calculation.

\begin{figure}[t]
\centerline{
\epsfxsize=8cm
\epsfbox{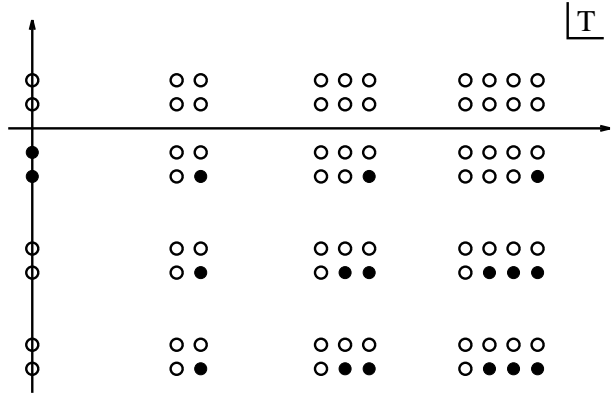}
}
\caption{Positions of saddle-points in complex $T$-plane.} 
\label{fig:s51-complex}
\end{figure}

The problem lies in the fact that we do not  know the contour which passes
all physical saddle-points that are distributed in complex-time plane.
Because of this, we cannot determine the weights of the saddle-points. 
In the case when $x_{\rm i}$ and $x_{\rm f}$
are in the forbidden region on the right side of the double-well
potential as depicted in Fig.~\ref{fig:s51-potshape}, saddle-points in complex
$T$-plane are given by 
\begin{eqnarray}
T_{\rm s} &=&\pm T(b_2,x_{\rm i}) \pm T(b_2,x_{\rm f}) + l T(a_1,b_1) + mT(b_1,a_2) +
nT(a_2,b_2)  
\\
 && l,m,n = 0,\pm 1, \pm 2,...\ , 
\end{eqnarray}
where 
\begin{eqnarray}
T(x,y) = 2 \int_{x}^{y} {dx'  \over  \sqrt {2 \left( E-V(x') \right)}}.
\end{eqnarray} 
Notice that $T(x,y)$ is pure imaginary if $(x, y)$ is in a forbidden
region. These saddle-points $T_{\rm s}$ are 
shown as solid circles in Fig.~\ref{fig:s51-complex}.
The unphysical saddle-points are also shown as open circles in
the same figure. We know of no contour
which passes only the physical ones and avoids the unphysical ones. 
Thus so far the complex-time method has not been worked out for 
the double-well and more complicated potentials.


\subsection{Green function and the WKB wavefunctions}
We shall examine the Green functions by using the 
complete orthonormal set of the eigenfunctions $\{ \ket{\psi_n}\}$ of $H$.
For example, the retarded Green function in (\ref{eq:s51-advancedresolv}),
(\ref{eq:s51-retardedresolv}) is written as
\begin{eqnarray}
G^R(x_{\rm i}, x_{\rm f}; E) = \sum_n \frac{\braket{x_{\rm f}}{\psi_n} \braket{\psi_n}{x_{\rm i}}
         }{ E + i\delta -\lambda_n}, 
\label{eq:s52-advg}
\end{eqnarray}
where the sum over the complete set is actually made of the integration
over the continuum spectrum and sum over the discrete bound states.

In order to study the Green functions in the asymptotic region,
we use the WKB approximation for the wavefunctions 
$\psi(x) = \braket{x}{\psi}$.
For a state with the eigenvalue $\lambda$ in the continuum spectrum,
the second-order WKB approximation yields the following;
\begin{eqnarray}
\psi_\lambda (x) = \frac{1}{\sqrt{2\pi p(x)}} \times \left\{\begin{array}{ll}
        \left(A\bl e^{i \int_x^{x_1} p(x') dx'} 
        + B\bl e^{-i \int_x^{x_1} p(x') dx'}\right)_,&{\rm for}         \ x\in{\rm  I},\\ 
        \left(C\bl e^{-i \int_{x_2}^x p(x') dx'} 
        + D\bl e^{i \int_{x_2}^x p(x') dx'}\right)_, 
                & {\rm for} \ x{\rm  \in II},\end{array}\right. 
\label{eq:s52-wavef}
\end{eqnarray}
where $p(x) = \sqrt{2(\lambda-V(x))}$.
We choose the limits $x_1$ and $x_2$ of the integrations 
to be the nearest turning points for definiteness.
Among the coefficients $A$, $B$, $C$, and $D$, only two 
are independent. Their inter-relation is linear due to the
superposition principle and can be written
as follows \cite{rf:ak};
\begin{eqnarray}
\left(\begin{array}{l}A\bl \\ B\bl \end{array}\right) = S(\lambda) \left(\begin{array}{l}C\bl \\ D\bl\end{array}\right)_. 
\end{eqnarray}
The 2$\times$2 matrix $S(\lambda)$ is determined by the 
shape of the potential $V(x)$ in the intermediate region between
$x_1$ and $x_2$.
The flux conservation law is satisfied for the Schr\"odinger equation;
\begin{eqnarray}
\frac{d}{d x} j = 0, \quad 
        j = \frac{i}{2} \left(\frac{d \psi^\ast }{ d x } \psi
            - \psi^\ast \frac{d \psi }{ d x }\right)_.  
\label{eq:s52-flux}
\end{eqnarray}
Substituting the WKB expression (\ref{eq:s52-wavef}) into the above, we find that
\begin{eqnarray}
 j = \frac{1}{ 2 \pi} (|A|^2 - |B|^2) = 
        \frac{1}{2 \pi} (|C|^2 - |D|^2). 
\label{eq:s52-abcd}
\end{eqnarray}
This results in the following relation between the matrix elements of 
$S(\lambda)$;
\def\se#1{S_{#1}}
\def\sen#1{|S_{#1}|^2}
\begin{eqnarray}
&&\sen{11}- \sen{21} = 1, \quad \sen{12} - \sen{22} = 1, \\ 
&&      \se{11}^\ast \se{12} - \se{21}^\ast \se{22} = 0. 
\label{eq:s52-sesen}
\end{eqnarray}
As a result, the matrix $S(\lambda)$ is parameterized by three real
functions $\alpha(\lambda), \beta(\lambda),$ and $\rho(\lambda)$ as follows;
\begin{eqnarray}
S(\lambda) = \left(\begin{array}{cc} e^{i\alpha} \cosh\rho & e^{i\beta} \sinh\rho \\
e^{-i\beta} \sinh\rho & e^{-i\alpha} \cosh\rho \end{array}\right)\ . 
\label{eq:s52-sabrho}
\end{eqnarray}
For a given eigenvalue $\lambda$, there are two independent eigenfunctions.
In order to construct them, we look at the inner-product of the
eigenfunction $\psi_\lambda (x)$ with coefficients 
($A_\lambda, B_\lambda, C_\lambda, D_\lambda$)
and another eigenfunction  $\tilde \psi_{\lambda'} (x)$ with
($\tilde A_{\lambda'}, \tilde B_{\lambda'}, \tilde C_{\lambda'}, 
\tilde D_{\lambda'}$).
Integration in the asymptotic regions determines the coefficient
of the delta function $\delta(\lambda - \lambda')$ completely, as
the delta function can come only from the infinite integrations.
[In order to calculate the finite terms we need to solve the 
Schr\"odinger equation completely. But this is not necessary for the
current purpose.]
In fact, integration in the region II yields,
\begin{eqnarray}
&&\int_{x_2}^\infty d x \ \psi^\ast_\lambda (x)\tilde \psi_{\lambda'} (x)
        = {1 \over 2\pi} \int_{x_2}^\infty 
                {d x \over \sqrt{p(x) p' (x)}}\nonumber\\
        &&\quad\times\Bigg( C_\lambda^\ast \tilde C_{\lambda'}
                e^{i \int_{x_2}^x ( p(x') -  p' (x') ) d x'}
        + D_\lambda^\ast \tilde D_{\lambda'}
                e^{-i \int_{x_2}^x ( p(x') - p' (x') ) d x'}\nonumber\\
        &&\quad\quad  + C_\lambda^\ast \tilde D_{\lambda'}
                e^{i \int_{x_2}^x ( p(x') + p' (x') ) d x'}
        + D_\lambda^\ast \tilde C_{\lambda'}
                e^{-i \int_{x_2}^x ( p(x') + p' (x') ) d x'}
        \Bigg),
\label{eq:s52-ortho}
\end{eqnarray}
As we are interested in the singularity of the above
for $\lambda = \lambda'$, we expand the exponent
using the following for $\lambda \sim \lambda'$, 
\begin{eqnarray}
p(x') -  p' (x') \simeq \frac{ (\lambda - \lambda') }{ p(x')}. 
\label{eq:s52-simp}
\end{eqnarray}
By using a new coordinate $y$ defined by
$ d y = d x / p(x)$, we find the contribution from 
the first two terms to the delta function;
\begin{eqnarray}
\int_{x_2}^\infty d x \ \psi^\ast_\lambda (x)\tilde \psi_{\lambda'} (x)
        &=& {1 \over 2\pi} \int_{y_2}^\infty d y
        \left( C_\lambda^\ast \tilde C_{\lambda}
        e^{i (\lambda-\lambda') (y-y_2)} +
        D_\lambda^\ast \tilde D_{\lambda} 
                e^{-i (\lambda-\lambda') (y-y_2)} \right)+ ... \nonumber\\
        &=& {1 \over 2} \left( C_\lambda^\ast \tilde C_{\lambda}+
        D_\lambda^\ast \tilde D_{\lambda} \right)
        \delta(\lambda - \lambda') + ... \ ,
\label{eq:s52-almost}
\end{eqnarray}  
where we have neglected all the finite terms.
Doing the similar calculation for the region I, we find
\begin{eqnarray}
\int_{-\infty}^\infty d x \psi^\ast_\lambda (x)\tilde \psi_{\lambda'} (x)
 = {1 \over 2} \left( 
        A_\lambda^\ast \tilde A_{\lambda}+
        B_\lambda^\ast \tilde B_{\lambda}+
        C_\lambda^\ast \tilde C_{\lambda}+
        D_\lambda^\ast \tilde D_{\lambda} \right)
        \delta(\lambda - \lambda'). 
\label{eq:s52-combined}
\end{eqnarray} 
Using the above result, we choose our orthonormal eigenfunctions
for a given $\lambda$ as the following;
\begin{eqnarray}
(A_\lambda, B_\lambda, C_\lambda, D_\lambda)^{(1)} &=& \left(
        {e^{i\alpha} \over \cosh \rho}, \ 0, \ 1, \ 
        -e^{i(\alpha - \beta)}\tanh\rho
        \right), 
\label{eq:s52-choicea}\\
        (A_\lambda, B_\lambda, C_\lambda, D_\lambda)^{(2)} &=&  \left(
        e^{i(\alpha + \beta)}\tanh\rho, \ 1, \ 0, \ 
        {e^{i\alpha} \over \cosh \rho}
        \right).
\label{eq:s52-choiceb}
\end{eqnarray}
[The above corresponds to two incoming states.
There are of course other choices, such as stationary states,
but all of those reads to the same result for the Feynman kernel.]

\begin{figure}[b]
\centerline{
\epsfxsize=8cm
\epsfbox{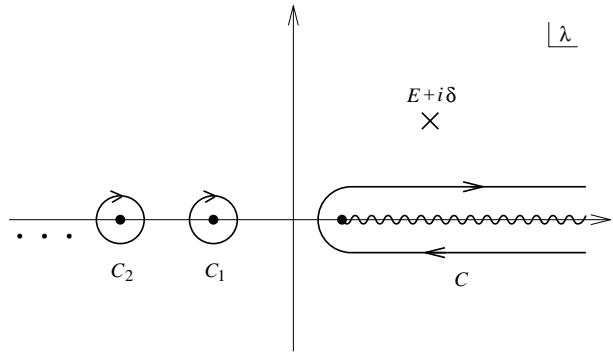}
}
\caption[The complex-plane of the eigenvalue $\lambda$.]{The complex-plane of the eigenvalue $\lambda$.
The cut on the real-axis is the continuous spectra and
the poles are the bound states.
}
\label{fig:s52-lambdapl}
\end{figure}

Let us first look at the  case when both $x_{\rm i}$ and $x_{\rm f}$ are in 
an allowed region in II.
From (\ref{eq:s52-choicea}) and (\ref{eq:s52-choiceb}), we find that
\begin{eqnarray}
&&\sum_{i=1,2}\psi_\lambda^{(i)}(x_{\rm f})\psi_\lambda^{(i)\ast}(x_{\rm i})
={1\over 2\pi \sqrt{p(x_{\rm i})p(x_{\rm f})}}\\
&&\quad\times \left[
e^{i \int_{x_{\rm i}}^{x_{\rm f}} p(x') dx'} 
 - e^{i \left(
\int_{x_2}^{x_{\rm i}} p(x') dx' +\int_{x_2}^{x_{\rm f}} p(x') dx' \right)} 
e^{i(\alpha - \beta)} \tanh \rho + ({\rm c.c.}) \right]_. \\  
\
\label{eq:s52-xxintwo}
\end{eqnarray}
The complex conjugate part can be understood as the  
same function below the cut on the real-axis of the complex 
$\lambda$-plane (see Fig.~\ref{fig:s52-lambdapl}).
This is guaranteed by the fact that 
as $\lambda$ moves below the cut, the role of the 
coefficients $A$ and $B$ is exchanged and likewise for $C$ and $D$,
and as a result, the phases $\alpha$ and $\beta$ change their signs, 
while $\rho$ does not.
Therefore the sum over the continuous spectra in the Feynman kernel (\ref{eq:s52-advg})
is written as the following $\lambda$-integration;
\begin{eqnarray}
&&G^R(x_{\rm i}, x_{\rm f};E) = 
{1\over 2\pi}  \int_C d\lambda \ 
{1 \over E + i\delta - \lambda}
{1 \over \sqrt{p(x_{\rm i})p(x_{\rm f})}}  \nonumber
\\
&&\ \times  
\left\{\begin{array}{l}
e^{i \int_{x_2}^{x_{\rm i}} p(x') dx'}
\left[e^{-i \int_{x_2}^{x_{\rm f}} p(x') dx'} 
- e^{i \int_{x_2}^{x_{\rm f}} p(x') dx'} 
e^{i(\alpha - \beta)} \tanh \rho\right] \ {\rm for}\  x_{\rm i} > x_{\rm f}, 
\vspace{1ex}
\\
e^{i \int_{x_2}^{x_{\rm f}} p(x') dx'}
\left[e^{-i \int_{x_2}^{x_{\rm i}} p(x') dx'} 
- e^{i \int_{x_2}^{x_{\rm i}} p(x') dx'} 
e^{i(\alpha - \beta)} \tanh \rho\right]\  {\rm for}\  x_{\rm f} > x_{\rm i}
. \end{array}\right.
\label{eq:s52-gxxintwo}
\end{eqnarray}
The integration contour $C$ in the complex $\lambda$-plane
is as in Fig.~\ref{fig:s52-lambdapl}.
In (\ref{eq:s52-gxxintwo}), we have chosen the integrand so that it
converges for $|\lambda| \rightarrow \infty$,
which we need later. 
As for the discrete spectrum, it is determined by the
absence of the diverging behavior of the wavefunction, 
which translates into the condition that for $C=0$, $A=0$.
From (\ref{eq:s52-sabrho}), we thus find that $\sinh \rho =0$ when $\lambda$
is equal to a discreet eigenvalue of $H$.
This allows us to write the sum over the discreet spectra
as pole integrations of the second term in (\ref{eq:s52-gxxintwo})
depicted in Fig.~\ref{fig:s52-lambdapl}.
Connecting all the contours and closing it at the 
infinity, we find that the whole contour enclose the 
pole at $\lambda=E + i \delta$.  Thus we end up with the expression
\begin{eqnarray}
&&G^R(x_{\rm i}, x_{\rm f};E)= 
 - {i \over \sqrt{p(x_{\rm i})p(x_{\rm f})}} \\
&& \ \times  
\left\{\begin{array}{ll}
e^{i \int_{x_2}^{x_{\rm i}} p(x') dx'}\left(e^{-i \int_{x_2}^{x_{\rm f}}
p(x') dx'} 
+ iR e^{i\int_{x_2}^{x_{\rm f}} p(x') dx'} \right)\ {\rm for}\ x_{\rm
i} > x_{\rm f}, 
\\
e^{i \int_{x_2}^{x_{\rm i}} p(x') dx'}\left(e^{-i \int_{x_2}^{x_{\rm f}}
p(x') dx'} 
+ iR e^{i\int_{x_2}^{x_{\rm f}} p(x') dx'y} \right)\ {\rm for}\ x_{\rm
f} > x_{\rm i},\end{array}\right. 
\label{eq:s52-ggxxintwo}
\end{eqnarray}
where we defined the ``reflection coefficient'' $R$ by the following,
\begin{eqnarray}
 R = i {S_{21} \over S_{22}} 
= i e^{i(\alpha - \beta)} \tanh \rho . 
\label{eq:s52-rdef}
\end{eqnarray}
We note that this expression (\ref{eq:s52-ggxxintwo}) is valid for
a general value of $E$ with proper analytic continuation of
the coefficients.
Therefore, if any end-points are in a forbidden region, 
simple analytic continuation of (\ref{eq:s52-ggxxintwo}) is appropriate.
We shall use this result in the next subsection.


\subsection{Reduction formula and its expansion}
In this subsection, we derive a formula for evaluating the Green
function in a system with an arbitrary potential to which we can apply
the WKB approximation. 
We consider a general potential which has arbitrary number of 
wells, and show that the Green function is given by
summing up all contributions of classical paths. 

\begin{figure}[b]
\centerline{
\epsfxsize=8cm
\epsfbox{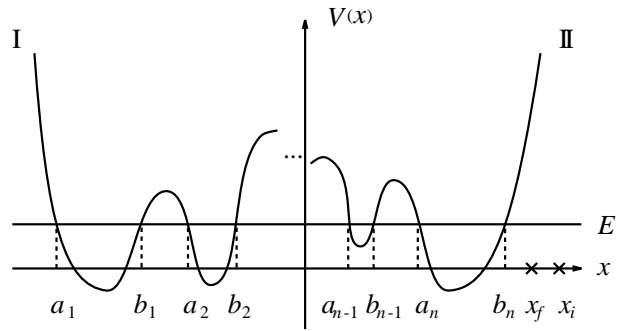}
}
\caption[A potential with $n$ wells]{A potential with $n$ wells we consider.
The turning points are denoted by $a_i$ and $b_i$,  for left 
and right of the $i$-th well, respectively ($i = 1 \sim n$).}
\label{fig:s53-one}
\end{figure}

Let us consider the case when $x_{\rm i}$ and  $x_{\rm f}$ lie in the forbidden 
region on the same sides of the wells. 
The potential which has $n$ wells is depicted
in Fig.~\ref{fig:s53-one}.
By the analytical continuation of (\ref{eq:s52-ggxxintwo}) in $E$, 
we find the retarded resolvent in the following form;  
\begin{eqnarray}
 G^R(x_{\rm i}, x_{\rm f}; E) = -\vert p(x_{\rm i}) p(x_{\rm f}) \vert^{-{1/2}}
 e^{- \Delta_i} \left( e^{ \Delta_f}+iR_n e^{-
 \Delta_f}\right)_,
\label{eq:s53-nwell}
\end{eqnarray}
where 
\begin{eqnarray}
\Delta_{i,f} \equiv \int_{b_n}^{x_{i,f}} dx |p(x)|. 
\end{eqnarray}
In order to specify the fact that this expression is for
$n$ wells, we attach the subscript $(n)$ to the coefficients
hereafter.

Due to the existence of the 
intermediate region $(b_{n-1},a_n)$, the matrix $S^{(n)}$,
which connects the regions I and II, can be written in terms of the
matrices $S^{(n-1)}$ that connects regions I and $(b_{n-1},a_n)$ and
$\tilde{S}$ for $(b_{n-1},a_n)$ and II. If we apply the linear WKB
connection formula for the latter region, 
which we can obtain from the saddle-point method for a linear
potential,  we obtain the following;
\begin{eqnarray}
S^{(n)}(E)&=&S^{(n-1)}(E) 
       \left(\begin{array}{cc} e^{- \Delta_{n-1}} & 0 
\\
                              0 & e^{ \Delta_{n-1}} \end{array}\right) 
       \left(\begin{array}{cc} {i \over 2} & -{i \over 2} \\
                         1   &   1 \end{array}\right)
\\
 &&\quad\quad\quad\times
       \left(\begin{array}{cc}e^{-i \left( W_{n}+{ \pi \over 2} \right) } & 0 \\
                0 & e^{i \left( W_{n}+{ \pi \over 2} \right )}  \end{array}\right)
       \left(\begin{array}{cc}{1 \over 2} &  i \\
                {1 \over 2} & -i \end{array}\right)
\\
 &=& S^{(n-1)}(E)
       \left(\begin{array}{cc}
      {1 \over 2} e^{- \Delta_{n-1}}\cos W_{n}
  &   e^{- \Delta_{n-1}} \sin W_{n} \\
     -e^{ \Delta_{n-1}} \sin W_{n}
  &   2 e^{ \Delta_{n-1}}\cos W_{n}\end{array}\right),
\end{eqnarray}
where 
$W_{n}$\  and $\Delta_{n-1}$ are defined by,
\begin{eqnarray}
 W_{n}&=& \int_{a_{n}}^{b_{n}}dx |p(x)| ,
\label{eq:s53-wn}
\\
 \Delta_{n-1}&=& \int_{b_{n-1}}^{a_{n}}dx |p(x)| .
\end{eqnarray}          
Therefore, we find the relations;
\begin{eqnarray}
 S_{21}^{(n)}&=&
  {1 \over 2 }S_{21}^{(n-1)} e^{- \Delta_{n-1}} \cos W_{n} 
 -S_{22}^{(n-1)} e^{ \Delta_{n-1}} \sin W_{n} \ , 
\label{eq:s53-cn}
\\
 S_{22}^{(n)}&=&
   S_{21}^{(n-1)} e^{- \Delta_{n-1}} \sin W_{n}
 + 2 S_{22}^{(n-1)} e^{ \Delta_{n-1}} \cos W_{n} \ .
\label{eq:s53-dn}
\end{eqnarray}
From (\ref{eq:s52-rdef}), (\ref{eq:s53-cn}) and  (\ref{eq:s53-dn}), we find that $R_{n}$ can be written as the 
following;
\begin{eqnarray}
 R_{n}  = {i \over 2}\ 
{- \sin W_{n}-{i \over 2} e^{-2 \Delta_{n-1}} R_{n-1} \cos W_{n}
 \over 
 \cos W_{n} -{i \over 2}e^{-2 \Delta_{n-1}} R_{n-1} \sin W_{n} }.
\end{eqnarray}
We find it most convenient to rewrite this to the following two
expressions;
\begin{eqnarray}
 R_{n} &=& {1 \over 2}\ {1-\tilde{R}_{n} e^{2iW_{n}} 
                               \over 1+\tilde{R}_{n} e^{2iW_{n}}},
\label{eq:s53-gnplusone}\\
 \tilde{R}_{n}&=&{
 {1-{1 \over 2}R_{n-1} e^{-2 \Delta_{n-1}}}
 \over
 {1+{1 \over 2}R_{n-1} e^{-2 \Delta_{n-1}}}
 }.  
\label{eq:s53-fn}
\end{eqnarray} 
Defined in this manner, the function $R_{n-1}$ corresponds to the 
reflection amplitude at the turning point $b_{n-1}$.

\begin{figure}[t,b]
\centerline{
\epsfxsize=12cm
\epsfbox{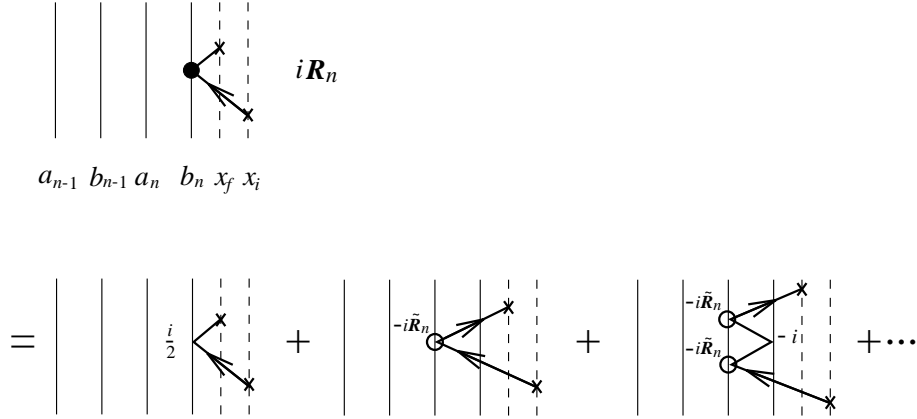}
}
\caption[Diagrammatic representation of the expansion of $iR_n$.]{Diagrammatic representation of the expansion of $iR_n$.
The information of the region left of $a_n$ is contained in $-i\tilde R_n$.}   
\label{fig:s53-two}
\end{figure}

\begin{figure}[h,t,b]
\centerline{
\epsfxsize=12cm
\epsfbox{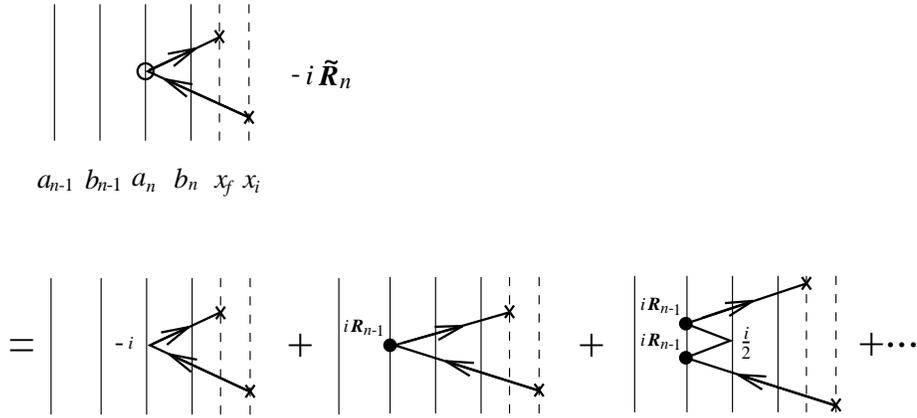}
}
\caption[Diagrammatic representation of $-i\tilde R_n$.]{Diagrammatic representation of the expansion of
$-i\tilde R_n$. The each path corresponds to a group of paths
with different number of oscillation in the forbidden region
$(b_{n-1}, a_n)$.}
\label{fig:s53-five}
\end{figure}

In order to see the correspondence between these expressions
and the saddle-point\linebreak method, 
let us expand
(\ref{eq:s53-gnplusone}) in an infinite series as follows;
\begin{eqnarray}
iR_{n} =  {i \over 2} + \left( -i\tilde{R}_{n} \right)\ e^{2iW_{n}} +
(-i\tilde{R}_{n})^2 (-i)\ e^{4iW_{n}} + ...  \quad .
\label{eq:s53-gnrec}
\end{eqnarray}
The convergence of this series is guaranteed by the implicit factor 
$i \delta$.
The retarded resolvent is then, 
\begin{eqnarray}
&&G^R(x_{\rm i}, x_{\rm f}; E) = -|p(x_{\rm i}) p(x_{\rm f})|^{-{1/2}} \nonumber  \\
&& \times \left[ e^{-\left( \Delta_i -\Delta_f
\right)} + \left\{{i \over 2}+\left( -i\tilde{R}_{n} \right)\ e^{2iW_{n}} +
\left( -i\tilde{R}_{n} \right)^2(-i)\ e^{4iW_{n}}
+ ... \right\} e^{-(\Delta_i +\Delta_f)} \right].\nonumber\\
\   
\label{eq:s53-grexpand}
\end{eqnarray}
This is the expression to be compared to the saddle-point expression (\ref{eq:s51-green}).
If we choose $T$ to be negative imaginary, $-i\tau$,
the factor in the exponent is,
\begin{eqnarray}
i(ET+S_{\cl}) = E\tau - S_E = -\int d\tau \left( {dx\over
d\tau}\right)^2 = -\int dx |p(x)|.
\end{eqnarray}
Therefore, the two factors $\Delta_{i,f}$ is equal to the above
quantity for the path from $x_{i,f}$ to the turning point $b_n$.
The first term in (\ref{eq:s53-grexpand}) corresponds
the contribution of the pure-imaginary-time path that starts from 
$x_{\rm i}$ and reaches $x_{\rm f}$ directly.
The factor $e^{-(\Delta_i +\Delta_f)}$ in the rest of the terms
is for the path from $x_{\rm i}$ to the turning point $b_n$ and then from
$b_n$ to $x_{\rm f}$.
The expansion of $iR_n$ is understood as contributions of
the paths that oscillate in the allowed region $(a_n, b_n)$.
Various factors have unique interpretations as factors coming
from the turning points.  This is most conveniently depicted in
Fig.~\ref{fig:s53-two}.
Similarly to (\ref{eq:s53-gnrec}), the expression (\ref{eq:s53-fn}) is expanded as the following, 
\begin{eqnarray}
-i\tilde{R}_{n} = -i + (iR_{n-1})\ e^{-2 \Delta_{n-1}} 
+ (iR_{n-1})^2 ({i \over 2})\ e^{-4 \Delta_{n-1}} + ...   \quad  .
\label{eq:s53-fnrec}
\end{eqnarray}
The corresponding diagrams are illustrated in Fig.~\ref{fig:s53-five}. 

We have so far derived the expressions that results from the 
WKB approximation for the $n$-th well.
This procedure can be applied recursively for the rest of the 
wells.  At the end, we are left with $R_1$, which is given by
\begin{eqnarray}
iR_1 =  {1  \over  2} \tan W_1  
       = {i \over 2} + (-i)\ e^{-2iW_1} + (-i)^3\ e^{-4iW_1}
+ ... \ ,  
\label{eq:s53-rtildeexpand}
\end{eqnarray}
where $W_1$ is defined by (\ref{eq:s53-wn}) by replacing $n$ with $1$. Therefore we
find that $R_1$ is expressed as a sum of the contribution of the
classical paths which evolve in the first well. 

Combining (\ref{eq:s53-grexpand}), (\ref{eq:s53-fnrec}), and (\ref{eq:s53-rtildeexpand}),
we find all complex-time paths are included in the 
expression of the resolvent $G^R$.
Therefore, the resulting Green function in a $n$-well potential is given by 
\begin{eqnarray}
G^R(x_{\rm i}, x_{\rm f}; E) = - |p(x_{\rm i}) p(x_{\rm f})|^{-1/2} \sum_{x_{\cl}}\ 
f(x_{\cl}) e^{i W(x_{\rm i},x_{\rm f}; E)} .
\label{eq:s53-grn}
\end{eqnarray}
where $f(x_{\cl})$ is determined by the number of reflections and
transmissions which the classical path contains, and $W(x_{\rm i},x_{\rm f}; E)$ is
determined by what wells the path crosses and what barriers it tunnels
through. The rules for calculating $f(x_{\cl})$ is the following: When a
path has a reflection in the allowed region, it obtains $-i$. In the
case of a reflection in the forbidden region, it gets $i / 2$.

\begin{figure}[b,t]
\centerline{
\epsfxsize=12cm
\epsfbox{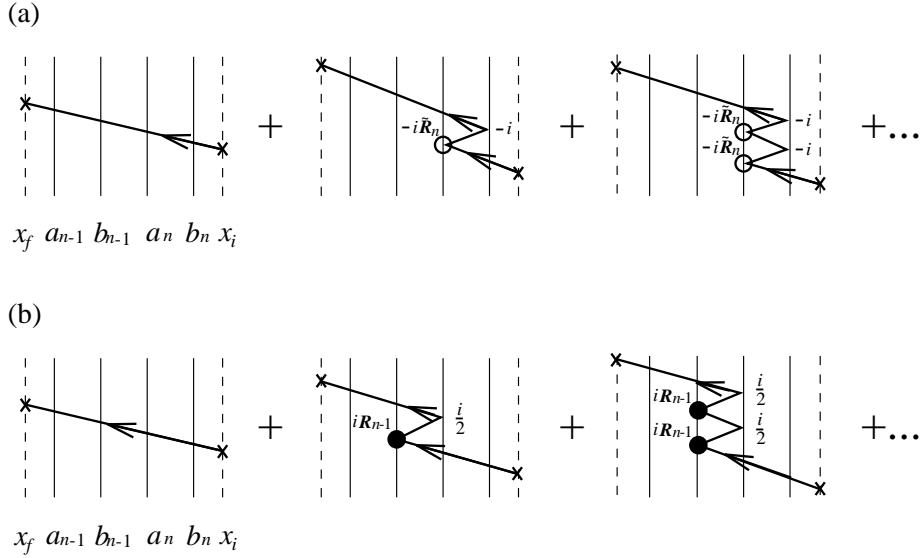}
}
\caption{Classical paths traversing in the $n$-th well 
and tunneling through the $(n-1)$-th barrier.
}
\label{fig:s53-eight}
\end{figure}

Let us next examine the case
when $x_{\rm i}$, $x_{\rm f}$ are on opposite sides of the wells. 
Applying the analysis similar to the one in the previous section ,
we find that the Green function is expressed as follows;
\begin{eqnarray}
 G^R(x_{\rm i}, x_{\rm f}; E) = -\vert p(x_{\rm i}) p(x_{\rm f}) \vert^{-{1/2}}  
T_n e^{- \Delta_i - \Delta_f}\ .
\end{eqnarray}
where $T_n$ is the (analytically-continued) transmission amplitude  
\begin{eqnarray}
 T_n = {1 \over S_{22}^{(n)}} \ .
\end{eqnarray}
Just as the previous case, we can express $T_n$ in terms of $T_{n-1}$
as in the following;
\begin{eqnarray}
 T_{n}  &=& {1 \over {S_{21}^{(n-1)}} e^{- \Delta_{n-1}} 
\sin W_n + 2 S_{22}^{(n-1)}  e^{ \Delta_{n-1}} \cos W_n} \\
 &=& {  T_{n-1} e^{- \Delta_{n-1} + i W_n} 
     \over 
       1 + {1 \over 2} R_{n-1} e^{-2 \Delta_{n-1}} + e^{2iW_n}\left(1- {1 \over
 2}R_{n-1} e^{-2 \Delta_{n-1}}\right)}  \\
 &=& { e^{- \Delta_{n-1}} \over 1 + {1 \over 2} R_{n-1} e^{-2 \Delta_{n-1}}}
   \ { e^{i W_n}  \over 1 + \tilde{R}_{n} e^{2i W_n}}
   \ T_{n-1}.
\label{eq:s53-tn}
\end{eqnarray}
In the above, we have used (\ref{eq:s52-rdef}), (\ref{eq:s53-dn}) and
(\ref{eq:s53-fn}). 
Let us show that $T_{n}$ consists of all contributions of classical
paths. The second factor in (\ref{eq:s53-tn}) is expanded in the
following way,      
\begin{eqnarray}
{e^{iW_n} \over  1+\tilde{R}_n e^{2iW_n}}=e^{iW_n} + 
(-i)(-i\tilde{R}_n)\ e^{3iW_n} + (-i)^2(-i\tilde{R}_n)^2\ e^{5iW_n}+...
\quad ,
\end{eqnarray}
which corresponds to the diagrams in Fig.~\ref{fig:s53-eight} (a).
The first factor
in (\ref{eq:s53-tn}) is expanded as,
\begin{eqnarray}
  {  e^{- \Delta_{n-1}} \over 1+{1 \over 2}R_{n-1}\ e^{-2
\Delta_{n-1}}} =
e^{- \Delta_{n-1}} + ({i \over 2})(iR_{n-1})\ e^{-3 \Delta_{n-1}} +  ({i \over
2})^2(iR_{n-1})^2\ e^{-5 \Delta_{n-1}} + ...  \ ,
\end{eqnarray}
which is illustrated in Fig.~\ref{fig:s53-eight} (b).
As the previous case, repeating this procedure, we reach $T_1$, which
is given by
\begin{eqnarray}
T_1 &=&{1 \over 2 \cos W_1}  \\
               &=&e^{iW_1} + (-i)^2\ e^{3iW_1} + (-i)^4\ e^{5iW_1}+... \quad .
\end{eqnarray}
Therefore we conclude that $T_n$ consists of all contributions of
classical paths. 

The above analysis can be applied to other cases with various
locations of $x_{\rm i}$ and $x_{\rm f}$. 
Thus we verify that thus constructed resolvent can be always
interpreted as sum over the complex-time classical paths.
 
\begin{figure}[b,t]
\centerline{
\epsfxsize=10cm
\epsfbox{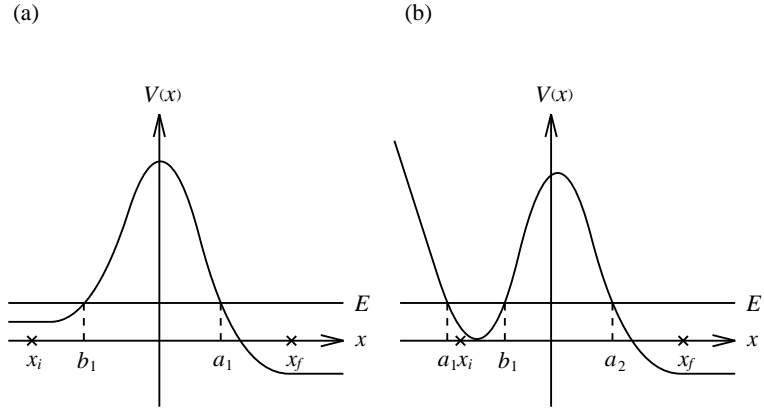}
}
\caption{A scattering process against a
potential barrier and a metastable system.}
\label{fig:s53-ten}
\end{figure}

This formalism is valid for a scattering process and metastable system 
(Fig.~\ref{fig:s53-ten})
as well as for a stationary system. It is apparent from
this derivation that this complex-time method reproduces the result
of the WKB approximation with the linear connection formula.    
The potential we consider is given in Fig.~\ref{fig:s53-ten} (b).
The turning points are labeled in the figure by $a_1$, $b_1$, and $a_2$.
First we consider the case when $x_{\rm i}$ is in the well and $x_{\rm f}$ is 
outside of the barrier. 
The analysis similar to the above yields the Green function as in the 
following;
\begin{eqnarray}
G^R(x_{\rm i}, x_{\rm f}; E) = -i |p(x_{\rm i}) p(x_{\rm f})|^{-1/2}  
 { \left( e^{-iW_i} -ie^{iW_i} \right)\ e^{iW_1- \Delta_1 +iW_f}
\over (1+\tilde{R} e^{2iW_1})(1+{1 \over 4} e^{-2 \Delta_1})},
\label{eq:s53-metaone}
\end{eqnarray}  
where $\tilde{R}$ is given by 
\begin{eqnarray}
-i\tilde{R}
 &=& -i\ {1-{1 \over 4}e^{-2\Delta_1} \over 1+{1 \over
4}e^{-2\Delta_1}} \\
 &=& -i + {i \over 2} e^{-2\Delta_1} +({i \over 2})^3 e^{-4\Delta_1}
+... \ .
\label{eq:s53-metatwo}
\end{eqnarray}
From (\ref{eq:s53-metaone}) and (\ref{eq:s53-metatwo}), we find again
that the resolvent is 
equal to the sum over the physical saddle-points.
We find that the poles of the Green function
are determined by the following;
\begin{eqnarray}
1+ e^{2iW_1} + {1 \over 4} e^{-2\Delta_1} \left( 1-e^{2iW_1} \right)=0.
\end{eqnarray}
Let us solve this equation iteratively. Then we obtain 
\begin{eqnarray}
W_1(E) = \left( n+{1 \over 2} \right) \pi -{i \over 4} e^{-2\Delta_1}.
\label{eq:s53-metathree}
\end{eqnarray}
From  (\ref{eq:s53-metathree}), we find an imaginary part of the
energy eigenvalue  
\begin{eqnarray}
{\rm Im}\ E_n =  - {i \over 2T(E_n)} e^{-2\Delta_1},
\label{eq:s53-imeexpress}
\end{eqnarray}
where $T(E_n)$ is the period of the classical path between the
turning points $a_1$, $b_1$
\begin{eqnarray}
T(E) = 2 \int_{a_1}^{b_1} {dx  \over  \sqrt {2 \left( E-V(x) \right)}}.
\end{eqnarray}
We find the decay rate of this metastable system.
We note that the factor $1/2$ in  (\ref{eq:s53-imeexpress}) comes from the
weight $1/2^n$ of the saddle-points with $n$ reflections in the forbidden
region.

Let us apply this method to a scattering process against a potential
barrier in Fig.~\ref{fig:s53-ten} (a). When $x_{\rm i}$ and $x_{\rm f}$ are separated by the
barrier, the Green function is given by 
\begin{eqnarray}
G^R(x_{\rm i}, x_{\rm f}; E) 
 &=&-i |p(x_{\rm i}) p(x_{\rm f})|^{-1/2} \tilde{T}e^{i(W_i+W_f)}  \nonumber\\
&=& -i |p(x_{\rm i}) p(x_{\rm f})|^{-1/2}
\sum^{\infty}_{k=1} e^{i(W_i+W_f)-(2k-1)\Delta_1} \left({i \over
2}\right)^{2(k-1)}_,
\label{eq:s53-transmission}
\end{eqnarray}  
where we have used the WKB expression for the transmission
coefficient $\tilde{T}$;
\begin{eqnarray}
 \tilde{T} =\  {e^{-\Delta_1} \over 1+{1 \over 4}e^{-2\Delta_1}}.
\end{eqnarray}
When $x_{\rm i}$ and $x_{\rm f}$ are on the same side, the Green function is given
by
\begin{eqnarray}
&&G^R(x_{\rm i}, x_{\rm f}; E) 
= -i |p(x_{\rm i}) p(x_{\rm f})|^{-1/2}
e^{iW_i}\left[e^{-iW_f}-i \tilde{R} e^{iW_f}\right] \nonumber\\
&&= -i |p(x_{\rm i}) p(x_{\rm f})|^{-1/2}
e^{iW_i}\left[e^{-iW_f}+{i \over 2}e^{iW_f}+\sum^{\infty}_{k=1}
e^{iW_f-2k \Delta_1} (-i)^{2k-1}\right]_,   
\label{eq:s53-reflection}
\end{eqnarray}  
where we have used the expression for the reflection amplitude,
\begin{eqnarray}
-i \tilde{R}=-i \ {1-{1 \over 4}e^{-2\Delta_1} \over 1+{1
\over 4}e^{-2\Delta_1}}.
\end{eqnarray}
The coefficients $\tilde{T}$ and $\tilde{R}$ satisfy the unitarity condition 
\begin{eqnarray}
|\tilde{T}|^2+|\tilde{R}|^2=1. 
\label{eq:s53-trunitary}
\end{eqnarray}
We again confirm the validity of the sum over the physical
saddle-points.
The weight $1/2^n$ is crucial for (\ref{eq:s53-trunitary}).

We have given the re-formulation of the complex-time method. In
quantum mechanics, using the connection formula for the wavefunctions, 
we constructed the
reduction formula in the number of the turning points 
for the Green function.
This yields series expansions, which can be understood as sum over 
in the classical complex-time trajectories.
This is understood as a sum over the
physical saddle-points with specific weights and phases in the path-integral
method.
This shows the validity of the method proposed before, in the 
context of the path-integral method.
We confirmed that this method yields results identical with
that of the WKB approximation.
Thus our construction gives solid basis for the starting point
of the complex-time method.

We must elaborate this complex-time
formalism so that it could be applied to field theory and 
two-particle scattering
process. Thus somewhat different analysis may be required.
One of them could be the combination of the
valley method and the current complex-time method.
The valley method can be used to identify the 
imaginary-time tunneling paths, which are
converted to collective coordinates.
Thus, while the incoming particle are expressed in
real-time expressions, the tunneling part
may be obtained as the imaginary part of the
complex-time development along the valley trajectory.
There are rather interesting possibilities along this line,
which should be further investigated.

\section{Asymptotic analysis of the perturbation theory}
The tunneling phenomenon has some subtle relation to
the large order behavior of the perturbative series of the theory.
Dyson argued\cite{rf:dyson} that the perturbative series in QED 
is an asymptotic series.
Later, more accurate estimates of the large order behavior
of the perturbative series were carried out in many
models.\cite{rf:large}
Although asymptotic series are divergent, it allows 
summation techniques, such as Borel summation and
Pad\'e approximation.
In theories with tunneling phenomena, the perturbative series,
however, are generally not Borel-summable.
In such a case, the Borel integral encounters singularities, which are
related to the existence of the nonperturbative tunneling
phenomena.  Therefore, only after the perturbative and nonperturbative
effects are combined the physical quantities are well-defined.

The consideration of the valley, described in detail in this
paper, provides an insight into this interplay between the 
perturbative and non-perturbative contributions to path-integral:
In the context of the the high energy behavior of the B and L 
violating amplitude considered in section 2,
we have noted that the contribution of the configuration which has
an instanton and an anti-instanton in small separation 
compared with their size become important.
Since this configuration merges into the vacuum continuously, 
it implies that
the perturbative and non-perturbative contributions transmute
into each other in this configuration. 
This situation casts a doubt against our intuitive understanding 
that there must be a clear separation between  the two. 
For the full understanding of the tunneling phenomena, 
we consider it necessary to understand this situation.
Approaches in this section provide some possible direction
for progress.
In the first subsection, we show that the valley calculation provides 
an easy way to evaluate the  large order behavior of the 
perturbative expansion in a model in quantum mechanics.\cite{rf:kiku}
In the second subsection, the asympton theory, which is 
an analysis of the perturbative functional, not just the coefficients, 
is presented.

\subsection{Borel function in the valley method} 
We consider the symmetric double-well potential
\be V(\phi; g) = {1\over 2}\phi^2(1 - g \phi )^2\ee
and evaluate the transition amplitude
\(\langle 0| e^{- HT} | 0\rangle \)
by the path-integral Eq.~(\ref{eq:s41pathintegral}).
We are interested in the perturbative expansion of 
\be 
Z(g^2) \equiv {\langle 0| e^{- HT} | 0\rangle \over 
\langle 0| e^{- H_0T } | 0 \rangle}, \label{eq:s2Z}\ee
(\(H_0 \equiv (1/ 2)[ p^2 + \phi^2 ]\)
the free part of the Hamiltonian)
or the  energy shift \(E(g^2)\) of the ground 
state due to the interaction, which is evaluated by 
\be E(g^2) = - \lim_{T\rightarrow \infty} {1\over T} \ln Z(g^2). 
\ee
The time \(T\) is taken sufficiently large, compared with
the inverse of the excitation energies.

The perturbative series of \(Z(g^2)\) in the present model
is a power series in \(g^2\)
\be Z_{\rm pert}(g^2) = 1 + \sum_{n=1}^\infty Z_n g^{2n}\label{eq:s2pZ}. \ee
The coefficient \(Z_n\) are evaluated by 
expanding the action in powers of \(g\),
\be e^{ - S[\phi; g] } = \sum_n g^n F_n[\phi], \quad
S[\phi; g] \equiv \int_0^T d\tau 
\left[ {1\over 2} \left(\dot \phi \right)^2 + V(\phi; g) \right] 
\label{eq:expac}\ee
and path-integrating each term,
\be Z_n  = { \int {\cal D}\phi
F_{2n}[\phi]\over  \int {\cal D}\phi e^{ - S_0[\phi] }},\ee
in the background \(\phi_{\rm B} \equiv 0\),
where \(S_0\) is the action for the harmonic oscillator
\be S_0 = \int_0^T d\tau {1\over 2} 
\left[ \left(\dot \phi \right)^2 + \phi^2 \right]. 
\ee
Note that the functional \(F_{n}[\phi]\) 
with odd \(n\) has odd powers of \(\phi\)
and gives no contribution.
Although \(Z_n\) are well-defined objects, 
the infinite series (\ref{eq:s2pZ}) does not  
converge;\cite{rf:asym,rf:brezin,rf:dyson,rf:benwu,rf:nonB,rf:tHftB}  
it is known to be asymptotic to  \(Z(g^2)\).

The Borel summation method is a powerful one to analyze 
the asymptotic series Eq.~(\ref{eq:s2pZ}).
To apply the method, we first calculate
 the Borel function \(B(t)\) defined by a series
\be B(t) = \sum_n {Z_n\over n!} t^n \label{eq:s2B}.\ee
\(Z(g^2)\) is recovered, at least formally,  by the integral
\be Z(g^2) = \int_0^\infty dt\, e^{-t} B( t g^2 ) 
= {1\over g^2}\int_0^\infty dt\, e^{-t/g^2} B(t)
\label{eq:s2ZinB}\ee
since 
\be \int_0^\infty dt\, e^{-t} t^n = n!. \ee
Note the series (\ref{eq:s2B}) has wider region of convergence
in the complex \(t\)-plane than the series (\ref{eq:s2pZ}) has
in \(g^2\)-plane.
This may give a chance to get a meaningful answer even from the 
non-convergent series. 

We intend to calculate the Borel function 
of the present model not
by the series expansion (\ref{eq:s2B}), but in a rather direct way
using the valley method.
It is well-known that the present model possesses the instanton (I)
and anti-instanton (\= I), the solutions of the Euclidean equation
of motion that travel between \(\phi=0\) and \(\phi= 1/g\).
Applying the proper valley equation, we have further found there
exist the valley trajectory that starts as a small fluctuation 
at the vacuum and reaches to the I--\=I pair at the
end.\cite{rf:pvalley}
We first approximate \(Z(g^2)\) to an integral over
the valley trajectory \(\phi_V(\alpha) \), where \(\alpha\) can
be any variable that parameterize the valley.
In another word, we adopt a line \(\phi_V(\alpha) \) 
in the functional
space as the background for calculating \(Z(g^2)\),
instead of using  a single point \(\phi_{\rm B}\) as in the
perturbative calculation.
The explicit expression as the collective coordinate integral is
obtained by the Faddeev-Popov (FP) method as explained in section 3.
To avoid the cumbersome \(g\) dependence in the background 
valley \(\phi_V(\alpha) \), 
we rescale \(S\) and \(\phi\) as
\bea \phi &\rightarrow& g\phi \\
      S[\phi; g] & \rightarrow & g^2 S[\phi]. \eea
With this transformation, \(g\) drops out from the action and
\(\phi_V(\alpha)\) becomes \(g\) invariant.
We now obtain
\be Z(g^2) = \int d\alpha\, e^{- S[\phi_V(\alpha)] / g^2 }
\left| {d S[\phi_V(\alpha)] \over d\alpha}\right| 
\left|{\delta S\over \delta \phi_V(\alpha)}\right|^{-1} {\cal Z},
\label{eq:s2Zvalley}\ee
where  
\bea {\cal Z} 
&=& {{\cal N}\over (\sqrt{2\pi}g)\langle 0| e^{- H_0T } | 0 \rangle } 
\int [d\varphi]'
\exp\left( - {1\over 2}(\varphi\cdot D(\alpha) \cdot \varphi)
- S'_{\rm int}[\varphi;g] \right) \nonumber\\
&&\times \left[ 1 - { g ( \varphi \cdot D(\alpha) \cdot 
d\phi_V(\alpha)/d\alpha)
\over d S[\phi_V(\alpha) ] / d\alpha}\right].
\label{eq:s2calZ}
\eea
In Eq.~(\ref{eq:s2calZ}), 
\(\varphi \equiv g (\phi - \phi_V(\alpha))\) represents
the fluctuations around \(\phi_V\),  \([d\varphi]'\)
is the functional measure except
the mode parallel to \(\delta S/\delta \phi\) that was
removed by the FP method,
\(1/ \sqrt{2\pi}g \) comes from this elimination,
\(D(\alpha)\) represents \(\delta^2 S/\delta \phi_V^2(\alpha)\) and
the abbreviation like \(\varphi\cdot D(\alpha) \cdot \varphi\)
stands for
\[
\int_0^T d\tau d\tau' {\delta^2 S\over \delta\phi_V(\alpha)(\tau)
\delta\phi_V(\alpha)(\tau')} \varphi(\tau)\varphi(\tau'),
\]
\(S'_{\rm int}\) is the remaining action after
the value at \(\phi_V\) and the quadratic 
term  \(\varphi D \varphi\) are subtracted.

The trick of extracting the Borel function from 
Eq.~(\ref{eq:s2Zvalley}) is 
to choose the  parameter \(\alpha\) as
the value \(t\) of the action of the valley,
so that \(t = S[\phi_V(t)]\).
Then Eq.~(\ref{eq:s2Zvalley}) is written as
\be Z(g^2) = \int_0^{S_{\rm cl}} dt\, e^{ - t/g^2} 
\left|{\delta S\over \delta \phi_V(t)}\right|^{-1}{\cal Z}, 
\label{eq:s2ZVB}\ee
where \(S_{\rm cl} = 1/3 \) is the twice of the instanton action
which the valley action cannot exceed.
Note the resemblance of this expression to Eq.~(\ref{eq:s2ZinB}).
The integrand 
\be B_V(t) \equiv 
\left|{\delta S\over \delta \phi_V(t)}\right|^{-1}{\cal Z} \ee
is the approximation for the Borel function in terms of the valley
method.

Although \(B_V\) is an approximation,
it suffices to predict the large order behavior of 
the perturbative expansion.
The bottom line is that the large order behavior is controlled
by singularities of the Borel function in the complex
\(t\)-plane and that the valley approximation
\(B_V\) also reproduces them.
The valley approaches asymptotically to the well-separated
I--\=I pair \(\phi_{\rm I\bar I}(R)\) with \(R\) their
separation. The action \(t\)  behaves 
\be t \simeq S_{\rm cl} - \mbox{const} \times e^{-R} 
\label{eq:s2t}\ee
at \(\phi_{\rm I\bar I}\).
There exists the normalized quasi-zero mode \(\phi_{\rm R}\)
that causes variation of \(R\), i.e. 
\(\phi_{\rm R} = (1/\eta)(d\phi_{\rm I\bar I}/dR)\), where \(\eta\)
is the Jacobian between R and \(\phi_{\rm R}\).
The functional gradient  is proportional  to
the quasi-zero mode according to the definition of
our valley equation, 
\(\phi_{\rm R} =  (\delta S / \delta \phi_V) /
|\delta S / \delta \phi_V |\). 
Since \( (\delta S / \delta \phi_V)\cdot(d\phi_{\rm I\bar I}/dR) 
= dt/dR\), 
we can  conclude
\be \left|{\delta S \over \delta \phi_V(t)}\right|^{-1} \simeq 
{1\over \sqrt{12}
(S_{\rm cl} - t ) }, \ee
where we have used \( \eta = \sqrt{1/12}\).
The calculation of  \({\cal Z}\) is also simple at 
the well-separated I--\=I pair; to the leading order in
\(g^2\) it is approximated to
the product of the square of the single instanton
determinant and the Jacobian for the overall translation,
\be {\cal Z} = {1\over 2\pi g^2 }{12 \over \sqrt 3} T. \ee
(The calculations of the determinant and  the Jacobian are
found in Ref.~\citen{rf:clec}.)
Combining these,
we obtain
\be B_V(t) \simeq {T\over \pi g^2}{ 1 \over S_{\rm cl} - t}. \ee
This simple pole controls the large order behavior of 
the perturbative expansion of \(Z(g^2)\).
In fact, inserting its expansion 
\( \sim (T/\pi g^2) \sum_n (S_{\rm cl})^{-n-1} t^n \)
into (\ref{eq:s2ZVB}) and integrating it over
to infinity (neglecting the  upper 
bound), we obtain 
\be Z_n \simeq {T\over \pi} {S_{\rm cl}}^{-n - 1} n!,\ee
or, for coefficients of perturbative expansion of \( E(g^2)\)
in powers of \(g^2\),
\be E_n \simeq -{1\over \pi} {S_{\rm cl}}^{-n - 1} n! \ee
which coincides with the known results.\cite{rf:nonB}

The singularity of \(B_V(t)\) at \(t = S_{\rm cl}\) indicates 
the non Borel-summability of the expansion.
There will be an ambiguity depending on how we turn
aside the pole when we perform a contour integral
corresponding to Eq.~(\ref{eq:s2ZinB})
if we only know \(B(t)\) from the perturbative expansion.
The advantage of using the valley method is that we now know
the ambiguity is, at least in the present model,
an artifact: the integral has the upper limit at \(t = S_{\rm cl}\).
This understanding may provide us 
to construct a specific prescription to get unambiguous predictions
from the perturbative expansion
by properly taking into account the non-perturbative 
contributions in a class of models which is known to
be non Borel-summable.\footnote{See Ref.~\citen{rf:kiku} 
for the detailed discussion on this point.}

\subsection{Asymptons} 
In this subsection, we look into the behavior of the 
perturbative functional , $F_n[\phi] $ in Eq.~(\ref{eq:expac}),
in the functional space of path-integral
and discuss it is dominated by a series of configurations,
which we name ``asymptons''. 
We will show their contribution indeed reproduces the large order 
behavior of the perturbative series, which shows the fictitious 
nature of the non Borel-summability and provides us with
the insight into the interplay between the perturbative and
the non-perturbative effects.

In the first part of this subsection we present a general formalism
for the analysis of the asymptons. Then a simple
double-well quantum mechanical model are treated in the next part.
The SU(2) gauge field theory is studied in the third part.
Then the analysis of the asympton on the valley line
is carried out.  This subsection ends with the
discussion on the consequences of the asymptons and on the summing methods.

\subsubsection{General formalism for the asympton analysis}
Let us use \(z_n\) for  each  coefficient in the perturbative expansion
of a given model (not necessarily the symmetric double-well as in the
previous section),
\begin{equation}
 z_n  = { \int {\cal D}\phi
F_{n}[\phi]\over  \int {\cal D}\phi e^{ - S_0[\phi] }}.
\end{equation}
We note that the coefficients \(z_n\) are known to grow
as a factorial of $n$ for large
$n$ in a wide range of models while 
the expansion (\ref{eq:expac}) is convergent
for a given configuration \(\phi\) as long as
the integrals \( \int d\tau (\phi)^m \) are convergent.
Our aim here is to understand how this can happen.
For this purpose we study the behavior of 
the functional $F_n[\phi]$ itself.
The asympton theory answers this question.\cite{rf:AA,rf:AT} 

The expression for \(F_n[\phi]\) is complicated in general. 
In the limit $n \rightarrow \infty$, however, a
compact expression can be obtained.
The functional $F_{n}[\phi]$ is expressed 
as a contour integral in the complex $g$-plane as follows;
\begin{equation}
 F_{n}[\phi]=\frac{1}{2\pi i} \oint \frac{dg}{g^{n+1}} 
e^{-S[\phi,g]}=\frac{1}{2\pi i} \oint \frac{dg}{g}
e^{-\tilde{S}[\phi,g]},
\label{eq6-2:a}
\end{equation}
where 
\begin{eqnarray}
\tilde{S}[\phi,g]=S[\phi,g]+n \log g.
\end{eqnarray}
We apply the saddle-point approximation to the $g$-integral (\ref{eq6-2:a})
valid for large $n$.

In this subsection we deal with theories whose
action is a second order polynomial of $g$,
which we express as 
$S[\phi,g]=c_{0}[\phi]-gc_{1}[\phi]+g^{2}c_{2}[\phi]$.
The saddle points of $\tilde{S}[\phi,g]$ are given by the
following;
\begin{eqnarray}
g_{\pm}=\frac{c_{1}}{4c_{2}}(1\pm\sqrt{D}),
\hspace*{1cm}
D\equiv 1-\frac{8nc_{2}}{c_{1}^{2}}.
\end{eqnarray}
In order to examine the behavior of the perturbative functional,
we divide the functional space of $\phi$ into two regions;
the one with $D>0$ and the one of $D<0$, for the
the expressions of  $F_{n}[\phi]$ obtained in the
saddle-point approximation are different depending on the
sign of $D$.  As was found in Ref.~\citen{rf:AA,rf:AT},
the region of negative $D$ is of importance.
We will concentrate on this case hereafter.

When $D<0$ the saddle points are complex, which we denote 
by $g_\pm$ as follows;
\begin{eqnarray}
g_{\pm}=|g|e^{\pm i\theta},
\hspace*{5mm}
|g|=\sqrt {\frac{n}{2c_{2}}},
\hspace*{5mm}
\cos \theta=\frac{c_{1}}
{\sqrt{8nc_{2}}},
\label{eq:6-22}
\end{eqnarray}
where $\theta\in [0,\pi/2]$. 
We obtain the following expressions at the saddle points $g_{\pm}$;
\begin{eqnarray}
S_{\pm} &=& c_{0}-n-\frac{n}{2}e^{\pm 2i\theta},\label{saddle action}\\
\tilde{S}''_{\pm} &=& \frac{2n}{|g|^{2}}\sin \theta(\sin\theta\pm i\cos\theta).
\end{eqnarray}
Since $\rm{Re}(\tilde{S}^{''}_{\pm})>0$ at both saddle points, 
we choose the integration contour 
to go through both saddle point along the real axes. 
The contour integration yields the following expression;
\begin{eqnarray}
F_{n}[\phi]=\frac{2}{ \sqrt{2 \pi}} 
\mbox{Im}\left[ \frac{1}{\sqrt{\tilde{S}''_- g_{-}^{2}}} 
\frac{\mbox{exp}(-S[\phi,g_{-}])}{g_{-}^n}\right].\label{eq:6-21}
\end{eqnarray}
We stress here that $g_-$ is a functional of $\phi$
and a function of $n$, defined by \eqa{6-22}.

Let us look for peaks of $F_{n}[\phi]$ in order to understand
its the large order behavior. 
Therefore we must solve the following equation: 
\begin{eqnarray}
\frac{\delta F_{n}[\phi]}{\delta \phi}
=\left.\frac{\partial F_{n}[\phi]}{\partial g_{-}}\right|_{\phi}
\frac{\delta g_{-}}{\delta \phi}
+\left.\frac{\delta F_{n}[\phi]}{\delta \phi}\right|_{g_{-}}=0.
\label{eq:6-27}
\end{eqnarray}
At the leading order of $\hbar$ (which we have neglected to write), 
the $g_{-}$ derivative in the first term acts on $\tilde S$ and is zero 
due to the saddle-point condition.
At the same order, we get the following expression by using (\ref{eq:6-21});
\begin{eqnarray}
\mbox{Im}\left [ \frac{\delta S[\phi,g_{-}]}{\delta \phi} 
e^{i\sigma}\right ]=0,
\label{eq:6-23}
\end{eqnarray}
where
\begin{eqnarray}
\sigma&=&\mbox{arg}\left[ \frac{1}{\sqrt{\tilde{S_{-}^{''}}g_{-}^{2}}} 
\frac{e^{-S[\phi,g_{-}]}}{g_{-}^{n}}\right]\nonumber\\
&=&n(\theta-\frac{1}{2} \sin 2\theta)+\frac{\theta}{2}+\frac{\pi}{4}.
\label{eq:6-26}
\end{eqnarray}
This expression (\ref{eq:6-23}) is similar to the ordinary equation of motion
except for the fact that the ``coupling constant'' $g_-$ and 
the phase $\sigma$ are functionals of $\phi$. 
The solutions can be identified by first 
solving the equation of motion (\ref{eq:6-23}) for arbitrary constant
$g_-$ and $\sigma$ and then solving the self-consistent equation (\ref{eq:6-22}) 
for $|g|$ and $\theta$.

\subsubsection{Asympton for the double-well quantum mechanics}
We first consider the quantum mechanics of the
double-well potential, whose action and perturbative functionals
are given by Eqs.~(\ref{eq:expac}) and (6.1).
Then the equation (\ref{eq:6-23}), which determine the peaks of $F_{n}[\phi]$, 
can be written as follows by using the scaled variable $\varphi\equiv |g|\phi$;
\begin{eqnarray}
-\frac{d ^{2}\varphi}{d \tau^{2}}
+\frac{\partial V_{\rm{eff}}(\varphi)}{\partial \varphi}=0.
\label{eq:6-2b}
\end{eqnarray}
The ``effective potential'' $V_{\rm{eff}}(\varphi)$ 
in the above is defined by the following;
\begin{eqnarray}
V_{\rm{eff}}(\varphi)=
\frac{1}{2}\varphi^{2}-\kappa_{3}\varphi^{3}+\frac{1}{2}\kappa_{4}\varphi^{4},
\end{eqnarray}
where 
\begin{eqnarray}
\kappa_{3}=\frac{\sin(\sigma-\theta)}{\sin\sigma},\label{eq:6-25a}\\
\kappa_{4}=\frac{\sin(\sigma-2\theta)}{\sin\sigma}.\label{eq:6-25b}
\end{eqnarray}
play the role of the ``coupling constants''.
The self-consistent equation (\ref{eq:6-22}) is now rewritten by
using the scaled variable as follows;
\begin{eqnarray}
|g|=\sqrt{\frac{R_{4}}{n}},\label{eq:6-211a}\\ 
\cos\theta=\frac{R_{3}}{2 R_{4}},\label{eq:6-211b}
\end{eqnarray}
where
\begin{eqnarray}
R_{m}=\int d\tau \varphi^{m}.
\label{eq:rmdef}
\end{eqnarray}
We note that the effective equation of motion (\ref{eq:6-2b}) 
is independent of $|g|$; it has only two free parameters, $n$ and $\theta$
(or alternatively $\kappa_3$ and $\kappa_4$).
Consequently, the only self-consistent equation we need to solve is 
Eq.~({\ref{eq:6-211b}), which determines the value of $\theta$
for a given $n$; 
Eq.~(\ref{eq:6-211a}) simply determines
the value of $|g|$, which will be used for the evaluation of 
$F_n[\phi]$.

The effective equation (\ref{eq:6-2b}) possess a bounce-like solution;
\begin{eqnarray}
\varphi=\frac{1}{\kappa_{3}+\sqrt{\kappa_{3}^{2}-\kappa_{4}}\cosh \tau}.
\label{eq:6-24}
\end{eqnarray}
where we have chosen the peak position to be at $\tau=0$.
This solution is apparently meaningful only for $\kappa_3^2 > \kappa_4$,
which is a valid assumption as we will see later.
Next, we must solve the self-consistent equation (\ref{eq:6-211b}) 
to determine the value of $\theta$.
Substituting the solution (\ref{eq:6-24}) into Eq.~({\ref{eq:rmdef}), 
we obtain the following equations for $R_{3,4}$; 
\begin{eqnarray}
R_{3}&=&-\frac{3\kappa_{3}}{\kappa^{2}_{4}}+
\frac{3\kappa_{3}^{2}-\kappa_{4}}{\kappa^{5/2}_4}
\mbox{arctanh}\frac{\kappa_{3}}{\sqrt{\kappa_{4}}},\\
R_{4}&=&\frac{-15\kappa_{3}^{2}+4\kappa_{4}}{3\kappa_{4}^{3}}
+\frac{\kappa_3(5\kappa_3^2-3\kappa_4)}{\kappa_4^{7/2}}
\mbox{arctanh}\frac{\kappa_{3}}{\sqrt{\kappa_{4}}}.
\end{eqnarray}
In order to find solutions, 
we find it convenient to choose $\theta$ and $\kappa_{4}$.
Using the relation $2\kappa_{3}\cos \theta=\kappa_{4}+1$,
which can be derived from Eqs.~(\ref{eq:6-25a}) and (\ref{eq:6-25b}),
we express the above functions $R_{3,4}$ as functions of 
the two parameters, $\theta$ and $\kappa_{4}$.
And then we solve two equations (\ref{eq:6-25b}) and (\ref{eq:6-211b})
for $\theta$ and $\kappa_{4}$.
(Because Eqs.~(\ref{eq:6-25b}) and (\ref{eq:6-211b}) 
are invariant under the reflection
$\theta\to\pi-\theta$ and $\phi\to-\phi$, we investigate only 
the region where $\theta\in[0,\pi/2]$.)
In Fig.~\ref{c62-5}, we plot 
the numerical solutions of these two equations for $n=15$;
the solid line satisfy Eq.~(\ref{eq:6-211b}), 
and the dashed line Eq.~(\ref{eq:6-25b}). 
The solutions of (\ref{eq:6-27}) are given by the cross
points of these lines. 
The number of the solutions are obtained from the following 
consideration:
The parameter $\sigma$ varies from $\pi/4$ to $(n+3/4)\pi$ 
when $\theta$ varies from $0$ to $\pi$.
Since 
$\kappa_{4}=\cos 2\theta-\sin 2\theta \cot\sigma$, 
there is always one solution for every region 
width $\pi$ in $\sigma$.
Therefore the number of solutions is either $n$ or $n+1$.
All these solutions are the configurations 
that are stationary points of the perturbative functional
$F_n[\phi]$, which we have named the ``asymptons''.

\begin{wrapfigure}{c}{7cm}
\epsfxsize=7cm
\epsfbox{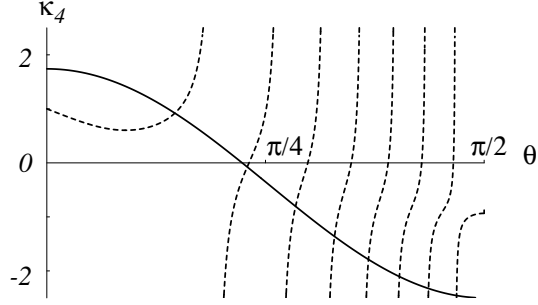}
\caption{The solid line satisfy Eq.~(\protect \ref{eq:6-211b}), 
and the dashed line Eq.~(\protect \ref{eq:6-25b})($n$=15).}
\label{c62-5}
\end{wrapfigure} 

In order to see which asympton dominates the path-integral,
we have evaluated the value of $F_{n}[\phi]$ at an asymptons, 
$\phi_{+}(\theta_{asym}\in [0,\pi/2]))$, and its partner (``anti-asympton'') 
$\phi_{-}\equiv -\phi_{+} (\pi-\theta_{asym}\in [\pi/2,\pi])$. 
The result is the following;
\begin{eqnarray}
F_{n}[\phi_{+}]&=&
\frac{1}{\sqrt{\pi n \sin\theta}}
\left ( \frac{n}{R_{4}}\mbox{exp}(-\kappa_{4}-2\sin^{2}\theta)\right)^{n/2}
\sin\sigma,\label{eq:6-28}\\
F_{n}[\phi_{-}]&=&(-)^{n}F_{n}[\phi_{+}].
\label{eq:fpmrel}
\end{eqnarray}
Most notable is the factor $n^{n/2}$ in $F_{n}[\phi]$.
This agrees with the leading term of the known large order behavior of the 
coefficients, \(z_n \sim n^{n/2}\).
In this sense the existence of the asympton explains
the non-Borel summability in the perturbation series. 
We note that Eq.~(\ref{eq:fpmrel}) shows that for odd $n$ 
the contribution of the asympton and the anti-asympton cancel.
This is in agreement with the fact that  $z_n=0$ for odd $n$.

The expression Eq.~(\ref{eq:6-28}) shows that 
the leading asympton is the one with
the largest $e^{-\kappa_{4}-2\sin^{2}\theta}/R_{4}$ 
($\equiv f(\kappa_{4},\theta)$).
Fig.~\ref{c62-3} shows the behavior of $f(\kappa_{4},\theta)$ 
in the $(\kappa_{4},\theta)$ space. 
The solid line satisfy Eq.~(\ref{eq:6-211b}). 
On this line, the values of $f(\kappa_{4},\theta)$ is represented 
by Fig.~\ref{c62-x}.
The maximum point is numerically $\theta\sim 0.6 (\kappa_{4}\sim 0.3)$.
The asympton, of which the value of $\theta$ 
is most close to that of the maximum point, is the leading asympton.

\begin{figure}[htb]
\parbox{\halftext}{
\centerline{\epsfxsize=7cm\epsfbox{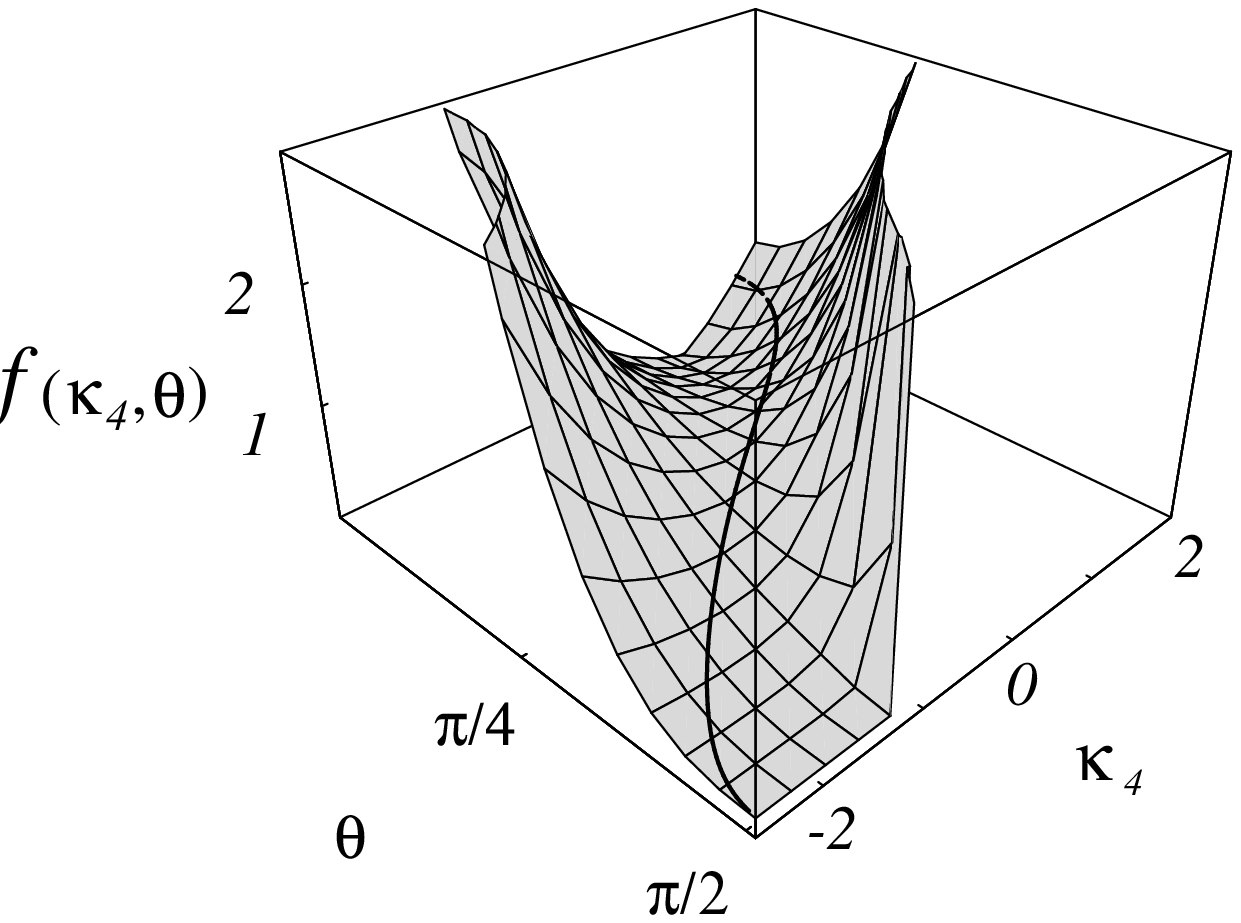}}
\caption{The function \protect $f(\kappa_{4},\theta)$ 
in the $(\kappa_{4},\theta)$ space.}
\label{c62-3}}
\hspace{4mm}
\parbox{\halftext}{
\centerline{\epsfxsize=7cm \epsfbox{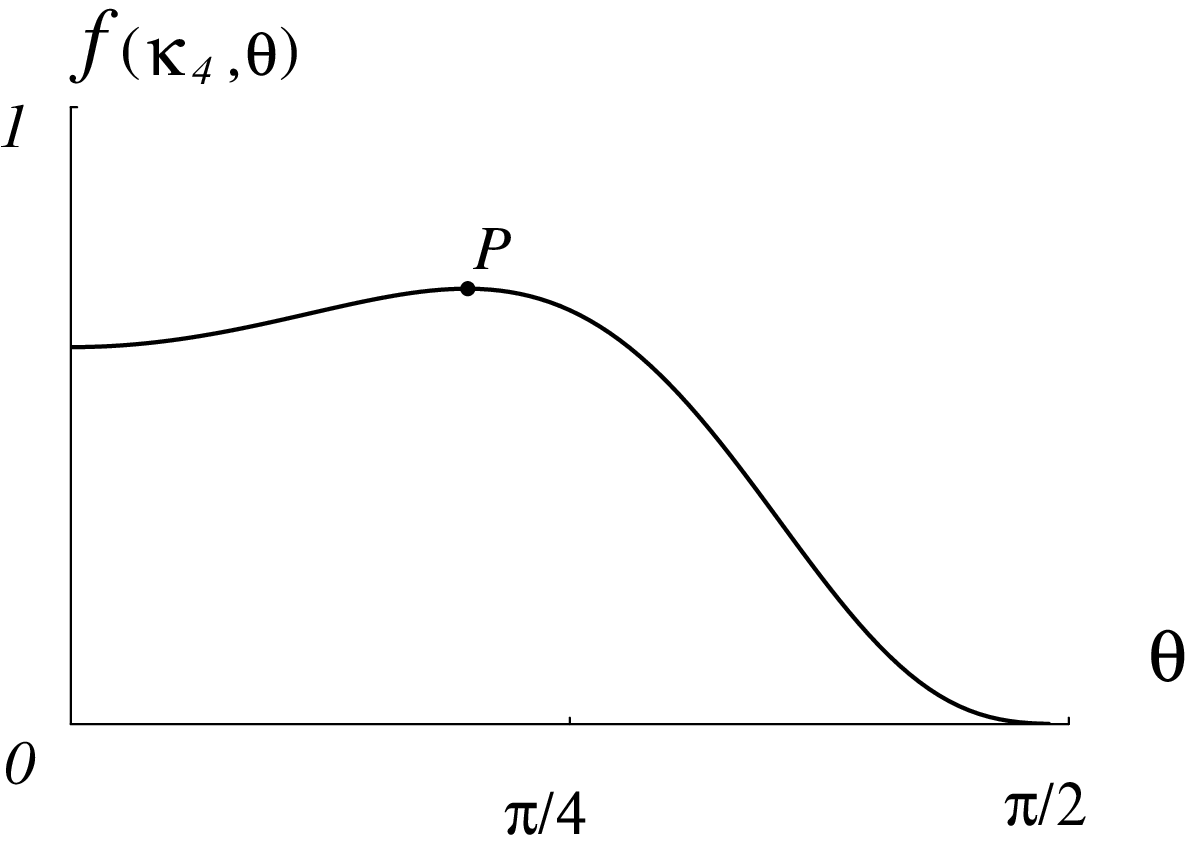}}
\caption{The values of \protect $f(\kappa_{4},\theta)$ 
on the line that satisfies Eq.~(\protect \ref{eq:6-211b})}
\label{c62-x}}
\end{figure}

\subsubsection{Asympton in the  SU(2) gauge field theory}
The analysis we have done for the quantum mechanics can be straightforwardly 
extended to the SU(2) gauge field theory.
This may provide the way to  predict the large order behavior of the 
model.\footnote{%
Actually, there is another cause of the non-summability in field
theories, called renormalons.  It is most plausible that renormalons
are not isolated configurations, but a sum of contribution from the
path-integration over a wide region of the functional space, unlike 
the asymptons. This is still an open question.}
The action of this model is 
\begin{eqnarray}
S[A_{\mu},g]=\frac{1}{2}\int d^{4}x \ \mbox{Tr} (F_{\mu\nu}F_{\mu\nu}),
\label{SU(2)}
\end{eqnarray}
where we define the field strength as,
\begin{eqnarray}
F_{\mu\nu}=\partial_{\mu} A_{\nu} - \partial_{\nu} A_{\mu}
-ig[A_{\mu},A_{\nu}].
\end{eqnarray}
In this model, the effective equation of motion (\ref{eq:6-23}) 
that determines the asymptons reads as follows;
\begin{eqnarray}
\mbox{Im}[(\partial_{\nu} F_{\mu\nu}+ig_{-}[F_{\mu\nu},A_{\nu}])e^{i\sigma}]=
0.
\label{eq:6-29}
\end{eqnarray}
We solve this equation (\ref{eq:6-29}) under the following ansatz;
\begin{eqnarray}
A_{\mu}(x)=\frac{-i}{\sqrt{48\pi^{2}}}
\frac{(\bar{\sigma}_{\mu}\sigma_{\nu}-\delta_{\mu\nu}) x_{\nu}}{x^2} s(\tau) 
\label{eq:6-210}
\end{eqnarray}
where $\tau = \log x^{2}$ ($x^2=\sum_{i=1}^4 x_i^2$) and 
using the Pauli matrices $\sigma_\mu$, 
\begin{eqnarray}
\bar{\sigma}_{\mu} = (-i,\sigma^{a}),\quad
\sigma_{\mu} = (i,\sigma^{a}).
\end{eqnarray}
This ansatz is motivated by the fact that the instanton,
anti-instanton, and instanton-anti-instanton pairs (valley)
that shares a common center $x_\mu=0$, can be described
using this ansatz. The existence of interplay between
perturbative and non-perturbative effects implies that
they share similar structure. We will see that it is
evident from the analysis of the asymptons.

By substituting (\ref{eq:6-210}) in (\ref{eq:6-29}), 
we find that (\ref{eq:6-29}) reduces to the following single equation;
\begin{eqnarray}
\mbox{Im}\left [\left ( \left ( \frac {d^{2} s}{d\tau^{2}}\right )
-2\tilde{g}_{-}^{2}s^{3}+3\tilde{g}_{-}s^{2}-s\right ) e^{i\sigma} \right ]=0,
\label{effective}
\end{eqnarray}
where $\tilde{g}_{-}=g_{-}/\sqrt{48\pi^{2}}$.
It is important to stress here that this is obtained not 
by narrowing our search for the asympton down to 
the subspace described by the
ansatz Eq.~(\ref{eq:6-210}); as long as  Eq.~(\ref{eq:6-210})
is satisfied, the full asympton equation Eq.~(\ref{eq:6-29})
is satisfied. Therefore if we find a solution to  Eq.~(\ref{eq:6-210}),
it is a peak of the perturbative functional $F_n[\phi]$
in the full functional space.
This equation can be written as follows 
by using the scaled variable $\tilde{s}(\tau)\equiv |\tilde{g}_{-}|s(\tau)$:
\begin{eqnarray}
-\frac{\partial^{2}\tilde{s} }{\partial \tau^{2}}
+\frac {\partial V_{\rm{eff}}(\tilde{s})}{\partial\tilde{s}}
=0,
\label{effective motion eq2}
\end{eqnarray}
where the effective potential $V_{\rm{eff}}(\tilde{s})$ 
is defined by 
\begin{eqnarray}
V_{\rm{eff}}(\tilde{s})=\frac{1}{2} \tilde{s}^{2} -\kappa_{3}\tilde{s}^{3}+\frac
{1}{2}\kappa_{4}\tilde{s}^{4}.
\end{eqnarray}
This effective equation of $\tilde{s}$ equals to 
that of the quantum mechanics (\ref{eq:6-2b}). 
Therefore we can use the result of the quantum mechanics. 
This way the asymptons are identified in the SU(2) gauge field theory.
Just as the asymptons in the quantum mechanical model were 
quite different from instantons or instanton pairs, and rather
like the bounces, the asymptons in the SU(2) gauge field theory
are not like instantons at all. 
It has non-zero field strength on a spherical shell,
and look like a merged concentric instanton-anti-instanton pair.

\subsubsection{Asympton on the Valley}
\newcommand{\rsim}{\,\lower0.9ex\hbox{$\stackrel{\displaystyle <}{\sim}$}\,}
From the above we have found that
asymptons are quite unlike the instanton configurations in 
general.  They are off the valley line studied extensively
in the previous sections.
On the other hand, we have seen in section 6.1 
that the consideration of valley provides an insight into
the summability properties of the perturbative series.
Thus we will next study the behavior of the perturbative functional
$F_n [\phi]$ on the valley.\footnote{Similar analysis
were presented in Ref.~\citen{rf:A}.
However, there was a mistake of factor two in one of the 
expression obtained in that paper.  This changes the 
behavior of $F_n[\phi]$ drastically.}
In order to simplify the analysis, 
we adopt the following approximation for valley
in the double-well quantum mechanical model;
\begin{eqnarray}
s(\tau)={1 \over \tilde g}{(1+e^{-\tau-d/2})(1+e^{\tau-d/2})},
\label{ansatz2}
\end{eqnarray}
In (\ref{ansatz2}), if we choose $d$ large enough, 
we obtain instanton-anti-instanton pair.  It gives a
reasonable approximation of the valley line parametrized by $d$. 

In the SU(2) gauge field theory,
the valley has the tensor structure (\ref{eq:6-210}).
If we substitute this expression into the action,
we find that 
\begin{equation}
S[A_{\mu},g]= \int_{-\infty}^\infty d \tau
\left[ {1 \over 2} \dot s^2 - {1 \over 2} s^2 (1-\tilde g s)^2 \right],
\end{equation}
where the $\tau$-integration comes from the integration 
over the four-dimensional radius from zero to infinity, 
and $\tilde g \equiv g / \sqrt{48 \pi^2}$.
Since the right hand side is equivalent to the
double-well quantum mechanical model with (imaginary) time $\tau$,
and the coupling constant $\tilde g$,
the following analysis is also valid for the SU(2) gauge field
theory by rewriting $g$ by $\tilde g$.

We first calculate the ratio of $F_{n}[A_{\mu}] g^{n}$ to $e^{-S}$ 
on the valley. 
Using (\ref{eq:6-22}) for Eq.~(\ref{eq:6-21}), 
we find it to be the following;
\begin{eqnarray}
\frac{F_{n}[\phi] g^{n}}{e^{-S}}=e^{\Delta}\sin \sigma,
\end{eqnarray}
where 
\begin{equation}
\Delta = 
\frac{c_{1}^{2}}{8c_{2}}+\frac{n}{2}-gc_{1}+g^{2}c_{2}
-\frac{n}{2}\log\frac{n}{2g^{2}c_{2}}
-\frac{1}{2}\log(\pi n\sin \theta).
\label{eq:dai}
\end{equation}
The magnitude of the perturbation coefficient is determined by $e^{\Delta}$. 
Therefore, we examine the behavior of $e^{\Delta}$ for large $d$.
The coefficients $c_{1,2}$ are approximated as follows for large $d$;
\begin{eqnarray}
\begin{array}{cc}
&c_{1}=\displaystyle\frac{1}{g^3} \left(d-3+\cdots\right),
\\[0.5cm]\label{eq:c1c2}
&c_{2}=\displaystyle\frac{1}{2g^4}\left(d-\frac{11}{3}+\cdots\right).
\end{array}
\end{eqnarray}
Using (\ref{eq:c1c2}) for $\Delta$ in (\ref{eq:dai}), we find that
\begin{eqnarray}
\Delta\sim -\frac{d}{4 {g}^{2}}+\frac{n}{2}+\frac{n}{2}
\log\frac{d}{n {g}^{2}}-\frac{1}{2}\log\pi n+
\frac{1}{4}\log\left( 1-\frac{d}{4 n {g}^{2}}\right).
\label{delta2}
\end{eqnarray}
From (\ref{delta2}), we find that $e^{\Delta}$ has the peak
on $d=2n {g}^{2}$ and 
that the magnitude of the peak is proportional to $2^{n/2}/\sqrt{n}$.
We will call this peak ``the valley asympton'', as it is the
asympton solution when limited to the valley line.
As $n$ increases, the valley asympton 
moves to the region of large $d$ and grows exponentially.  
The numerical analysis shows that the 
main contribution of the asympton is in the following region: 
\begin{eqnarray}
0.46 n {g}^{2}+\frac{11}{3}\rsim d \rsim 5.34 n {g}^{2}+
\frac{11}{3}.\label{region}
\end{eqnarray} 
We can obtain this result by solving the equation $\Delta =0$ numerically.  
The perturbative functional $F_{n}[\phi]$ on the valley is 
plotted in Fig.~\ref{c63-pe}.

\begin{figure}[htb]
\parbox{\halftext}{\centerline{\epsfxsize=7cm\epsfbox{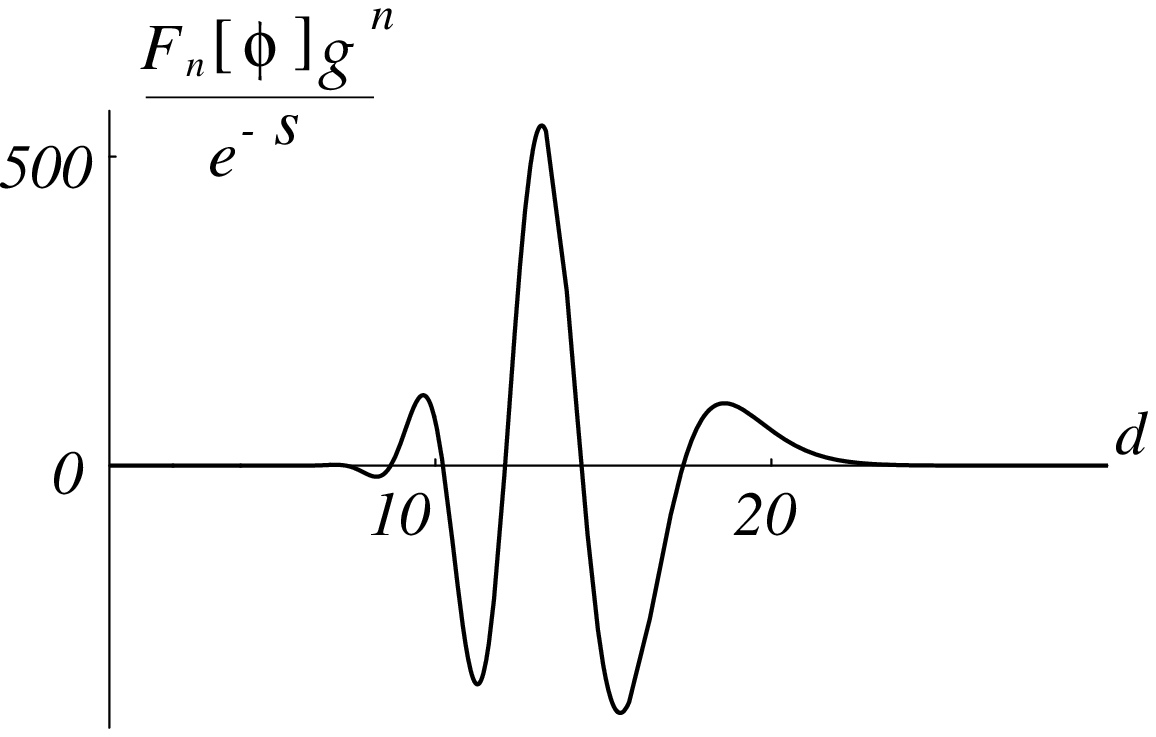}}
\caption{The behavior of \protect $F_{n}[\phi]g^{n}/e^{-S}$ for $n=30$,
$g=0.4$.}
\label{c63-pe}}
\hspace{4mm}
\parbox{\halftext}{\centerline{\epsfxsize=7cm\epsfbox{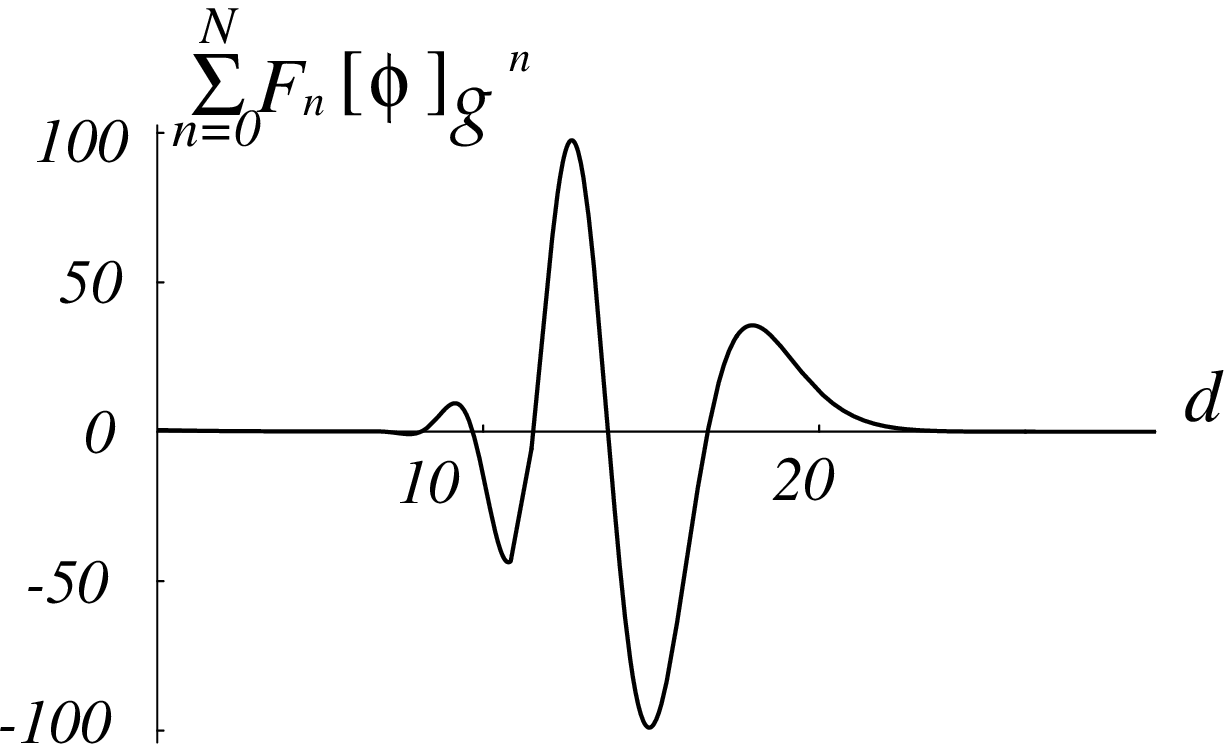}}
\caption{The behavior of \protect $\sum_{n=0}^{N}F_{n}[\phi]g^{n}$ for $N=30$,
$g=0.4$. }
\label{c63-all}}
\end{figure}

The numerical comparison of the partial sum, $\sum_{n=0}^{N} F_{n}[\phi] g^{n}$ 
with  $e^{-S}$ is plotted in 
Figs.~\ref{c63-all}, \ref{c63-k1} (for $N=30$), and 
\ref{c63-k2} (for $N=50$). (Note the difference in the vertical scales
between the first two and the latter.)

\begin{figure}[htb]
  \parbox{\halftext}{
       \centerline{\epsfxsize=7cm\epsfbox{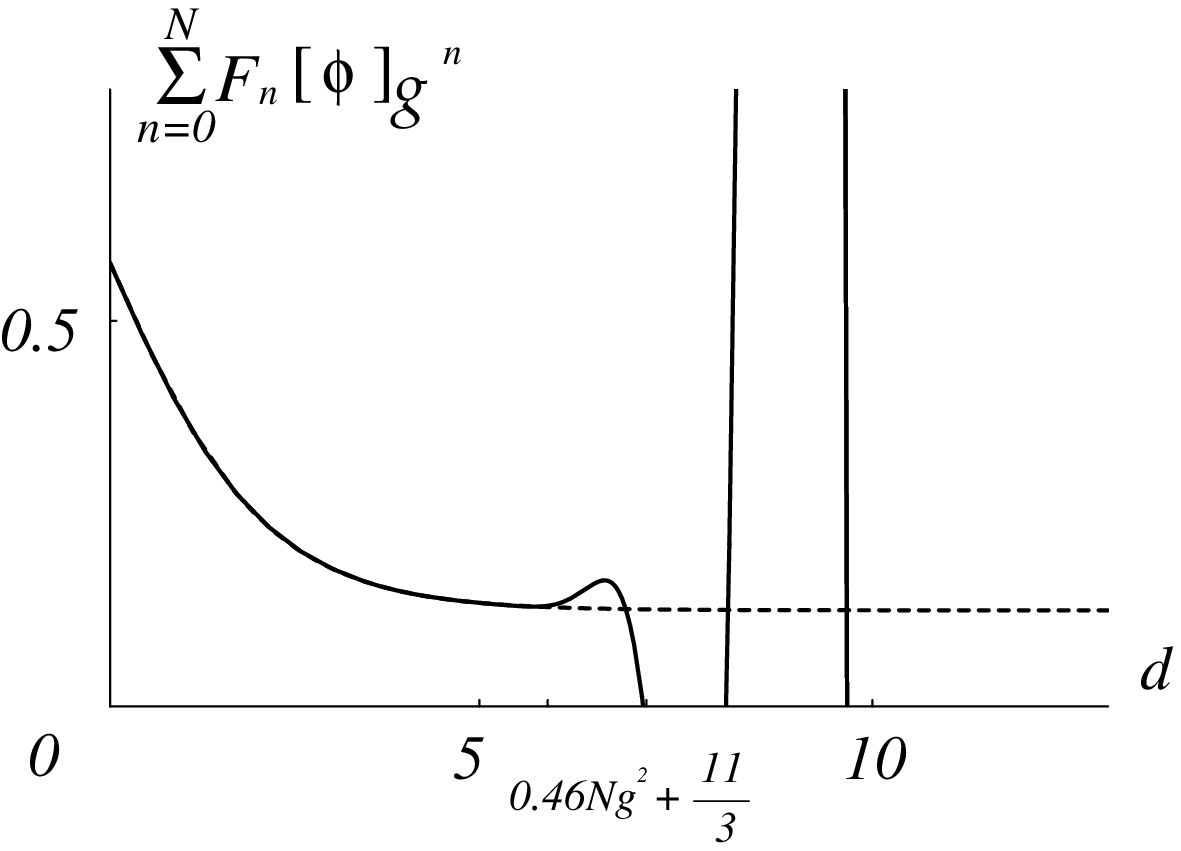}}
       \caption{The dashing line is \protect $e^{-S}$. The solid line 
is $\sum^{N}_{n=0}F_{n}[\phi]g^{n}$ for $N=30$, $g=0.4$.}
\label{c63-k1}}
\hspace{4mm}
  \parbox{\halftext}{
      \centerline{\epsfxsize=7cm\epsfbox{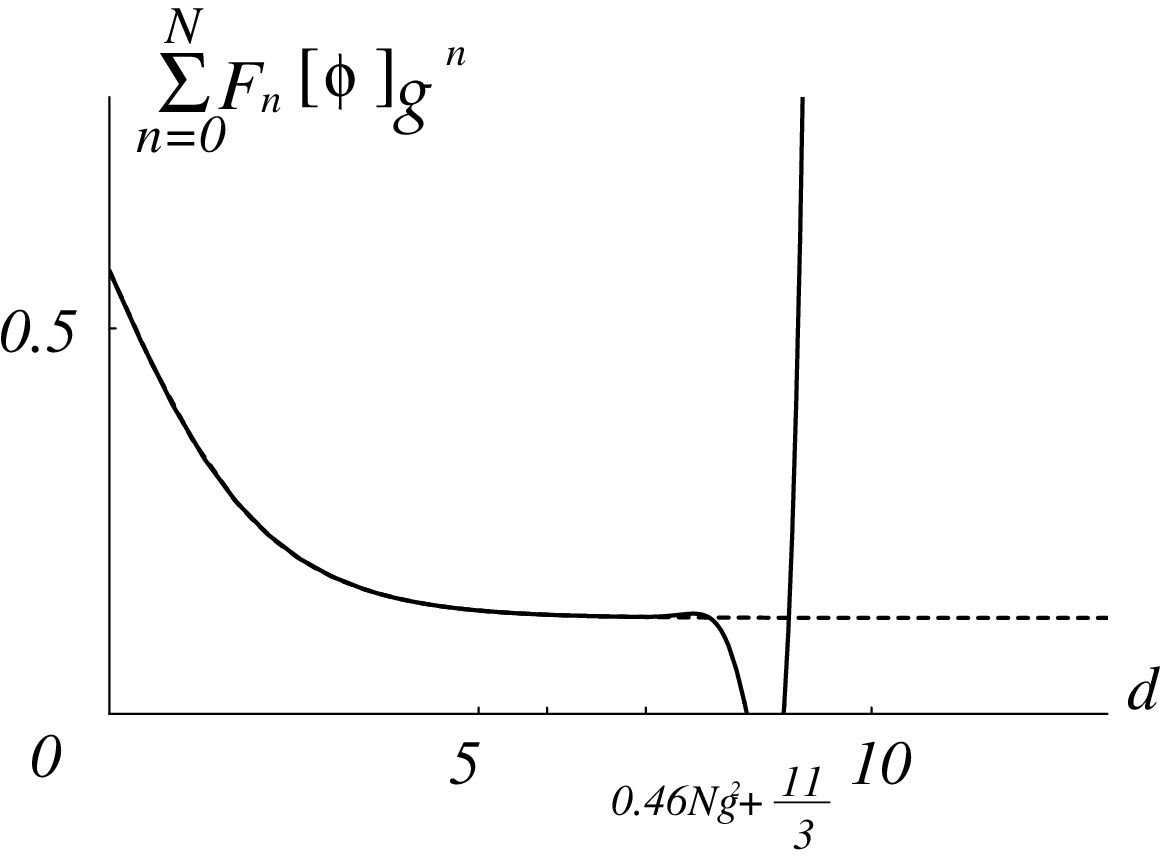}}
      \caption{The dashing line is \protect $e^{-S}$. The solid line 
is $\sum^{N}_{n=0}F_{n}[\phi]g^{n}$ for $N=50$, $g=0.4$.}
\label{c63-k2}}
\end{figure}

From Figs.~\ref{c63-all}, \ref{c63-k1}, 
we find that $\sum_{n=0}^{N} F_{n}[\phi] g^{n}$
deviates from $e^{-S}$ where 
$F_{n}[\phi]$ has the large contribution. 
For $d\rsim 0.46N{g}^{2}+\frac{11}{3}$, 
we find that $\sum_{n=0}^{N} F_{n}[\phi] g^{n}$ (the solid line) 
is very close to $e^{-S}$ (the dashed line).
This means that as far as the valley is concerned
the $n$-th order perturbation is valid only in the region
$d\rsim 0.46 N {g}^{2}+\frac{11}{3}$. 
The usual perturbation suffers from non-Borel summability
due to the valley asympton, even when limited to the valley line.

\subsubsection{Discussion}
We have so far identified the causes
of the Borel non-summability as asymptons and valley asymptons.
the next problem is use this knowledge to look into
the interplay of the perturbative series and the non-perturbative terms.
In the double-well potential model, Aoyama and Tamra\cite{rf:AT,rf:A,rf:S}
proposed a modified perturbation theory, obtained by 
dividing the functional space into two
fundamental regions of the symmetry $\phi \rightarrow 1/g - \phi$,
and carrying out the functional integral in only one of them.
The original path-integral is recovered by multiplying two 
on that integral, as the contribution from the fundamental
regions are the same.
As the order of the perturbation exceeds O($1/g^2$),
the asympton moves out from a fundamental region into another.
Thus the modified perturbation series is free from the asympton
at higher orders.  What remains is the anti-asympton, which
have the property (\ref{eq:fpmrel}).
This still renders the series, non-convergent.
But due to the disappearance of the asympton, the cancelation
for $n=$ odd no longer holds, and the modified perturbative
coefficient $\tilde{z}_n$ behaves as,
\begin{equation}
\tilde{z}_n \sim (-1)^n n^{n/2},
\end{equation}
for $n \rightarrow \infty$. Thus this series is a 
Borel-summable series,
which is quite desirable for many purposes.

This method, however, works only for the finite time-interval
(or finite temperature).
When the time-interval is infinity, the boundary of the
fundamental region is at infinity and the asympton never
moves out of the original region.
Furthermore, we do not know how to make it work
in the SU(2) gauge theory.

The valley asympton gives us a hope for possible resolution of the
problem.
It shows that (at least on the valley line)
the perturbative series suffers from non-summability
due to the configurations (valley asymptons) that live in 
$d > O(1)\times n/g^2$ region.
This means that if we introduce a cutoff on $d$ at O(1),
the perturbative coefficients begins to converge for  $n < O(1)/g^2$.
Since this is the order when the usual leading term, $n^{n/2}g^n$, 
is of O(1), this perturbative series with the cutoff never 
suffers from any large terms.
On the other hand, the region excluded by this cutoff
is where the instanton and the anti-instanton is separated well.
Thus the non-perturbative calculation (valley method)
provides a  quite effective evaluation of the path-integral.

What is to be done is to find a way to apply this kind of the cutoff method 
to the whole region with all the virtues seen on the valley.
This is a fascinating possibility, but remains as
an open question at this moment.

\section*{Acknowledgments}
We would like to thank our colleagues at Kyoto University 
for encouragements and
discussions at various stages of this work.
H.~Aoyama's work is supported in part by the Grant-in-Aid
for Scientific Research (C)-07640391 and (C)-08240222.
T.~Harano and S. Wada's work is supported in part 
by the Grant-in-Aid for JSPS fellows.
H.~Kikuchi's work is supported in part by the Grant-in-Aid
for Encouragement of Young Scientists, 08740215.

\newpage
\addcontentsline{toc}{section}{References}
%
%


\begin{thebibliography}{99}

\bibitem{rf:manton} N.~S.~Manton, \PR{D28,1983,2019};
F.~R.~Klinkhamer and N.~S.~Manton, \PR{D30,1984,2212}.
\bibitem{rf:thooft} G.~'t Hooft, \PRL{37,1976,8}; \PR{D14,1976,3432}.
\bibitem{rf:kuz} V.~A.~Kuzmin, V. A. Rubakov, and M. E. Shaposhnikov,
\PL{B155,1985,36}.
\bibitem{rf:ag} H.~Aoyama and H.~Goldberg, \PRL{188B,1987,506}.
\bibitem{rf:ringwald} A.~Ringwald, \NP{B330,1990,1}.
\bibitem{rf:espinosa} O.~Espinosa,  \NP{B343,1990,310}.
\bibitem{rf:ii} H.~Aoyama and H.~Kikuchi, \PL{B247,1990,75};
\PR{D43,1991,1999};\IJMP{A7,1992,2741}.
\bibitem{rf:rubsha} V.~A.~Rubakov and M.~E.~Shaposhnikov,
hep-th/9603208.
\bibitem{rf:clec} S.~Coleman, in {\it The Why's of Subnuclear 
Physics}, ed.~A.~Zichichi (Plenum, New York, 1979).
\bibitem{rf:Aff} I.~Affleck, \NP{B191,1981,429}.
\bibitem{rf:pvalley} H.~Aoyama and H.~Kikuchi, \NP{B369,1992,219}.
\bibitem{rf:silv} P.~G.~Silvestrov, \JL{Sov.~J.~Nucl.~Phys.,51,1990,1121}.
\bibitem{rf:originals}
D.~J.~Rowe and A.~Ryman, J.~Math.~Phys.\ {\bf 23} (1982), 732.
\bibitem{rf:oldoriginals}
C.~A.~Cayley, Philos.~Mag.\ {\bf 18} (1859), 264;
J.~C.~Maxwell, Philos.~Mag.\ {\bf 40} (1870), 421. 
\bibitem{rf:quapp} W.~Quapp, Chemical Physics Letters {\bf 253} (1996), 286.
\bibitem{rf:balitsky} I.~I.~Balitzky and A.~V.~Yung, \PL{B168,1986,113};
\NP{B274,1986,475}.
\bibitem{rf:boy} D.~Boyanovsky, R.~Willey and R.~Holman, \NP
{B376,1992,599}.
\bibitem{rf:coleman} S.~Coleman, \PR{D15,1977,2929}.
\bibitem{rf:callancoleman} C.~Callan and S.~Coleman, \PR
{D16,1977,1762}.
\bibitem{rf:bl} D.~E.~Brahm and C.~L.~Y.~Lee, \PR{D49,1994,4094}.
\bibitem{rf:ahsw} H. Aoyama, T.~Harano, M.~Sato and S.~Wada,\
Mod.~Phys.~Lett.~{\bf A11} (1995), 43; \NP{B466,1996,127}.
\bibitem{rf:aq} H.~Aoyama and H.~R.~Quinn, \PR{D31,1985,885};
\andvol{D34,1986,662}.
\bibitem{rf:aw} H.~Aoyama and S.~Wada, \PL{B349,1995,279}
\bibitem{rf:mattis} P.~B.~Arnold and M.~P.~Mattis, \PRL{66,1991,13}.
\bibitem{rf:yung} A.~V.~Yung, \NP{B191,1981,47}.
\bibitem{rf:khozering} V.~V.~Khoze and A.~Ringwald, \PL{B259,1991,106}.
\bibitem{rf:asym} L.~N.~Lipatov, \JL{JETP Lett.,25,1977,104};
\JL{Sov.~Phys.~JETP,45,1977,216}.
\bibitem{rf:brezin} E.~Br\'ezin, J.~C.~Le Guillou and J.~Zinn-Justin,
\PR{D15,1977,1544},1558.
\bibitem{rf:parisi} G.~Parisi, \PL{B66,1977,167}.
\bibitem{rf:fy} Y.~Frishman and S.~Yankielowicz, \PR{D19,1979,540}.
\bibitem{rf:hs} T.~Harano and M.~Sato, {\sl Kyoto University preprint,}
KUNS-1391 HE(TH)96/04.
\bibitem{rf:hk} A.~Hosoya and K.~Kikkawa, \NP{B101,1975,271}.
\bibitem{rf:mcl} D.~W.~McLaughlin, \JMP{13,1972,1099}.
\bibitem{rf:bender} I.~Bender, D.~Gomez, H.~Rothe and K.~Rothe,
\NP{B136,1978,259}.
\bibitem{rf:levi} S.~Levit, J.~W.~Negele and Z.~Paltiel, \PR
{C22,1980,1979}.
\bibitem{rf:patra} A.~Patrascioiu, \PR{D24,1981,496}.
\bibitem{rf:lap} A.~Lapedes and E.~Mottola, \NP{B203,1982,58}.
\bibitem{rf:weiss} U.~Weiss and W.~Haeffner, \PR{D27,1983,2916}.
\bibitem{rf:carlitz} R.~D.~Carlitz and D.~A.~Nicole,
\JL{Ann.~Phys.,164,1985,411}.
\bibitem{rf:khleb} S.~Y.~Khlebnikov, V.~A.~Rubakov and P.~G.~Tinyakov,
\NP{B367,1991,334}.
\bibitem{rf:son} V.~A.~Rubakov, D.~T.~Son and P.~G.~Tinyakov,
\PL{B287,1992,342}.
\bibitem{rf:rubakoc} D.~T.~Son and V.~A.~Rubakov, \NP{B424,1994, 55}
\bibitem{rf:ah} H.~Aoyama and T.~Harano, Mod.~Phys.~Lett.~{\bf A10} 
(1995),1136; \NP{B446,1955,315}.
\bibitem{rf:ak} H.~Aoyama and M.~Kobayashi, \PTP{64,1980,1045}.
\bibitem{rf:dyson} F.J.~Dyson, \PR{85,1952,631}.
\bibitem{rf:large} J.C.~Le~Guillou and J.~Zinn-Justin, 
ed.~{\it Large-order behavior of perturbation theory} 
(North Holland, Amsterdam, 1990) and references contained therein.
\bibitem{rf:kiku} H.~Kikuchi, \PR{D45,1992,1240}.
\bibitem{rf:benwu} C.~Bender and T.~Wu, \PR{D7,1973,1620}.
\bibitem{rf:nonB} E.~Br\'ezin, G.~Parisi, and J.~Zinn-Justin, 
\PR{D16,1977,408}.
\bibitem{rf:tHftB} G.~'t~Hooft, in {\it The Whys of Subnuclear
Physics} ed.~A.~Zichichi (Plenum, New York, 1979).
\bibitem{rf:AA} H.~Aoyama, Mod.~Phys.~Lett.~{\bf A7} (1992), 1337.
\bibitem{rf:AT} H.~Aoyama and A.~M.~Tamra, \NP{B384,1992,229}.
\bibitem{rf:A} H.~Aoyama, \PL{B329,1994,285}.
\bibitem{rf:S} H.~Suzuki,  Mod.~Phys.~Lett.~{\bf A11} (1996), 19.
\end{thebibliography}
\end{document}